\documentclass[12pt]{iopart}
\usepackage{iopams}  
\usepackage{graphicx}
\usepackage{color}
\usepackage{epsfig}
\usepackage{tabularx}
\usepackage[numbers]{natbib}
\usepackage{hyperref}
\usepackage{todonotes}
\usepackage{rotating}
\graphicspath{{./figures/}}

\begin{document}

\title[Connecting RIB facilities with the cosmos]{$r$-Process Nucleosynthesis: Connecting Rare-Isotope Beam Facilities with the Cosmos}

\author{C J Horowitz$^{1,38}$, A Arcones$^{6,36,38}$, B C\^ot\'e$^{4,31,38}$, I Dillmann$^{10,11,38}$, W Nazarewicz$^{4,23}$, I U Roederer$^{26,38}$, H Schatz$^{4,30,38}$, A Aprahamian$^{5,38}$,  D Atanasov$^7$, A Bauswein$^8$,  J Bliss$^6$, M Brodeur$^{5,38}$, J A Clark$^{9,38}$, A Frebel$^{12,38}$, F Foucart$^{13}$, C J Hansen$^{14}$, O Just$^{37,15}$, A Kankainen$^{16}$, G C McLaughlin$^{3,38}$, J M Kelly$^5$, S N Liddick$^{17,30,38}$, D M Lee$^{12,18,19}$, J Lippuner$^{33,34,35,38}$, D Martin$^6$, J Mendoza-Temis$^{20,21}$, B D Metzger$^2$, M R Mumpower$^{22,38}$,  G Perdikakis$^{23,24,38}$, J. Pereira$^{30,38}$, B W O'Shea$^{4,32,38}$,  R Reifarth$^{25}$, A M Rogers$^{27}$, D M Siegel$^2$, A Spyrou$^{4,30,38}$, R Surman$^{5,38}$, X Tang$^{28}$, T Uesaka$^{29}$, M Wang$^{28}$}

\address{$^1$ Department of Physics and Center for the Exploration of Energy and Matter, Indiana University, Bloomington, IN 47405, USA}
\address{$^2$ Columbia Astrophysics Laboratory, Columbia University, New York, NY, 10027, USA}
\address{$^3$ Department of Physics, North Carolina State University, Raleigh, NC, 27695-8202, USA}
\address{$^4$ Department of Physics and Astronomy, Michigan State University, East Lansing, MI, 48864, USA}
\address{$^5$ Physics Department, University of Notre Dame, Notre Dame, IN 46556}
\address{$^6$ Institut f\"{u}r Kernphysik, Technische Universit\"{a}t Darmstadt, Schlossgartenstra\ss e 2, 64289 Darmstadt, Germany}
\address{$^7$ Max-Planck Institute for Nuclear Physics, Sapfercheckweg 1, 69117 Heidelberg, Germany}
\address{$^8$ Heidelberg Institute for Theoretical Studies, Schloss-Wolfsbrunnenweg 35, 69118 Heidelberg, Germany}
\address{$^{9}$ Physics Division, Argonne National Laboratory, Argonne, IL  60439,  USA}
\address{$^{10}$ TRIUMF, 4004 Wesbrook Mall, Vancouver BC V6T 2A3, Canada}
\address{$^{11}$ Department of Physics and Astronomy, University of Victoria, Victoria BC V8W 2Y2, Canada}
\address{$^{12}$ MIT, USA}
\address{$^{13}$ Department of Physics, University of New Hampshire, Durham, New Hampshire 03824, USA}
\address{$^{14}$ Dark Cosmology Centre, University of Copenhagen, Juliane Maries Vej 30, 2100 Copenhagen, Denmark}
\address{$^{15}$ Max-Planck-Institut f\"ur Astrophysik, Karl-Schwarzschild Str. 1, 85748 Garching, Germany}
\address{$^{16}$ University of Jyv\"askyl\"a, P.O. Box 35, FI-40014 University of Jyv\"askyl\"a, Jyv\"askyl\"a, Finland}
\address{$^{17}$ Department of Chemistry, Michigan State University, East Lansing, MI, 48864, USA}
\address{$^{18}$ Department of Physics, Fisk University, Nashville, TN 37208}
\address{$^{19}$ Department of Physics \& Astronomy, Vanderbilt University, Nashville, TN 37212, USA}
\address{$^{20}$ Facultad de Fisica, Universidad Veracruzana, Xalapa; Mexico}
\address{$^{21}$ Instituto de Ciencias Nucleares y Centro de Ciencias de la Complejidad, Universidad Nacional Autonoma de M\'{e}xico, Circuito Cultural de Ciudad Universitaria S/N, 04510 Cd. de M\'{e}xico, M\'{e}xico}
\address{$^{22}$ Theoretical Division, Los Alamos National Laboratory, Los Alamos, NM 87545, USA}
\address{$^{23}$ FRIB Laboratory, Michigan State University, East Lansing, Michigan 48824, USA}
\address{$^{24}$ Department of Physics, Central Michigan University, Mt. Pleasant, MI 48859, USA}
\address{$^{25}$ Goethe University Frankfurt, Frankfurt, Germany}
\address{$^{26}$ Department of Astronomy, University of Michigan, 1085 S.\ University Ave., Ann Arbor, MI 48109, USA}
\address{$^{27}$ Dept. of Physics and Applied Physics, University of Massachusetts Lowell, Lowell, MA 01854}
\address{$^{28}$ Institute of Modern Physics, Chinese Academy of Sciences, Lanzhou, Gansu 730000, P. R. China}
\address{$^{29}$ RIKEN Nishina Center for Accelerator-Based Science, Wako, Saitama 351-0198, Japan}
\address{$^{30}$ National Superconducting Cyclotron Laboratory, Michigan State University, East Lansing, Michigan 48824, USA}
\address{$^{31}$ Konkoly Observatory, Research Centre for Astronomy and Earth Sciences, Hungarian Academy of Sciences, Konkoly Thege Miklos ut 15-17, H-1121 Budapest, Hungary}
\address{$^{32}$ Department of Computational Mathematics, Science, and Engineering, Michigan State University, East Lansing, MI, 48864, USA}
\address{$^{33}$ CCS-2, Los Alamos National Laboratory, P.O. Box 1663, Los Alamos, NM 87545, USA}
\address{$^{34}$ Center for Nonlinear Studies, Los Alamos National Laboratory, P.O. Box 1663, Los Alamos, NM 87545, USA}
\address{$^{35}$ Center for Theoretical Astrophysics, Los Alamos National Laboratory, P.O. Box 1663, Los Alamos, NM, 87545, USA}
\address{$^{36}$ GSI Helmholtzzentrum f\"{u}r Schwerionenforschung GmbH, Planckstr. 1, Darmstadt 64291, Germany}
\address{$^{37}$ Astrophysical Big Bang Laboratory, RIKEN, Saitama 351-0198, Japan}
\address{$^{38}$ Joint Institute for Nuclear Astrophysics Center for the Evolution of the Elements JINA-CEE}
\date{\today}
\begin{abstract}
This is an exciting time for the study of $r$-process nucleosynthesis. Recently, a neutron star merger GW170817 was  observed in extraordinary detail with gravitational waves and electromagnetic radiation from radio to $\gamma$ rays.  The very red color of the associated kilonova suggests that neutron star mergers are an important $r$-process site. Astrophysical simulations of neutron star mergers and core collapse supernovae are making rapid progress. Detection of both, electron neutrinos and antineutrinos from the next galactic supernova will constrain the composition of neutrino-driven winds and provide unique nucleosynthesis information. Finally FRIB and other rare-isotope beam facilities will soon have dramatic new capabilities to synthesize many neutron-rich nuclei that are involved in the $r$-process. The new capabilities can significantly improve our understanding of the $r$-process and likely resolve one of the main outstanding problems in classical nuclear astrophysics.  However, to make best use of the new experimental capabilities and to fully interpret the results, a great deal of infrastructure is needed in many related areas of astrophysics, astronomy, and nuclear theory. We will place these experiments in context by discussing astrophysical simulations and observations of $r$-process sites, observations of stellar abundances, galactic chemical evolution, and nuclear theory for the structure and reactions of very neutron-rich nuclei. This review paper was  initiated at a three-week International Collaborations in Nuclear Theory program in June 2016 where we explored promising $r$-process experiments and discussed their likely impact, and their astrophysical, astronomical, and nuclear theory context.

\end{abstract}

\pacs{00.00, 20.00, 42.10}
\submitto{\jpg}
%
%
\maketitle

{\tableofcontents{}}
\title[Connecting RIB facilities with the cosmos]{$r$-Process Nucleosynthesis: Connecting Rare-Isotope Beam Facilities with the Cosmos}

%






\section{Introduction}


How were the elements from Iron to Uranium made?  The influential National Academy of Science report ``Connecting Quarks to the Cosmos'' identified this question as one of eleven questions at the intersections of astronomy and physics that are of deep interest and are ripe for answering \cite{NAP10079}.  Ever since the pioneering works of Burbidge, Burbidge, Fowler and Hoyle \cite{BBFH} and Al Cameron \cite{Cameron+57} we think these elements are made predominantly by both the slow neutron capture process (or $s$-process) \cite{Sprocess} and by the rapid neutron capture process (or $r$-process) \cite{Arn07}, where seed nuclei capture neutrons more rapidly than many $\beta$-decays.     

One of the major grand challenges of our day is the determination of the site or sites for the $r$-process and therefore the identification of the origin of more than half of all the elements heavier than iron. The answer is complex and highly intermingled between the astrophysics that provides a description of the conditions of the relevant scenarios and the physics of nuclei that operates in those scenarios. Connecting rare isotopes to the Cosmos is an ambitious, yet a feasible, attempt to infer the nature of the extreme stellar environments where the $r$-process occurs by determining important properties of very neutron-rich heavy nuclei that can be produced at the Facility for Rare Isotope Beams (FRIB) and other radioactive ion accelerators.  This nuclear information can then be incorporated in detailed astrophysical simulations to make predictions for the elemental and isotopic abundances produced and for the emitted gravitational waves, neutrinos and electromagnetic radiations. 

Our understanding of the $r$-process has recently taken a dramatic turn with the extraordinary multi-messenger observations of gravitational wave and $\gamma$-ray, x-ray, ultra-violet, visible, infrared and radio radiations from the neutron star merger GW170817 \cite{GW170817}.  The very red color of this event that was observed two or more days after the merger, peaking in the infrared, has been interpreted as evidence for the production of lanthanides via the $r$-process \cite{2017ApJ...848L..19C}.  Furthermore, the amount of material ejected and the rate of neutron star mergers suggest that these mergers are a, perhaps dominating, site of $r$-process nucleosynthesis \cite{Kasen:2017sxr}. This is in line with earlier observations of strongly $r$-process enriched stars in the dwarf galaxy Reticulum II that are best explained by a neutron star merger $r$-process \cite{ji16nat}, the detection of live interstellar $^{244}$Pu archived in terrestrial reservoirs like deep-sea crusts  \cite{Wallner2015} that indicates a rare prolific $r$-process production site such as neutron star mergers, and the finding that earlier arguments against a neutron star merger $r$-process based on theoretical galactic chemical evolution models (e.g. \cite{argast04}) are less constraining than originally assumed (for example \cite{shen15,2015ApJ...804L..35I}). Observations of mergers are discussed in Sec. \ref{sect_grav_waves_nsms}, while Sec. \ref{subsection_GCE_grav_wave} discusses the role of mergers for galactic chemical evolution, and simulations of neutron star mergers are discussed in Sec. \ref{sec.mergers}. The improved observational constraints on the $r$-process site make the need for accurate nuclear data even more pressing as the nuclear physics increasingly becomes the major missing piece in the puzzle of the origin of the elements. For example, nuclear physics will be needed to infer the physical conditions in neutron star mergers that lead to the observed $r$-process features, to disentangle contributions from different ejecta components in neutron star mergers, and to identify the role that alternative $r$-process sites may still play. 

This review is organized as follows: Section \ref{sect_observations} discusses a large variety of observations related to the $r$-process.  Galactic chemical evolution simulations are reviewed in Sec. \ref{subsection_GCE_MS}, while Sec. \ref{section_AS}  reviews astrophysical simulations of $r$-process sites.  Given conditions present in a site, one can perform detailed nuclear reaction network simulations to predict nucleosynthetic yields.  Here, an accurate understanding of the relevant nuclear physics is required for any comprehensive theory of heavy element formation. Nuclear physics enables the calculation of the characteristic abundance patterns produced in a particular astrophysical $r$-process model and is thus a prerequisite for a full understanding of all the elements a particular site may produce, and for validating the site model through comparison with abundance observations.  Nuclear physics is also essential to use nucleosynthesis observations to obtain information and constraints on the extreme environment of the nucleosynthesis site, such as temperature and density evolution, neutron-richness, hydrodynamic mixing processes, or neutrino physics. Last but not least, only with reliable nuclear physics will one ultimately arrive at an understanding of the actual mechanism of element formation. The challenge for understanding the $r$-process is that the relevant nuclei are very neutron-rich and knowledge of their properties and reactions therefore extremely limited. 

The sensitivity of nucleosynthesis yields to nuclear physics and its uncertainties is discussed in Sec. \ref{section_NSS}. Nuclear theory has made great strides in describing the properties of neutron-rich nuclei, as discussed in Sec. \ref{sec_theory}, but is still far from making predictions of structure and reactions with  accuracy that is needed for astrophysical applications. Experimental data are therefore essential for the key nuclear physics ingredients in $r$-process calculations, and techniques and approaches to obtain these data are reviewed in Sec. \ref{sec_exp}. Experimental data on neutron-rich isotopes can also guide the development of quantified theoretical models capable of making  reliable predictions for all nuclei involved in the $r$-process. Obtaining the critical experimental data will be an iterative process that requires close interaction between experimentalists, nuclear theorists, and astrophysicists. As knowledge of nuclear physics, astrophysics, and astronomical observables evolves, sensitivity studies linking nuclear physics with observables will change, in turn changing experimental and theoretical nuclear physics priorities. Making this interaction cycle between astrophysicists and nuclear physicists as efficient and straightforward as possible will be key for success.   Also key for success are the radioactive beam facilities that have capabilities to perform $r$-process experiments as reviewed in Sec. \ref{sec_facilities}.  Finally Sec. \ref{sec_summary} presents a summary and outlook.


\section{Observations}
\label{sect_observations}

\subsection{Observations of stellar abundances}
\label{sect_obs}


The heavy elements in the atmospheres of most late-type (F-G-K) stars, which have effective temperatures of $\approx$~4000--7000~K, reflect the stars' natal compositions, and are untouched by the products of nuclear burning in the interior. Each star thus retains a chemical memory of the content of one piece of the interstellar medium (ISM) at the time and location of its formation. Collectively, many stars record the fossil record of the varied and changing composition of the ISM across cosmic time. Astronomers refer to this concept as ``Galactic archaeology.''

The overall metallicity\footnote{The term ``metallicity'' describes the overall metal content, and astronomers usually refer to elements heavier than He
as ``metals.'' Here, metallicity is quantified explicitly as [Fe/H] using the standard definition of abundance ratios:\
for elements X and Y, the logarithmic abundance ratio relative to the Solar ratio is defined as [X/Y]~$\equiv \log_{10} (N_{\rm X}/N_{\rm Y}) - \log_{10} (N_{\rm X}/N_{\rm Y})_{\odot}$.} of the ISM generally increases as time passes and more stars contribute freshly-produced metals. Stars with the lowest metallicity, commonly known as metal-poor stars, formed in regions of the ISM polluted by relatively few enrichment events. When the observed abundance patterns reflect so few events, there is an opportunity to isolate the chemical signatures of individual ones. Some metal-poor stars show large overabundances of elements produced by $r$-process nucleosynthesis. These stars are not the sites where the $r$-process occurred. Rather, these stars formed from material enriched by metals from earlier generations of stars, including material ejected from the $r$-process production site.

\subsubsection{Measuring detailed $r$-process abundance patterns from Galactic halo field stars}
\label{abundpatterns}

Late-type stars are the only sites beyond the Solar System where detailed chemical abundance patterns for large numbers of elements can be derived. When metal-poor stars that are highly enhanced in $r$-process elements are identified, detailed abundance analyses based on high-resolution spectra, model atmospheres, and atomic transition data can be used to derive the abundance pattern of elements present. Only a small fraction of stars in the sky are metal-poor, and only a small fraction of these stars ($\approx$~3\%) are highly enhanced in $r$-process elements \citep{barklem05}. The first highly $r$-process enhanced star discovered was $\mbox{CS~22892-052}$ \citep{sneden94}, and subsequent high-resolution spectroscopic follow-up observations have confirmed an additional $\approx$~30 stars with [Eu/Fe]~$> +1.0$, where Eu is taken as a representative element produced by the $r$-process \citep{hill02,hill17,christlieb04,honda04,aoki07,aoki10,francois07,frebel07,lai08,hayek09,mashonkina10,mashonkina14,roederer14rpro,roederer16ret2,ji16ret2,placco17,sakari18}.

\begin{figure*}
\begin{center}
\centerline{\includegraphics[angle=270,width=5.0in]       {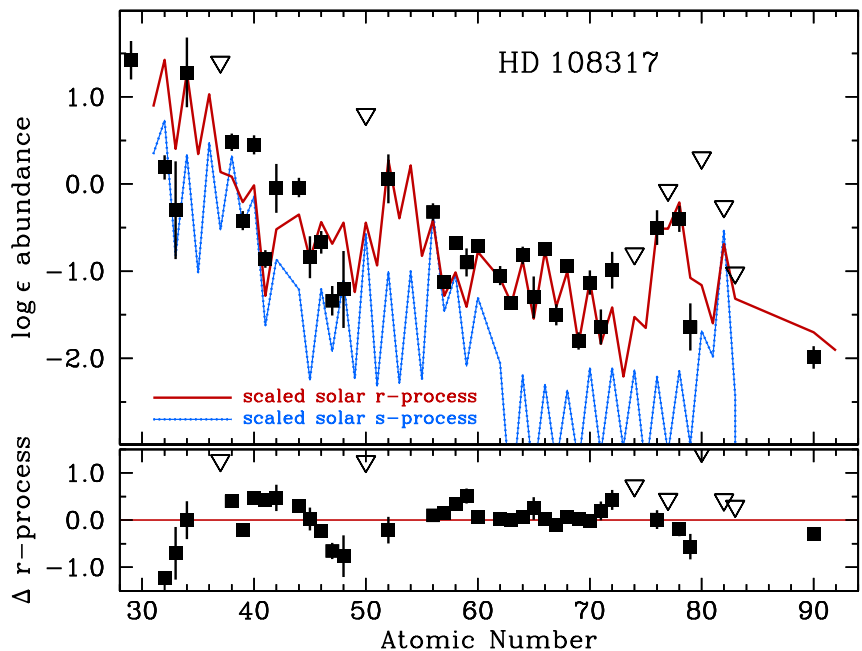}}
\caption{\label{hd108plot} 
The heavy element abundance pattern in the metal-poor star \mbox{HD~108317}, which is moderately enhanced in $r$-process material. The black squares indicate elements that are detected, and the open triangles indicate upper limits derived from non detections. The red and blue curves mark the Solar System elemental abundance patterns, scaled downward to match the Eu ($Z =$~63) or Ba ($Z =$~56) abundances in \mbox{HD~108317}. The bottom panel shows the residuals compared to the (red) scaled Solar System $r$-process pattern. 
The abundance pattern in this star is a close match to the scaled $r$-process pattern. Many of the elements shown here, including Ge ($Z =$~32), As ($Z =$~33), Se ($Z =$~34), Cd ($Z =$~48), Te ($Z =$~52), Lu ($Z =$~71), Os ($Z =$~76), Pt ($Z =$~78), and Au ($Z =$~79) could only be detected in the UV part of the spectrum. 
(Figure from \cite{roederer14hst}) 
}
\end{center}
\end{figure*}

At the temperatures and pressures found in late-type stellar atmospheres, not all heavy elements present absorption lines that can be detected with confidence in the optical and near-infrared portions of the spectrum ($\approx$~3030--25000~\AA) accessible to ground-based telescopes. Additional elements can be detected in the near-ultraviolet (UV) portion of the spectrum ($\approx$~1900--3030~\AA), but this requires that the star is bright enough ($V$ magnitude $\lesssim$~10 or so) to be observed with the \textit{Hubble Space Telescope} (\textit{HST}). When UV and optical spectra are obtainable, however, more than 30~elements produced by the $r$-process can be detected in a single star \citep{cowan02,cowan05,roederer10hst,sneden03,barbuy11,roederer12te,roederer12apjs,siqueiramello13,roederer14hst}. Figure~\ref{hd108plot} shows one example. The star shown in Figure~\ref{hd108plot}, \mbox{HD~108317}, is only moderately enhanced in $r$-process material ([Eu/Fe]~$= +0.5$), yet 34 elements heavier than Zn ($Z =$~30) have been detected. 

The elements at the three $r$-process peaks (Se, $Z =$~34; Te, $Z =$~52; and Os, Ir, and Pt, $Z =$~76--78) are among those that are best detected in the UV using \textit{HST} (Hubble Space Telescope), and these elements have only been detected together in three stars at present \citep{roederer12b,roederer16ipro,roederer14hst}. Elements at the first and second $r$-process peaks have not yet been observed in a highly $r$-process enhanced star, unfortunately, because no known star is sufficiently bright, although observational work to identify such stars is underway. The relative abundances and locations (in mass) of the $r$-process peaks are sensitive to the conditions (e.g., \cite{kratz1993}), so they are especially valuable probes of the $r$-process. Many of the radioactive progenitor nuclei for elements at the first and second $r$-process peak can be produced by current radioactive beam facilities in sufficient quantities to measure some of their relevant nuclear properties, so the nuclear uncertainties are somewhat smaller than for predictions of other $r$-process elements. The relationship between observed abundance patterns derived from stellar spectroscopy, nuclear properties measured by RIB facilities, and models of candidate $r$-process sites is clear:\ if the first two are known, the third can be deduced. Thus access to the UV spectral domain after \textit{HST} has been decommissioned is critical to maximize the impact of future RIB facilities, which can produce radioactive progenitors for the second and even the third abundance peak.   

Finer levels of abundance detail---isotopic abundances of heavy elements---cannot readily be extracted from stellar spectra. The wavelength shifts of spectral lines of different isotopes of a given element are smaller than the typical line widths ($\approx$~4-7~km~s$^{-1}$) in late-type stellar spectra, which are set by convective and turbulent motions in the stellar atmospheres. Studies over the last few decades have demonstrated the limits of isotopic measurements for heavy elements \citep{magain95,lambert02,sneden02,aoki03,lundqvist07,roederer08,collet09,gallagher15}, and these are unlikely to be improved in the foreseeable future.

\subsubsection{Deviations from the $r$-process pattern}
\label{lightrpro}

One key observational result is that the $r$-process pattern is robust from one star to another, and agrees well with the Solar System $r$-process residuals, for elements at and between the second and third $r$-process peaks (Ba to Au, e.g., \cite{roederer12te,roederer12b}). Sometimes this so-called ``universality'' extends to lighter $r$-process elements and the actinides, but not always. 

Among stars with high and moderate amounts of $r$-process enhancement, elements between the first and second $r$-process peaks sometimes show element-to-element variations and star-to-star variations in their overall level relative to the scaled Solar System $r$-process pattern (e.g., \cite{johnson02,truran02,aoki05,barklem05,roederer10ubiq,hansen12,siqueiramello14,spite18}). The bottom panel of Figure~\ref{hd108plot} illustrates one example of these variations, where a smoothly-varying deviation of abundances from the scaled Solar System $r$-process residuals is found for 32~$\leq Z \leq$~48.

Combining this result with the low star-to-star scatter observed among metal-poor stars for $\alpha$- and iron-group elements, like [Mg/Fe], poses a challenge for Galactic chemical evolution models (Section~\ref{subsection_GCE_low_Z_MW}). It is clear that the main formation site producing Mg and Fe at low metallicity (i.e., supernovae) cannot be the (only) source for the $r$-process elements. More and different formation sites are needed to account for the star-to-star scatter in the $r$-process element abundances. Theoretical studies have shown that almost all the chemical patterns of metal-poor stars can be explained by a combination of two formation processes:\ a heavy ``H-event'' explaining the robust $r$-process pattern (beyond the second $r$-process peak) and a light ``L-event'' (e.g., \cite{qian07,Hansen.etal:2014}). The L-event shows a larger internal abundance scatter and is therefore less robust. This L-event could be assigned to a number of formation sites and processes, not all of which may be known at present.  

Observations of the element Ge ($Z =$~32), which sits at the transition between the iron-group and neutron-capture elements, indicate that this element does not follow the scaled Solar System $r$-process residuals. Furthermore, it does not correlate with overall $r$-process enhancement \citep{cowan05}. This result demonstrates that Ge is likely not produced in the $r$-process (cf.\ \cite{roederer12ge}), but instead could be a product of proton-captures during explosive nucleosynthesis in supernovae \citep{frohlich06,pignatari10}. 

The actinides Th ($Z =$~90) and U ($Z =$~92) are another example of deviations from the $r$-process pattern. A subset of highly $r$-process enhanced stars show radioactive $^{232}$Th and $^{238}$U enhanced relative to the elements in the Rare Earth domain and third $r$-process peak \citep{hill02,honda04,hayek09,siqueiramello14,mashonkina14}, the so-called ``actinide boost'' \citep{schatz02}. This phenomenon appears to be limited to elements beyond the third peak \citep{roederer09th}, but its physical origin is unknown at present. Future observational work to identify larger samples of actinide boost and non-actinide boost stars should help to clarify the matter.

$r$-process investigations could be better served by attempting to reproduce the full range of $r$-process patterns across cosmic history, and not just the Solar System $r$-process residuals. Increasing the observed chemical inventories (Section~\ref{abundpatterns}) is one important step toward that goal. Identifying and characterizing the known deviations from the $r$-process pattern represent another.

\subsubsection{Environmental constraints on the $r$-process from stars in ultra-faint dwarf galaxies}

Stars like \mbox{HD~108317} or \mbox{CS~22892-052} are located in the field, unaffiliated with any known stellar cluster, stream, or galaxy. This limits their utility in terms of constraining the site of the $r$-process based on its environment. Recently, the lowest-luminosity galaxies known---also called ultra-faint dwarf galaxies, or UFDs---have been identified as sites of limited chemical evolution where the imprints of single nucleosynthesis events can be observed in the present-day stars (e.g., \cite{bovill11,frebel12,frebel14,frebel15}).

The recent discovery of the UFD galaxy Reticulum~II (Ret~II; \cite{bechtol15,koposov15a,koposov15b,simon15,walker15}) has enabled new insights into the astrophysical site of the $r$-process by providing additional information of the star forming environment \citep{ji16nat,ji16ret2,roederer16ret2}. A single, rare, and prolific $r$-process event must have taken place in Ret~II, leading to most of the stars in this galaxy being highly $r$-process enhanced. Simulations can be used to estimate the star-forming gas mass that may have been present in Ret~II \citep{beniamini16,safarzadeh17ret2}, which enables calculations of the mass of $r$-process ejecta from this single event. The results exclude the low total yields expected to arise from neutrino-driven winds in supernovae. Instead, the much larger yields from neutron star mergers (or other events producing similarly large yields, like jet-driven supernovae) are consistent with the observations of Ret~II. Furthermore, the delayed enrichment by a neutron star merger can be accommodated, and its ejecta retained, as  simulations of the first galaxies have shown \citep{safarzadeh17dtd,beniamini17}. 

It has become clear that considering the galactic environment in which $r$-process stars form is crucial for progress in understanding the astrophysical site and conditions of the $r$-process. Given the old age of the UFD galaxies (e.g., \citep{brown14}), and the similarity between the highly $r$-process enhanced stars in Ret~II and the Galactic halo field, another implication is that $r$-process stars found in the halo likely originated in systems similar to Ret~II. This offers the opportunity to use metal-poor $r$-process enhanced stars in dwarf galaxies as well as the halo to directly predict $r$-process yields for comparison with various theoretical works.

Other observational work has identified the presence of distinct levels of $r$-process enhancement in more luminous dwarf galaxies \citep{tsujimoto14,tsujimoto15dra,tsujimoto15sne,tsujimoto17dra}. From this, the occurrence rate and yields of $r$-process events can be quantified. These results from the Draco dwarf galaxy, for example, suggest two distinct sites of $r$-process nucleosynthesis, possibly a magneto-rotational supernova and a neutron star merger. A rare neutron star merger outcome---and thus high levels of $r$-process enhancement in present-day stars, perhaps diluted by Fe production---may also become inevitable in more massive systems like Draco \citep{roederer16ret2}.

A second low-luminosity system, Tucana~III (Tuc~III), containing at least one $r$-process enhanced star has been identified \citep{hansen17tuc3}, although the level of $r$-process enhancement in Tuc~III is more modest and consistent with that in globular cluster stars (e.g., \cite{gratton04}). Future work on this system and the ensemble of dwarf galaxies and stellar systems around the Milky Way will help to place further environmental constraints on the nature of the $r$-process.

\subsubsection{Evidence from elements not produced by the $r$-process}

Another observational approach to identify the astrophysical site(s) of the $r$-process is to consider the light elements that could be produced along with the $r$-process. Are the abundance ratios among elements from C to Zn (6~$\leq Z \leq$~30) statistically different in stars with high levels of $r$-process enhancement and those without? No significant differences are found, either among highly $r$-process enhanced stars in the field \citep{roederer14rpro} or the \mbox{Ret~II} UFD galaxy \citep{roederer16ret2}. For the field stars, the average differences are constrained to be $\approx$~3.5\% or less. 

This result may be interpreted to indicate that the site responsible for producing the high level of $r$-process enhancement did not produce any light elements. This could occur because the site of the $r$-process (e.g., neutron star mergers) is physically unassociated with the site of light element production (normal supernovae) or because the site of the $r$-process (e.g., a jet) is decoupled from the regions of supernovae where light element production occurs. This evidence supports the recent association of neutron star mergers as a viable site of $r$-process nucleosynthesis (see Sec.~\ref{sect_grav_waves_nsms}).

\subsubsection{Stars with low levels of $r$-process material}
\label{Sec:lowlevelr}

Highly $r$-process enhanced stars, as highlighted in the previous sections, comprise only a few percent of the local Galactic halo field population, and only a small fraction of UFD galaxies boast large numbers of highly $r$-process enhanced stars. In these stars, elements produced by the $r$-process are still in the minority:\ the ratio of number densities of individual $r$-process elements to hydrogen rarely exceed $10^{-10}$. In all other stars, the heavy elements are even less abundant.

Perhaps surprisingly, trace amounts of heavy elements are found in virtually all stars that have been studied \citep{roederer13}, including stars in nearly all dwarf galaxies. This would seem to imply that the products of neutron-capture nucleosynthesis were produced frequently, perhaps even in the first stars \citep{roederer14cempno}, and widely dispersed. In these cases, heavy elements other than Sr and Ba are rarely detected, but their non-detection may simply be a consequence of their trace abundance, not true absence. Occasionally, elements like Eu and Yb are also detected when high-quality spectra are obtained. With such limited abundance information, however, it is difficult to distinguish the nature of the neutron-capture nucleosynthesis responsible for their production \citep{casey17,roederer17}. 

These heavy elements did not necessarily originate in the same kind of astrophysical site that enriched \mbox{Ret~II} and provided the high $r$-process enhancement in field stars like \mbox{CS~22892-052}. Some supernovae or massive stars (prior to their final explosions) could have produced small amounts of these elements, whether by a weak $r$-process or some other neutron-capture nucleosynthesis mechanism \citep{boyd12,aoki13,cescutti13,wanajo13,qian14,komiya16}. This is reminiscent of the ``L-event'' noted previously (Section~\ref{lightrpro}). The presence of massive stars in all star-forming regions would provide a natural explanation for the apparent ubiquity of heavy elements in nearly all stars observed today. Some of the elemental ratios favor an $r$-process origin, rather than an $s$-process origin. There is some consensus, however, that normal massive-star supernova models could not produce sufficiently low $Y_e$ to reach the $A \sim$~170 mass region needed to explain these observations with a weak $r$-process, so the nucleosynthesis mechanism and production site remain an open question at present.  The neutrino-driven winds from magnetized, rotating proto-neutron stars could provide one source of moderate quantities of light r-process nuclei associated with core collapse supernovae at low metallicity \cite{Vlasov2017,0004-637X-659-1-561}.



\subsection{Multi-messenger observations of possible $r$-process events}

Observations of energetic astronomical events with not just photons, but also neutrinos or gravitational waves, can provide especially important information on the $r$-process.  This is because neutrinos and gravitational waves come from deep within an astronomical event and may directly probe the extreme conditions that generate the many neutrons needed for the $r$-process. These observations may locate the site of the $r$-process, provide information on conditions there and could even help determine the electron fraction $Y_e$, one of the most important parameters for nucleosynthesis.  

\subsubsection{Neutrinos from core collapse supernovae}
\label{subsec.SNnu}

\begin{figure}[!htb]
	\centering
		\includegraphics[width=0.75\textwidth]{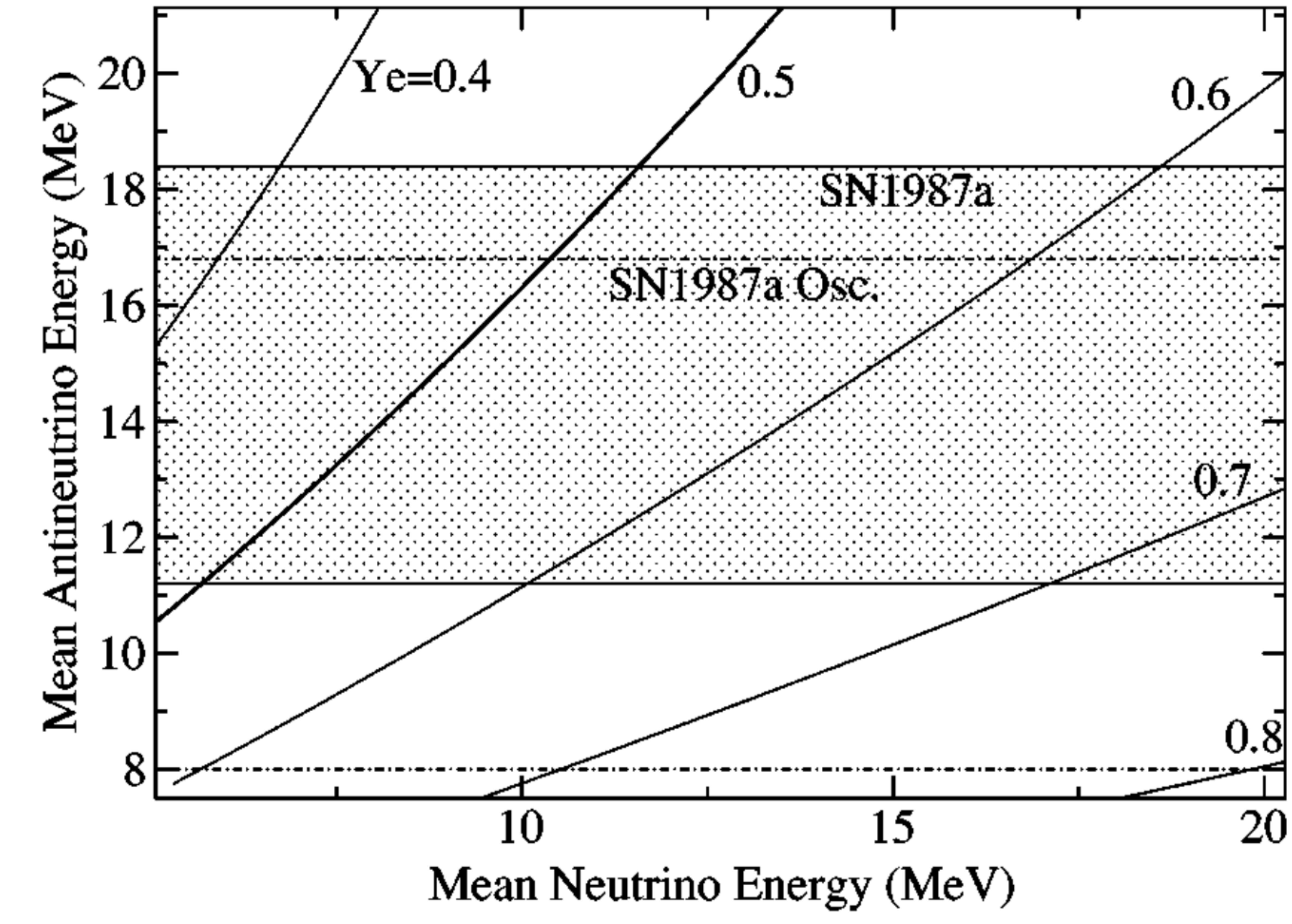}
	\caption{Mean antineutrino $\bar\nu_e$ energy versus mean neutrino $\nu_e$ energy for a core collapse supernova. Contours of electron fraction $Y_e$ are indicated for the neutrino-driven wind.  The wind is very neutron-rich only in the upper left corner of the figure.  The solid band shows the approximate mean $\bar\nu_e$ energy inferred from the $\approx 20$ events detected from SN1987a \cite{PhysRevD.65.083005}. }
	\label{fig:SN}
\end{figure}

One frequently studied theoretical $r$-process site is the neutrino-driven wind during a core collapse supernova (CCSN).  Here intense neutrino and antineutrino fluxes blow baryons off of the protoneutron star and determine the ratio of neutrons to protons in the wind.  Antineutrinos capture on protons to make neutrons
\begin{equation} \label{Eq.nuep}
\bar\nu_e + p \rightarrow n + e^+\, ,
\end{equation}
while neutrinos destroy neutrons
\begin{equation} \label{Eq.nuen}
\nu_e + n \rightarrow p + e\, .
\end{equation}
Therefore the relative rates of these two reactions determine the ratio of neutrons to protons in the wind. These cross sections grow with energy, therefore the wind will be neutron-rich if $\bar\nu_e$ are very energetic or if $\nu_e$ have very low energies.  Figure \ref{fig:SN} plots the electron fraction $Y_e$ expected in the wind for mean energies of $\bar\nu_e$ (on the y axis) and $\nu_e$ (on the x axis).  Thus simple robust neutrino physics determines the $Y_e$ of the wind.       

Essentially all supernova simulations, over the last fifteen years, find that energies for $\bar\nu_e$ are not much larger than $\nu_e$ energies.  Therefore $Y_e$ is expected to be near 0.5 and the wind is not predicted to be very neutron-rich. Although this wind can make lighter $r$-process nuclei this strongly suggests that the wind is not the site of the main $r$-process.

Neutrino oscillations could change either $\bar\nu_e$ or $\nu_e$ energies and therefore the $Y_e$ of the wind.  In general one expects oscillations of $\nu_x\rightarrow \nu_e$ to increase the energy of the $\nu_e$ more than oscillations of $\bar\nu_x\rightarrow \bar\nu_e$ to increase the energy of $\bar\nu_e$.  Here x represents either $\mu$ or $\tau$ flavors.  This is because most simulations find $E_{\nu_x}\approx E_{\bar\nu_x}>E_{\bar\nu_e}>E_{\nu_e}$ and therefore $\bar\nu_x$ and $\bar\nu_e$ have more similar energies than $\nu_x$ and $\nu_e$ before oscillations.  As a result, most neutrino oscillations only make the wind less neutron-rich and thus an even more unlikely site for the main $r$-process.     

Exotic neutrino physics could perhaps help.  For example, if there is a new sterile neutrino $\nu_s$ that lacks conventional weak interactions and has appropriate properties then $\nu_e\rightarrow\nu_s$ oscillations could take place while $\bar\nu_e\rightarrow\bar\nu_s$ oscillations do not.   In this case the wind could be very neutron-rich and produce main $r$-process elements \cite{1999PhRvC..59.2873M,Caldwell2000}. 

A fundamental observable to probe both conventional and unexpected neutrino physics is to measure detailed neutrino $\nu_e$ and antineutrino $\bar\nu_e$ spectra from the next galactic core collapse supernova.  About 20 
$\bar\nu_e$ antineutrino events were detected from supernova SN1987A.  For the next galactic core collapse SN, we expect many thousands of events in a number of large neutrino detectors \cite{Mirizzi:2015eza}.  
SNO+ will be a new kilo-tonne scale liquid scintillator detector $\approx$ 2~km underground in VALE's Creighton mine near Sudbury, Ontario, Canada. Although its main focus is the neutrinoless $\beta\beta$-decay, Galactic Supernova neutrinos and antineutrinos can be also detected. It is part of the SuperNova Early Warning System (SNEWS) network.
Super-Kamiokande (Japan) is an existing large water Cherenkov detector that is very good at observing antineutrinos $\bar\nu_e$ and should provide detailed information on the $\bar\nu_e$ spectrum (y axis in Fig. \ref{fig:SN}).  The Deep Underground Neutrino Experiment (DUNE) is a large liquid Ar detector that is being built in the Homestake gold mine \cite{dune}.  This detector should be able to measure the $\nu_e$ spectrum very well (x axis in Fig. \ref{fig:SN}).  Together the expected detailed $\bar\nu_e$ and $\nu_e$ spectra could suggest new neutrino physics or confirm our present expectations and help infer the electron fraction of the wind and therefore the expected nucleosynthesis.  

\subsubsection{Gravitational waves from neutron star mergers}
\label{sect_grav_waves_nsms}

The first detections of gravitational waves (GW) by the Advanced LIGO detectors were powered by mergers of binary black hole systems~\cite{GW150914,GW151226,GW170104,GW170608}.  These historic observations opened the field of gravitational wave astronomy.
A subsequent detection of merging black holes by both LIGO and Virgo~\cite{GW170814} then demonstrated that a three-detector network can provide tighter constraints on the localization of the merging objects, thus facilitating electromagnetic follow-up of gravitational wave events. 

Binary black hole systems, however, are probably not important for nucleosynthesis. 
From that point of view, the recent detection of gravitational waves from the merger of a binary neutron star system, GW170817~\cite{GW170817} is much more significant. An extensive electromagnetic follow-up campaign of GW170817~\cite{EM170817} allowed for the determination of its host galaxy (NGC4993), the observation of a delayed (possibly off-axis) gamma-ray burst~\cite{GRB:GW170817}, and, most importantly from a nucleosynthesis point of view, of UV/optical/infrared emission consistent with the radioactive decay of the ashes of the $r$-process in a few percents of a solar mass of material ejected by the merger~\cite{GW170817:DynEj,Cowperthwaite:2017}. This electromagnetic signal, called a {\it kilonova}, is discussed in Sec.~\ref{sect_electro_kilo}.

The very approximate event rate derived from this first observation indicates that, when Advanced LIGO reaches its designed sensitivity, we can expect an order of one to a few observations per month of neutron star - neutron star mergers, and perhaps some neutron star - black hole mergers as well.  Over the next five years, GW observations will more accurately determine the rate of neutron star and neutron star-black hole mergers and provide information on the mass distribution of merging systems.  This is fundamental information to determine the possible $r$-process nucleosynthesis contributions from these systems.  Simulations of these mergers are discussed in Sec. \ref{section_AS}.

\subsubsection{Electromagnetic observations and Kilonovae}
\label{sect_electro_kilo}
\newcommand{\be}{\begin{equation}}
\newcommand{\ee}{\end{equation}}

Nuclei which are freshly synthesized by the $r$-process are radioactive.  As matter in the expanding ejecta of a neutron star merger decays back to stability, the energy released via $\beta$-decays, $\alpha$-decays,
and fission can power transient thermal emission lasting days to weeks, known as a `kilonova'
\citep{Li&Paczynski98,Metzger:2010sy,Roberts+11,Grossman+14,Barnes+16}.  Kilonovae provide a unique probe to directly observe and quantify the production of $r$-process nuclei.  Their brightness, duration, and colors are diagnostic of the quantity of $r$-process matter, as well of physical processes during the merger and its aftermath (see \cite{FernandezMetzger16} for a review). 
In general, kilonovae are promising electromagnetic counterparts to GW signals detected by Advanced LIGO, because their emission is approximately isotropic (compared to the relativistically beamed emission of a gamma-ray burst) and the kilonova is detectable at optical wavelengths, where sensitive searches are possible \citep{Metzger&Berger12,Cowperthwaite&Berger15}. 

Because the photon opacity of the merger ejecta is dominated by Doppler-broadened atomic line transitions, the colors of kilonovae are diagnostics for the types of nuclei synthesized in the merger ejecta.  If the ejecta contains lanthanide or actinide nuclei (atomic mass number $A \gtrsim 145$), then the optical opacity is very high due to the complex atomic structure of the f-shell valence electrons of these elements, resulting in kilonova emission which peaks at near-infrared wavelengths \citep{Kasen+13,Tanaka&Hotokezaka13}.   On the other hand, ejecta containing only lighter $r$-process elements ($A \lesssim 145$) with d-shell valence electrons will possess a lower opacity, and hence will also produce a bluer component to the emission at early times \citep{Metzger&Fernandez14,Kasen+15}.  The latter can be produced by ejecta with higher electron fractions ($Y_e \gtrsim 0.25$), which may be produced either by shock heating at the interface of the merging neutron stars (e.g., \cite{Wanajo+14}), or due to the effects of neutrinos on outflows from the remnant accretion disk (e.g., \cite{Surman+06}, \cite{Just+15}, \cite{Roberts+16}).  An $\sim$hour-long UV transient could also be produced by the decay of free neutrons in the outer layers of the ejecta if they expand sufficiently rapidly for the $r$-process to freeze-out prematurely \citep{Metzger+15}.

Electromagnetic follow-up of GW170817 revealed an optical/infrared signal most naturally explained by a two-component model: a red emission due to neutron-rich ejecta producing a significant fraction of lanthanides or actinides, and a bluer emission due to ejecta with a higher initial $Y_e$~\cite{Cowperthwaite:2017}. 
Before GW170817, a more tentative candidate kilonova had already been discovered following the short duration gamma-ray burst GRB 130603B \citep{Berger+13,Tanvir+13}. 
Reproducing the peak luminosity of this event required the ejection of $\sim
0.03-0.08M_{\odot}$ of neutron-rich material \citep{Berger+13,Tanvir+13,Hotokezaka+13,Barnes+16}.

In the future, we are likely to have at our disposal a population of neutron star mergers detected through both GW and electromagnetic emission.
Additionally, improved theoretical models derived from numerical simulations of mergers, from nuclear theory, and from experimental nuclear data will help us relate the parameters of the merging binary, the properties of the ejected material, and the outcome of nucleosynthesis. 
Finally, absorption or emission lines from individual $r$-process elements could be detected in a post-maximum near-infrared spectrum \citep{Kasen+15}. This would provide critical information about the elements produced in a merger event. First tentative line features consistent with light r-process elements Cs and Te were reported in the kilonova spectrum from GW170817 \cite{Smartt2017}. Conclusive data will likely require future 30~m-class telescopes. 

\subsection{Isotopic abundances in meteoritic grains, crusts and sediments}

In addition to stellar abundances, meteoritic abundances can also provide clues for the astrophysical site of the $r$-process. Radioactive nuclei produced by the $r$-process and incorporated into the early solar system can serve as a clock for measuring the time interval between the last $r$-process production event and solar system formation. The decay products and remaining parent nuclei of these radioactive nuclides can be found today in primitive meteorites that are largely unaltered relics from the early solar system. From the timescale thus obtained, we can constrain the frequency of $r$-process events. This argument is based on the comparison of the ratio of unstable to stable isotopes between meteoritic abundances and the theoretical production yields. The commonly-used isotope ratios are $^{247}{\rm Cm}/^{235}{\rm U}$, $^{129}{\rm I}/^{127}{\rm I}$, $^{244}{\rm Pu}/^{238}{\rm U}$ where the half-lives of these unstable isotopes are $t_{1/2}$= 15.6 Myr, 15.7 Myr, and 80 Myr, respectively.
Several works thus far converge to a last $r$-process event time of about 100 Myr (e.g. \citep{Dauphas2005,Lugaro2014}). This timescale is relatively long when compared to the meteoritic signatures of core-collapse supernova injection of shorter-lived radionuclides such as $^{26}{\rm Al}$ ($t_{1/2}$= 0.72 Myr) and $^{60}{\rm Fe}$ ($t_{1/2}$= 2.62 Myr). The implied far lower frequency of $r$-process events, compared with core-collapse supernovae points to either neutron star mergers, or a relatively rare type of supernova as the $r$-process site. Similar conclusions can also be drawn from recent detection measurements of radioactive nuclides in deep sea crusts and sediments. The current abundance of $^{244}{\rm Pu}$ in deep sea reservoirs, which are expected to be effective continuous collectors of interstellar dust, is found to be much lower than in the early solar system, and much lower than expected for a continuous production of actinides in standard galactic supernovae at event rates of 1-2/100 yr \citep{Wallner2015,2015NatPh..11.1042H}. This again points to rare, prolific $r$-process sites such as neutron star mergers or rare types of supernovae. 
 
The analysis of meteoritic and deep sea sediment data from long-lived radioactive $r$-process isotopes relies on the theoretical prediction of production yields and on reliable half-lives. $r$-process calculations still involve large uncertainties in nuclear physics and astrophysical conditions, which lead to the prediction of a wide range of production yields such as $3.19\times 10^{-3}$ -- $1.46\times 10^{-1}$~M$_\odot$ for $^{244}${\rm Pu}\cite{GorielyJanka2016}. 

\section{Galactical chemical evolution simulations}
\label{subsection_GCE_MS}

While observations serve to constrain the source and nature of the $r$-process, galactic chemical evolution (GCE) can be thought of as the process by which galactic evolution is convolved with the underlying production mechanisms of heavy elements.  To reproduce the variety of chemical signatures observed in galaxies, galactic chemical evolution models need to follow the star formation history of individual galaxies from their birth to the present time while also taking into account their mass assembly history (i.e., galaxy mergers and accretion of matter from the intergalactic medium) as well as the gas circulation processes within and in between their interstellar and circumgalactic media \cite{2015ARA&A..53...51S,2017ARA&A..55...59N}.  Those models also need to follow the evolution of multiple stellar populations, the associated enrichment of the galactic gas, and the resulting modification of the composition of new generations of stars \cite{1980FCPh....5..287T,matteucci14}. Although stars and interactions between their remnants (e.g., compact binary mergers) are at the origin of the enrichment process, the Milky Way and local dwarf galaxies show different chemical evolution trends \cite{2009ARA&A..47..371T}, suggesting that stellar abundances can also be used to trace the formation history of galaxies.

The ultimate goal of galactic chemical evolution is to understand the origin of all elements across cosmic time as well as to obtain insights into what drives the formation and evolution of galaxies. In that regard, it is important to be able to disentangle the role of nuclear astrophysics and galaxy evolution when analyzing and interpreting stellar abundances. Galactic chemical evolution simulations are powerful tools to study this distinction, as they serve to better understand the complex interplay between the physical processes that give rise to the galactic structures, interactions, stellar populations, and chemical abundances we observe today. Finding a consistent picture of star formation in a galaxy evolution context that agrees with chemical evolution and galactic dynamics is the ``holy grail'' of galactic chemical evolution studies.  This section aims to highlight the multidisciplinary nature of galactic chemical evolution, describes how galactic chemical evolution models can be used to constrain the astrophysical site(s) of the $r$-process, and discusses the role played by nuclear data and experiments in the interpretation of stellar abundances using galactic chemical evolution.
 
\subsection{Inputs to include $r$-process sites}
\label{subsection_GCE_ingredients}
Galactic chemical evolution can be done in several ways, from simple analytical models to complex hydrodynamic simulations.  Simple models are fast to compute and are designed to reproduce the global averaged chemical evolution trends \cite{ferrini92,travaglio99,lanfranchi03,fenner2006,kobayashi15,andrews_w_s16,2017ApJ...835..128C}, while more sophisticated simulations also enable the study of the spread in the stellar abundances \cite{kobayashi11,vandevoort15,wehmeyer15}, the origin of radial metallicity gradients in galaxies \cite{chiappini01,pilkington12,minchev13,miranda16}, and the role of the mass assembly history of galaxies via mergers \cite{2012ApJ...745...50W,romano13,romano15}.  In spite of the multitude of numerical approaches, every simulation needs to input nucleosynthetic yields along with the properties of their astrophysical site in order to drive the chemical evolution process, regardless of the complexity and nature of the model used.

Nucleosynthetic yields are the foundation of all galactic chemical evolution simulations. They represent the amount and isotopic composition of the mass ejected by individual enrichment sources such as stellar winds, supernova explosions, and interactions involving stellar remnants (white dwarfs, neutron stars, and black holes). Those yields are usually dependent on the initial composition and mass of the progenitor stars, and typically include light elements up to germanium for core-collapse and Type Ia supernovae \cite{thielemann86,woosley95,iwamoto99,chieffi04,travaglio04_Ia,nomoto06,heger10}. Yields for low- and intermediate-mass asymptotic giant branch (AGB) stars typically include s-process nucleosynthesis \cite{bisterzo10,karakas14,cristallo15,2016ApJ...825...26K,2016ApJS..225...24P,2017arXiv170908677R}.  Although some sets of yields for massive stars do provide neutron-capture elements \cite{2012A&A...538L...2F,2016MNRAS.456.1803F,2016ApJS..225...24P,2016ApJ...821...38S,2017arXiv170908677R}, they usually do not include a proper treatment of the $r$-process. In galactic chemical evolution models, the $r$-process abundances pattern returned into the interstellar medium by a given source is either taken from the solar residuals \cite{Arn07}, or taken from theoretical calculations that properly focus on $r$-process nucleosynthesis (see Section~\ref{section_AS}).

In order to include an $r$-process site inside galactic chemical evolution simulations, the adopted $r$-process yields must be convolved with the occurrence frequency (or rate) of the considered site \cite{matteucci14_r}.  For example, if core collapse supernovae are used as a site, the $r$-process yields will be injected in the galactic gas following the rate of core collapse supernovae, with a certain correction if not all massive stars are believed to host the $r$-process. If neutron star mergers or black hole-neutron star mergers are used, the $r$-process yields will be convolved with a delay-time distribution (DTD) in the same manner as one would include the contribution of Type~Ia supernovae. Two different approaches are typically used to define this delay-time distribution: 1) the merger rate follows the core collapse supernovae rate, but is shifted to later times by a constant delay time; and, 2) the delay-time distribution is taken from short gamma-ray burst observations \cite{2015MNRAS.448.3026W} or from the predictions of population synthesis models \cite{2012ApJ...759...52D,2017arXiv170807885C}. The final input parameter that needs to be provided is the total mass of material ejected by each $r$-process event, which is often used to normalize the adopted $r$-process abundances pattern (e.g., the solar residuals). Once all of this information is  implemented, galactic chemical evolution simulations can be performed to test the contribution of different $r$-process sites to the chemical evolution of heavy elements in galaxies.

\subsection{Metal-poor stars and the early Universe}
\label{subsection_GCE_low_Z}
Low-metallicity stars are interesting targets for observation since they capture the first moments of enrichment in the early universe.  Their chemical signatures are likely to have been generated by only one or a few astrophysical events.  This brief window of time is thus ideal to study the origin of $r$-process elements \cite{qian07}, as abundances patterns can directly be compared with theoretical nucleosynthesis calculations.

\subsubsection{The Galactic halo}
\label{subsection_GCE_low_Z_MW}
The chemical evolution of $r$-process elements in the Milky Way have been investigated in the past using core collapse supernovae and neutron star mergers as potential sites \cite{travaglio99,argast04,cescutti06,matteucci14_r,2014A&A...564A.134M,cescutti15,2015ApJ...814...41H,2015ApJ...804L..35I,shen15,vandevoort15,wehmeyer15,2016ApJ...830...76K,2016A&A...589A..64M,2017ApJ...836..230C,2017arXiv170703401N}. The scatter in the observed $r$-process abundance ratios at low metallicity is larger than the scatter measured for $\alpha$ elements \cite{Sneden.etal:2008,Hansen.etal:2014}. This implies that $r$-process events are rare, occurred stochastically in the early universe, and that the spread in abundances can be used to constrain the general properties of $r$-process site(s) \cite{2008A&A...481..691C}. Because the level of scatter  (in, e.g., [Eu/Fe]) depends on the rate and amount of $r$-process ejecta of the considered sites, non-uniform mixing models can test different input parameters and provide insights into the nature of neutron star mergers and core collapse supernovae \cite{argast04,cescutti15,wehmeyer15}. However, hydrodynamic simulations of dwarf and Milky Way-like galaxies have shown that the spread in neutron-capture elements relative to Fe is also sensitive to the resolution and the adopted metal diffusion prescriptions \cite{shen15,vandevoort15,2017ApJ...838L..23H,2017arXiv170703401N}, suggesting that gas mixing and metal recycling processes in the early stages of galaxies need to be understood in order to efficiently constrain the properties of $r$-process sites.

Another observational indicator that can be used to probe $r$-process sites with galactic chemical evolution is the high level of $r$-process enrichment found in some of the most metal-poor stars in our Galaxy. This requires a prompt enrichment source in the early universe. Because neutron star mergers and black hole-neutron star mergers originate from the coalescence of the remnants of massive stars, the enrichment timescale of compact binary mergers is naturally longer than for core collapse supernovae. From that argument alone, core collapse supernovae seem to be the perfect candidates to explain those observations. Indeed, numerical simulations have shown that allowing some core collapse supernovae to produce the $r$-process can explain the early appearance of $r$-process enriched stars \cite{argast04,matteucci14_r,cescutti15,wehmeyer15}.
But from a theoretical point of view, as described in Section~\ref{section_AS}, neutron star mergers and black hole-neutron star mergers are more likely to synthesize the full $r$-process including the 2nd and 3rd peaks, while core collapse supernovae are more likely to only synthesize the 1st peak. If this is the case and compact binary mergers are indeed the main $r$-process sites, the question is how to introduce their ejecta into the most metal-poor stars, given the relatively long delay times of these events.

One solution to this challenge is to account for non-uniform mixing of metals in the interstellar medium, which can be seen in the scatter of the age-metallicity relationship of stars in the Milky Way \cite{1980ApJ...242..242T,2014A&A...562A..71B,2014A&A...565A..89B,2017MNRAS.471.3057M}. This indicates that stars with the same metallicity (expressed as [Fe/H] in this particular problem) can have different ages. This non-linearity between [Fe/H] and stellar age allows compact binary mergers to enrich stars with ejecta from very-low metallicity stars, even if they occur later than core collapse supernovae. Indeed, 3D hydrodynamic galaxy simulations that self-consistently track non-uniform mixing and stochastic processes succeeded in incorporating neutron-capture elements in the most metal-poor stars with neutron star mergers only \cite{shen15,vandevoort15,2017arXiv170703401N}.

Another solution to allow neutron star mergers to enrich very-low metallicity stars is to vary the star formation efficiency (SFE) throughout the formation history of our Galaxy. In galactic chemical evolution models, the SFE defines the amount of gas in which heavy elements are deposited and can thus control the rate at which the galactic gas is  enriched (i.e., how fast [Fe/H] is increasing) \cite{andrews_w_s16,2017arXiv171006442C}. From a cosmological structure formation perspective, massive galaxies like the Milky Way assembled in time from gas accretion and from galaxy mergers that occurred in the past \cite{1984Natur.311..517B,2005Natur.435..629S,2011ApJ...740..102K,2015MNRAS.446..521S,2017arXiv170302970P}. Within this framework, the low-metallicity stars present in our Galaxy likely formed in many distinct low-mass ``building block" galaxies at high redshift. If we assume that low-mass galaxies have lower SFE than massive galaxies (which is supported by both observations and theory \cite{2015ApJ...808L..30V}), this assembly scenario suggests that the enrichment process was slower in the early phases of our Galaxy, and that neutron star mergers could more easily enrich stars at low [Fe/H]. This solution has been suggested and tested with a simple galactic chemical evolution model \cite{2015ApJ...804L..35I} and thereafter confirmed using hydrodynamic simulations and semi-analytical models \cite{2015ApJ...814...41H,2016ApJ...830...76K,2017MNRAS.466.2474H}. We note that varying the amount of Fe ejected by core collapse supernovae can also alter the pace at which [Fe/H] increases in the early universe.

Using galaxy evolution arguments to explain the presence of $r$-process elements in metal-poor stars can be difficult, as there is a lack of observational constraints regarding galaxy properties in the early Universe.  Although it is observationally possible to recover the entire aggregate star formation history of a galaxy, it is not possible to measure how that galaxy's gas evolved in the past.  This is, nevertheless, an important piece of information for galactic chemical evolution because it sets the metal concentration of the galactic gas, the rate of early enrichment, how much mass is recycled into stars, and how much mass is lost from the galaxies into the circumgalactic and intergalactic media. In addition, observations provide few constraints on the number and mass of the progenitor galaxies that merged together to form the Milky Way halo---quantities that can significantly impact the local rate of star formation and gas dilution.  Future telescopes such as the James Webb Space Telescope will facilitate exploration of the early stages of galaxy formation.  In the mean time, hydrodynamic and cosmological simulations of galaxy evolution can provide insights into galactic inflows and outflows, non-uniform mixing, and on the importance of galaxy mergers.

To summarize, the level of scatter and the presence of $r$-process elements in metal-poor stars contain valuable information regarding the nature of $r$-process site(s), but to interpret those features, one needs to simultaneously consider $r$-process nucleosynthesis calculations, gas mixing processes, and galaxy formation mechanisms, as some galactic chemical evolution observables can be reproduced in various ways with numerical models.  Identifying the sources of $r$-process enrichment in our Galaxy thus represents a significant challenge that needs to be addressed in a multidisciplinary framework.

\subsubsection{Local dwarf galaxies}
\label{subsection_GCE_low_Z_dwarf}
Local dwarf spheroidal and ultra-faint galaxies are interesting targets for probing astrophysical $r$-process sites, as the great variety of $r$-process enrichment levels from one system to another directly probes the stochasticity and rarity of the underlying $r$-process events. In addition, compared to the Milky Way halo, which likely contains a complex mixture of several disrupted satellite galaxies \cite{2008A&ARv..15..145H,2016ARA&A..54..529B}, local dwarf galaxies are relatively pristine systems where the $r$-process enrichment can be studied in a much simpler context. In particular, cosmological simulations have suggested that ultra-faint dwarf galaxies may not have been involved in galaxy mergers throughout their lifetime \cite{2016arXiv161100759G}. In addition, some galaxies such as Reticulum~II might even have hosted only one neutron star merger, which allows hydrodynamic simulations to systematically explore the impact of the location where the neutron star merger occurs (relative to the center of the galaxy) \cite{safarzadeh17ret2}, and the impact of explosion energies and the gas density in which neutron star merger ejecta is injected \cite{beniamini17}.

We recall that the detection of prolific $r$-process enrichments in dwarf galaxies does not necessarily guarantee that they originate from a neutron star merger. Any other astrophysical site that can release a similar amount of $r$-process material is compatible \cite{roederer16ret2}. In fact, when considering the contribution of compact binary mergers in dwarf systems, one needs to address the probability of retaining the $r$-process ejecta within the star-forming region \cite{beniamini16,beniamini17}.  Given the low escape velocity of dwarf galaxies, the natal kick imparted to neutron stars soon after the supernova explosions could expel the binary system outside the galaxy before they have time to merge. However, using the low proper motions derived observationally from $\sim$\,10 binary neutron star systems, a previous study suggested that neutron star mergers could occur within low-mass galaxies like Reticulum~II \cite{beniamini16}. But calculating the fraction of neutron star mergers that stay within a galaxy and participate to the $r$-process enrichment depends on the adopted distribution of natal kicks imparted to neutron star binaries \cite{safarzadeh17dtd}. These input distributions can vary substantially based on model assumptions  \cite{2002ApJ...572..407B,2013ApJ...776...18F,2014ApJ...792..123B,2016MNRAS.456.4089B}.


\subsection{Enrichment history and the local Universe}
\label{subsection_GCE_local}
While old low-metallicity stars contain valuable information regarding the rise of $r$-process elements in the early universe, the low-redshift (Local) universe contains information regarding the total integrated $r$-process production through the lifetime of galaxies, and thus represents the complex endpoint of galactic chemical evolution. One of the main challenges when interpreting chemical abundances of relatively young [Fe/H]~$\sim$~0 stars is to define which astrophysical sites, between core collapse supernovae and compact binary mergers, have contributed the most to their $r$-process abundances. Galactic chemical evolution simulations are powerful tools to address this challenge since they can systematically test the contribution of different sites and provide predictions to be compared with a wide range of observations.

\subsubsection{$r$-process content of the Milky Way}
\label{subsection_GCE_r_content}
Calculating the current mass of $r$-process elements $M_\mathrm{r,tot}$ found inside the Milky Way represents the first step toward constraining the origin of $r$-process elements. As a first-order approximation, $M_\mathrm{r,tot}$ can be estimated by multiplying the $r$-process mass fraction found in the Solar system with the total baryonic mass of the Milky Way. This assumes that stars and gas on average have a Solar composition. For any given astrophysical site, one can derive the total number of events required to recover this Galactic $r$-process content by dividing $M_\mathrm{r,tot}$ by the average mass ejected per event. By distributing the events across the lifetime of our Galaxy ($\sim$\,13\,Gyr), which assumes a constant star formation history (SFH) for the Milky Way, the required number of events are converted into rates (e.g., neutron star merger rate) and can then be compared with the ones inferred from observations \cite{2015MNRAS.448..928K}.  Such an analytical approach has been used to calculate the Galactic neutron star merger and black hole-neutron star merger rates needed to reach $M_\mathrm{r,tot}$, assuming that compact binary mergers are the main source of $r$-process elements \cite{2000ApJ...534L..67Q,Metzger:2010sy,2013ApJ...773...78B,2014ApJ...795L...9B,2015NatPh..11.1042H}.

There are some limitations to the approach described in the previous paragraph. According to hydrodynamic galaxy simulations, the SFH of the Milky Way is not constant but should rather peak at early time during the first few Gyr of evolution and then decrease \cite{shen15}. The current (low-redshift) compact binary merger rates inferred using such varying SFH should then be lower compared to when using a constant SFH \cite{2015NatPh..11.1042H}, as binary mergers are more concentrated at earlier times. Studies that only account for the baryonic mass found in the Galactic discs likely underestimate $M_\mathrm{r,tot}$, as a significant amount of metals (and thus $r$-process elements) are found in the circumgalactic medium (CGM) surrounding our Galaxy \cite{2014ApJ...786...54P,2016ARA&A..54..529B,2017ARA&A..55..389T}. Accounting for this extra gas reservoir in analytical calculations would likely increase the derived neutron star-neutron star and black hole-neutron star merger rates required to explain the amount $r$-process elements observed in the Milky Way. For analytical calculations that do include the circumgalactic medium, a correction should be included to account for the fact that this hot gas reservoir has on average a lower metallicity than the Solar value \cite{2016ARA&A..54..529B}.

An alternative approach to calculate the amount and distribution of $r$-process elements in the Milky Way is to use GCE simulations. The advantage of this approach is the opportunity to track the production of heavy elements within a context that includes a time-dependent SFH and a proper treatment of the gas circulation processes inside and outside galaxies. However, the use of complex models leads to the inclusion of several sources of uncertainties.

\subsubsection{Gravitational waves and kilonovae}
\label{subsection_GCE_grav_wave}
From a nucleosynthesis point of view, because theoretical models of compact binary mergers are more likely to synthesize elements in the 2nd and 3rd $r$-process peaks compared to core collapse supernovae (see Section~\ref{section_AS}), they are excellent candidates for being the dominant $r$-process site in the Milky Way.  However, to validate this scenario, the merger rate derived using analytical calculations and GCE simulations needs to be consistent with the rates inferred by observations.  These observed rates can be estimated from binary pulsars and short-duration gamma-ray bursts \cite{Abadie:2010,2014ARA&A..52...43B,2015MNRAS.448..928K}. While such observations may not be direct evidences of neutron star mergers or black hole-neutron star mergers, they can be used for both the Milky Way \cite{2015MNRAS.448..928K} and for other galaxies in the Local universe \cite{2014ApJ...792..123B}.

Gravitational wave measurements provide direct evidences of compact binary mergers, although the origin and location of the source can be difficult to isolate if no electromagnetic emission is detected. Recently, the LIGO/Virgo collaboration has detected for the first time the gravitational wave coming from a neutron star merger (GW170817, see Section~\ref{sect_grav_waves_nsms} for details) \cite{GW170817}. This provided a direct constraint on the local merger rate density (1540$^{+3200}_{-1220}$\,Gpc$^{-3}$\,yr$^{-1}$). The electromagnetic emissions that followed this important event enabled the host galaxy (NGC\,4993) to be identified, and it confirmed that gamma-ray bursts can be associated with neutron star mergers. The analysis of the light curves of GW170817 demonstrated that neutron star mergers can indeed synthesize and eject significant amounts of $r$-process material that includes elements in the 2nd and 3rd abundance peaks (see Section~\ref{sect_electro_kilo}). 

Shortly after the announcement of these new constraints, a series of studies investigated whether neutron star mergers can be the main $r$-process site.  Work based on analytical calculations similar to the ones described in Section~\ref{subsection_GCE_r_content} concluded that, if GW170817 is a typical event, neutron star mergers are frequent enough and can eject enough material to explain the origin of the total $r$-process mass currently observed in the Milky Way \cite{2017arXiv171005836T,Cowperthwaite:2017,2017ApJ...848L..19C,2017arXiv171005442G,2017Natur.551...80K,2017arXiv171005445R,2017arXiv171005850T,2017arXiv171005805W,2018arXiv180101141H}.  In parallel, a framework has been set to connect GCE simulations with LIGO/Virgo measurements \cite{2014A&A...564A.134M,2017ApJ...836..230C}. Convolving those galaxy evolution simulations with the cosmic star formation history, in order to generate a merger rate density in units of Gpc$^{-3}$\,yr$^{-1}$, demonstrated that the rates defined by GW170817 are consistent with the ones needed in GCE studies to reproduce the evolution of Eu (an $r$-process tracer) in the Milky Way \cite{2017arXiv171005875C}. In summary, GW170817 supports the idea that neutron star mergers are the main $r$-process site, but at present the statistics are too poor, and the uncertainties regarding the ejecta of neutron star mergers are too large, to come to a definitive conclusion \cite{2017arXiv171005875C}.  In addition, there is a difficulty in reproducing the exact shape of the chemical evolution trend of Eu in the Milky Way using canonical delay-time distributions for neutron star mergers \cite{2016ApJ...830...76K,2017ApJ...836..230C,2018arXiv180101141H}. black hole-neutron star mergers, which hopefully will be detected in the next combined LIGO+Virgo observing run, could also play a role in the production of $r$-process elements in our Galaxy \cite{2014A&A...564A.134M,2016A&A...589A..64M,2017ApJ...836..230C}.

\subsection{Nuclear physics and chemical evolution}
\label{subsection_nuc_exp_GCE}

Yields are one of the most important input parameters used in GCE simulations (see Section~\ref{subsection_GCE_ingredients}).  Any uncertainty associated with those yields will propagate and affect GCE predictions.  To limit the impact of such uncertainties, the solar $r$-process residuals \cite{Arn07} 
can be used to represent the abundance pattern of the ejecta of all $r$-process sites in GCE simulations.  This widely-used approach is convenient to test different astrophysical sites, but it only provides limited information regarding how $r$-process elements are made.  For now, using theoretical $r$-process yields (see Section~\ref{section_AS}) may introduce a significant amount of uncertainty in numerical predictions \cite{2017arXiv171005875C}, but it offers a real opportunity to bridge nuclear astrophysics with the interpretation of stellar abundances in galaxies.

As described in Section~\ref{lightrpro}, metal-poor stars sometimes show variations in their abundance patterns for elements between the 1st and 2nd $r$-process peaks.  Such features are ideal targets for investigating the possible multiple origins of the light $r$-process elements.  These chemical signatures can be directly compared with theoretical nucleosynthesis yields for core collapse supernovae and compact binary mergers, without having to incorporate them into GCE calculations if one assumes that the target stars have been polluted by one or a limited number of $r$-process events.  Once theoretical yields are in agreement with observations, they can be implemented in inhomogeneous GCE models to verify if they can self-consistently reproduce the scatter seen in the observational data \cite{cescutti15}.  This GCE approach can put constraints on the probability (or frequency) of meeting a particular $r$-process condition (e.g., electron fraction, entropy) in the early universe as a function of its associated $r$-process abundance pattern.

While matching abundance patterns of individual metal-poor stars do not necessarily require GCE simulations, the (still-to-come) theoretical yields used to describe these stars should also be consistent with the $r$-process abundances seen in the Sun and in solar-metallicity stars.  To reach these observables, GCE simulations are required to keep track of the gas circulation and recycling processes and to consistently combine the different theoretical $r$-process yields according to the general properties of their astrophysical sites (e.g., total mass ejected per event, number of event per unit of stellar mass formed).

Suggested theoretical yields from various nucleosynthesis studies can also play a critical role that has perhaps gone under-appreciated by both astronomers and nuclear physicists: that the exclusion of a large, non-trivial part of parameter space for proposed yields should exist and that efforts should be made to make complementary ``exclusion'' or incompatibility GCE models and simulations that have observable consequences. If we cannot (currently) pin down the $precise$ nuclear network models (theoretical yields) that underlie astrophysical observations then we should at least attempt to first exclude those nuclear network models that are incompatible with those observations. Furthermore, both nuclear physicists and astronomers should aim to constrain yield predictions from both areas of expertise---fundamental yield constraints from theory and observational yield constraints from chemical abundance patterns and distributions. Currently, the authority on yield constraints has leaned more towards theory. Astronomers should help nuclear (astro)physicists to use astrophysical observations to infer the yields from the application of GCE models (see, e.g., \cite{lee2013,cescutti15}) and simulations and observed data to constrain their own nuclear network modeling efforts.

Chemical evolution is a convoluted process that involves all scales from nuclear physics to cosmological structure formation. To fully understand the origin of heavy elements and the $r$-process signature of all stars, a joint effort between experimental physics, nuclear astrophysics, GCE, and cosmology is needed. In such a framework, progress in the field of nuclear physics and $r$-process nucleosynthesis can be connected and compared to the abundances observed on the surface of stars.  Nuclear experiments conducted at large facilities such as FRIB will significantly improve our ability to interpret the $r$-process signatures observed in our local universe, as they will allow to constrain and better refine $r$-process yields calculated from first principles, which will represent the building block of all GCE simulations.  This will lead to a more precise quantification of the contribution of core collapse supernovae, neutron star mergers, and black hole-neutron star mergers on the overall production of all elements involved in the $r$-process.

\section{Astrophysical simulations}
\label{section_AS}


\begin{sidewaystable}
\centering
\begin{tabular}{lcccc}
\hline
{\bf Proposed Astrophysical Site}  & &  \multicolumn{3}{c}{{\bf Primary Conditions}}  \\
& Description & $S$/$k_B$ & $\tau$ (s) & $Y_e$ \\
\hline
\multicolumn{5}{l}{{\bf Compact object mergers}}\\
Tidal ejection \cite{Lattimer+74,Meyer89,Freiburghaus+99,Just+15,Korobkin:2012uy,Goriely:2011vg,hotokezaka:13,Foucart:2013a,Kawaguchi:2015,Roberts+16} & cold, fission & $\sim$1$-$10 & $\sim 10^{-4}-10^{-3}$ & 0.01$-$0.1 \\
Shocked ejecta \cite{Goriely:2011vg,Wanajo+14,Just+15,Sekiguchi:2015,Radice:2016dwd} & cold, hot &$\sim$30 & $\sim 10^{-4}-10^{-3}$ & 0.05$-$0.4  \\
Polar post-merger outflows \cite{Surman+08,Metzger:2010sy,Perego:2014fma,Just+15,Sekiguchi:2015,Martin:2015hxa,Foucart:2015} & cold, hot & $\sim$10$-$50 & $\sim 0.01-0.03$ & 0.2$-$0.5 \\
Equatorial post-merger outflows\cite{Surman+08,Metzger:2010sy,Perego:2014fma,Just+15,Martin:2015hxa} & cold, hot & $\sim$10$-$30 & $\sim 0.01-0.05$ & 0.1$-$0.3\\ 
\hline
\multicolumn{5}{l}{{\bf Core Collapse Supernovae}}\\
$\nu$-driven wind \cite{Meyer+92,Woosley+94,Arcones+07,Fischer+10,Hudepohl+10,Roberts+12,Arcones.Thielemann} 
& hot & $\sim$30$-$100  & & 0.4$-$0.55 \\
$\nu$-driven wind with sterile neutrinos \cite{1999PhRvC..59.2873M,2012JCAP...01..013T,2014PhRvD..89f1303W} & hot & $\sim$50$-$100 & & 0.1$-$0.4 \\
$\nu$-heated winds from rotating proto-magnetars \cite{Vlasov2017} & hot & & $\sim10^{-3}$  & 0.4$-$0.5 \\
\hline
\multicolumn{5}{l}{{\bf Various}}\\
Magneto-hydrodynamical jets from CCSN \cite{LeBlancWilson70,Winteler+12,Nishimura:2015nca} & cold, hot & $\sim$10$-$30 & & 0.1$-$0.4 \\
Collapsar magneto-hydrodynamical jets \cite{Nakamura:2013jda}    & hot & $\sim$10$-$10000   & & 0.1$-$0.5   \\
Accretion disks from CCSN \cite{Pruet+03,McLaughlin:2004be,Surman+06} & hot & $\sim$20$-$60  & & 0.2$-$0.4 \\
$\nu$ oscillations in disk winds \cite{Malkus:2012ts,Malkus:2015mda,PhysRevD.96.123015} & hot & $\sim$20$-$60  & & 0.1$-$0.4 \\
$\nu$ induced spallation in CCSN \cite{Banerjee+11} & cold &  & $-$ \\
\end{tabular} 
\caption{\label{tab:environments} Proposed astrophysical sites for the main $r$-process and their corresponding conditions.}
\end{sidewaystable}

\normalsize

There are many proposed sites for the $r$-process (see Tab.~\ref{tab:environments}). A common requirement is an environment that produces a large ratio of neutrons to seed nuclei.  This is necessary to produce the heaviest $r$-process elements.  Furthermore, one may require  an even larger ratio of neutrons to seed nuclei in order to insure a robust abundance pattern.  Otherwise, if there are just barely enough neutrons, the resulting abundance pattern may be very sensitive to small changes in the number of neutrons. In addition, the neutron density has to be high to drive the nucleosynthesis path away from stability to produce the prominent abundance peaks at $A=80$, $A=130$, and $A=195$ when the reaction sequence crosses the corresponding $N=50$, $N=82$, and $N=126$ shell closures. 

\subsection{Physics of neutron-richness}
In this section we discuss the physics that may make an environment neutron rich, or at least have a large ratio of neutrons to seed nuclei.  We consider the following possibilities: (1) weak equilibrium in large neutrino fluxes, (2) electron capture at high densities, (3) liberation of neutrons from neutron-rich stable nuclei via $(\alpha, n)$ reactions, and (4) decreasing the number of seed nuclei to increase the ratio of neutrons to seeds.

\subsubsection{Weak equilibrium}
\label{Sec:weak_equilibrium}
In large neutrino fluxes, the ratio of neutrons to protons is set by the relative rates of neutrino and antineutrino capture reactions (see Eqs. \ref{Eq.nuep} and \ref{Eq.nuen}). 
These rates depend on the neutrino and antineutrino fluxes, and because the cross sections grow with energy, the rates also depend on the mean energies of neutrinos and antineutrinos.  Furthermore, there are a number of important corrections to the cross sections for these reactions due to the mass difference between neutrons and protons, weak magnetism \cite{weakmag} and the binding energy shift of protons in a neutron-rich medium \cite{bindingenergyshift}. These corrections change the cross section for neutrino absorption relative to the cross section for antineutrino absorption so they impact the equilibrium ratio of neutrons to protons and the electron fraction. The electron fraction $Y_e = n_p / (n_n + n_p)$, where $n_p$ and $n_n$ are the number densities of protons and neutrons (either free or inside nuclei), respectively, is the most important parameter that determines the $r$-process nucleosynthesis outcome. $Y_e = 0.5$ would be symmetric matter (same number of neutrons and protons) and anything with $Y_e < 0.5$ is neutron rich. For the standard $r$-process to happen, we need the ejecta to be neutron rich, i.e.~$Y_e < 0.5$ (though depending on other conditions, an $r$-process is in principle also possible for $Y_e > 0.5$ \cite{Meyer2002}).

Neutrino oscillations or new neutrino physics could change $Y_e$.   In core collapse supernovae, oscillations typically increase the neutrino energy more than they increase the antineutrino energy so the net effect is often to make the wind less neutron rich.  However nonstandard oscillations for example involving sterile neutrinos could make the wind very neutron rich. Absent new neutrino physics, most supernova simulations find that $Y_e$ is close to 0.5.  We discuss nucleosynthesis in the neutrino-driven wind of core collapse supernovae in Section \ref{sec.neutrinowind}.

\subsubsection{Electron capture at high densities}  At great densities, electron capture drives matter neutron rich because of the large electron Fermi energy.  Indeed neutron star matter is expected to be very neutron rich.  However neutron stars have very large gravitational binding energies of order 200 MeV per nucleon.  Therefore, to take advantage of the neutron-richness of high density matter for nucleosynthesis one needs a way to eject some matter.  Furthermore, this must be done relatively quickly and in such a way that the weak interactions do not reset the ratio of neutrons to protons.      

In neutron star mergers, see Sec. \ref{sec.mergers}, matter can be ejected gravitationally from the tip of the tidal tail(s), as a result of high pressures induced by the collision, by neutrino heating in the form of thermal winds, by viscous or magnetically driven effects, or by other mechanisms.  Alternatively, neutron-rich matter can be ejected by magnetohydrodynamic effects in jet driven supernovae, perhaps with the formation of a magnetar.  We discuss this in Sec. \ref{Sec:MHD_SN}. Independent of the ejection mechanism, a common question is to what extent weak interactions reset the neutron to proton ratio as the matter is ejected.  These weak interactions will tend to decrease somewhat the ratio of neutrons to protons.

\subsubsection{Liberating neutrons from nuclei with nuclear reactions}
Free neutrons required for neutron-capture processes can also be produced via helium burning through ($\alpha$,n) reactions on slightly neutron-rich isotopes such as $^{13}$C or $^{22}$Ne (like in the $s$-process). The presence of neutron-rich seed isotopes, and therefore the neutron-richness of the helium burning environment stems from prior episodes of hydrogen burning, where $\beta^-$ decays produce neutron-rich nuclei from $N=Z$ seeds - for example via $^{12}$C(p,$\gamma$)$^{13}$N($\beta^-$)$^{13}$C or the CNO cycle producing $^{14}$N, which is later converted into $^{22}$Ne by helium induced reactions. One of the earliest proposed $r$-process scenarios was indeed explosive helium burning in core collapse supernovae, where a shockfront passes through the helium rich outer layer of a star \citep{Hillebrandt1977,Truran1978}. However, the consensus today is that the neutron densities that can be achieved in this scenario fall by far short of what is required for an $r$-process. Rather, non-explosive and explosive helium burning are the widely accepted scenarios to explain s-process nucleosynthesis \cite{Sprocess}, and possibly an i-process \cite{Herwig2011}. 

A variant of this scenario using neutrino induced nuclear reactions has recently been proposed as a possible $r$-process scenario in low metallicity supernovae, where lower neutron densities are needed to achieve a high neutron-to-seed ratio \citep{Banerjee2011}. In this scenario, the intense neutrino flux during the core collapse liberates neutrons in the helium shell via $^4$He($\nu$,$\nu$n)$^3$He(n,p)$^3$H($^3$H,2n)$^4$He and $^4$He($\nu$,$\nu$p)$^3$H($^3$H,2n)$^4$He. 

\subsubsection{Reducing the number of seed nuclei}  For a given number of neutrons, one can synthesize heavier nuclei, if the number of seed nuclei is reduced. Most $r$-process scenarios create their own seeds. In such models, seed production can be suppressed by very rapid expansion timescales leaving less time for the slow seed producing reactions, or by destroying most seed nuclei with a higher entropy. 

\subsection{Neutrino-driven winds}
\label{sec.neutrinowind}


Core-collapse supernovae and their neutrino-driven ejecta are an interesting nucleosynthesis site for the production of heavy elements.  After the successful launch of a supernova explosion, the proto-neutron star cools by emitting neutrinos. These neutrinos deposit enough energy (via absorption and scattering reactions) to power a baryonic outflow of matter with supersonic velocities. This is known as neutrino-driven wind. Already in 1957, core-collapse supernovae were suggested to be the astrophysical site for the $r$-process~\cite{BBFH, Cameron+57}. In core-collapse supernovae, neutron stars form and matter is rapidly ejected, therefore the conditions looked promising for the $r$-process. However, neutrino absorptions on neutrons increase the electron fraction preventing a succesful $r$-process (see Sec.~\ref{Sec:weak_equilibrium}). In the 1990s, delayed neutrino-driven explosions were simulated following the supernova evolution up to several seconds after the explosion. Woosley and collaborators \cite{Meyer+92, Woosley+94} found high entropy and low electron fraction ejecta where the $r$-process produced heavy elements. However, these results could not be reproduced  by similar simulations of other groups \cite{Takahashi.etal:1994}, or by analytic and steady-state models \cite{Qian.Woosley:1996, hoffman.woosley.qian:1997, Otsuki.Tagoshi.ea:2000, Thompson.Burrows.Meyer:2001, wanajo13}


The first few seconds of a core-collapse supernova are characterized by only small changes of neutron star mass, radius, neutrino luminosities, and neutrino energies. Therefore, one can neglect time dependencies and approximate the hydrodynamic problem with steady-state equations. This is only valid for the neutrino-driven wind when assuming spherically symmetric outflow of matter. However, the conclusions found with these kind of steady-state models can be extended to other neutrino-driven ejecta. Steady-state models (see e.g., \cite{Otsuki.Tagoshi.ea:2000, Thompson.Burrows.Meyer:2001}) played a critical role in identifying the appropriate conditions for the $r$-process to occur in core-collapse supernovae. Such conditions are defined by three wind parameters: entropy, expansion time scale, and electron fraction. 
High entropies and short expansion time scales lead to high $Y_{n}/Y_{\mathrm{seed}} \gtrsim 100$ and thus to an $r$-process that can form elements up to uranium. When the neutron-to-seed ratio is low ($Y_{n}/Y_{\mathrm{seed}} \lesssim 1$), i.e. in only slightly neutron-rich winds ($0.45 \lesssim Y_{e} < 0.5$), the weak $r$-process synthesizes heavy elements below the second $r$-process peak by $\alpha$, neutron, and proton capture reactions as well as $\beta$ decays. Since the expansion in the wind is much faster than the $\beta$-decay time-scale, charged-particle reactions become important for moving matter towards heavier nuclei. Proton-rich winds are another possibility favored by current supernova simulations. In this case the  $\nu p$-process \cite{Pruet.etal:2006,frohlich06,Wanajo:2006} can produce elements heavier than $^{64}$Ge along proton-rich nuclei. This is not an $r$-process, but a rapid proton capture process accelerated by (n,p) reactions that bypass slow $\beta$-decay rates using neutrons created by neutrino capture on protons. It is currently not clear if supernova ejecta are slightly neutron-rich or proton-rich \cite{Roberts+12}, or both at different times. Therefore nucleosynthesis studies need to be performed for both conditions in order to explore the potential impact of neutrino-driven winds on the origin of the heavy elements.

The most recent core-collapse supernova simulations (see e.g., \cite{Harris.etal:2017, Eichler.etal:2018, Wanajo.etal:2018}) indicate that the conditions in neutrino-driven winds are most likely proton rich, with some neutron-rich clumps ejected promptly in low mass progenitors \cite{Wanajo.etal:2011}. 
There are still open questions, and the answers may change the prediction of the conditions in supernova ejecta. For example, neutrino oscillations can dramatically change the electron fraction (see Sec.~\ref{subsec.SNnu}), magnetic fields may also affect the wind nucleosynthesis \cite{Thompson2017,Vlasov2017}, and there are still uncertainties in the determination of neutrino matter interactions that affect the electron fraction evolution (for example \cite{RobertsReddy2017}).

Taking into account current simulations and considering their uncertainties, especially for predicting the electron fraction and neutron-richness, core-collapse supernovae may contribute to the production of lighter heavy elements from gallium up to the second $r$-process peak. The existence of an additional $r$-process site for just these elements is supported by observations from ultra metal-poor stars that show a robust abundance pattern for elements with $A\geq130$, whereas they exhibit a large star-to-star scatter below the second peak (see e.g., \cite{Sneden.etal:2008} and Sec.~\ref{lightrpro}). Moreover, the solar system abundances of these elements may also require nucleosynthesis processes in addition to the main r- and various s-processes. The missing process was tentatively called LEPP (lighter element primary process) \cite{Montes.etal:2007,Travaglio.etal:2004}. Therefore, there may exist at least two $r$-processes: one that produces a robust abundance pattern and synthesizes elements up to uranium, and another one that only contributes to the so-called lighter heavy elements (at least Sr-Ag, maybe Ga - Cd). Core-collapse supernovae and their neutrino-driven winds are a possible production site for these lighter heavy elements \cite{qian07, Hansen.etal:2014}.

Given the high frequency of core collapse supernovae and the robust prediction of significant neutrino-driven outflows, neutrino-driven winds are an important nucleosynthesis site and their contribution to the chemical history of our universe must be understood. Advances of our understanding and numerical treatment of neutrino transport in high density matter are needed to reliably predict the neutron-richness evolution of the winds. For a given neutron-richness, reliable $r$-process nuclear physics is needed to predict the range of elements created. As conditions in the ejecta are most likely only slightly neutron or proton rich, the nuclear reactions involved proceed likely closer to stability compared to the main $r$-process, and can be more easily obtained experimentally. Once nuclear physics uncertainties have been reduced and controlled, one can use abundance observations to constrain the conditions in core-collapse supernova and learn about the neutrino spectra and luminosities and thus about the explosion. 

\subsection{Neutron star mergers}
\label{sec.mergers}

The first detection by LIGO and Virgo of gravitational waves (GWs) powered by merging neutron stars~\cite{GW170817}, and subsequent observations of that system in $\gamma$-rays, x-rays, UV, optical, infrared and radio bands~\cite{EM170817} was a remarkable breakthrough for both gravitational wave astrophysics and nuclear astrophysics. In the coming years, we expect the LIGO and Virgo detectors to observe many more mergers of binary black holes~\cite{LigoRate2016} and binary neutron stars, and to start observing mixed black hole--neutron star (BHNS) binaries~\cite{Abadie:2010}.

Theoretical results have long indicated that, in the presence of a neutron star, gravitational wave and electromagnetic observations of binary mergers can put useful constraints on uncertain nuclear physics, including the equation of state of neutron stars~\cite{Flanagan:2008,Lackey:2012,Read:2013} and the origin of heavy elements synthesized through $r$-process nucleosynthesis~\cite{Lattimer+74,Li&Paczynski98,Korobkin:2012uy,Wanajo+14}. Electromagnetic counterparts to the gravitational wave signal can also provide additional information about the properties of the merging objects, and the environment in which the merger occurs~\cite{metzger:11,Metzger:2014,Yu:2013,Metzger:2014b,Siegel2016a,Siegel2016b}. Finally, binary neutron star and black hole-neutron star mergers are generally assumed to be the engine beyond the observed population of short-hard gamma-ray bursts~\cite{Mochkovitch:1993}. 

GW170817, together with the electromagnetic signals observed in its aftermath, largely confirmed the potential of binary neutron star mergers as probes of a wide range of physical processes. In particular, as far as nucleosynthesis is concerned, the observed optical and infrared emission is consistent with the radioactive decay of the ashes of $r$-process nucleosynthesis in a few percent of a solar mass of neutron-rich material ejected by the merger~\cite{GW170817:DynEj,Cowperthwaite:2017}. This observation significantly increases the likelihood that binary neutron star mergers are (one of) the dominant source(s) of $r$-process elements in the Universe. An overview of the three different merger types, their ejected mass and the LIGO detection rate is given in Table~\ref{intro:table1}.

\begin{table}[!htb]
\caption{\label{intro:table1} Material ejected by the three types of compact binary mergers observable by Advanced LIGO, and estimated merger rates within the volume observable by advanced LIGO at design sensitivity (defined as a NS-NS merger detection range of 200 Mpc).}
\footnotesize
\begin{tabular}{@{}lcccc}
\br
Merger type & $r$ process & M$_{ej}$ (M$_\odot$) & Grav. waves & LIGO det. rate/yr \\
\mr
\textbf{NS-NS} & yes & yes & yes & 10$-$200$^{a}$\\
dynamical ejecta &  & $\sim10^{-3}-$0.02 &  & \\
post-merger ejecta &  & $\sim0.01-$0.05 &  & \\
\mr 
\textbf{NS-BH} &  & & yes & 0.2$-200^{b}$\\
low BH spin, high BH mass, small NS$^{d}$& no & no & & \\
high BH spin, low BH mass, large NS$^{d}$ & yes & yes & & \\
- dynamical ejecta & & $\sim10^{-2}$-0.2 &  & \\
- post-merger ejecta &  & $\sim$0.01$-$0.05 &  & \\
\mr
\textbf{BH-BH} & no & no & yes & 36$-$800$^{c}$ \\
\br
\end{tabular}\\
$^{a}$ Rates from~\cite{Abadie:2010}, rescaled for the volumetric rates inferred after the detection of GW170817~\cite{GW170817}.\\
$^{b}$ Rates from~\cite{Abadie:2010} for the advanced LIGO detectors at design sensitivity.\\
$^{c}$ Rate from~\cite{Abadie:2010} for the advanced LIGO detectors at design sensitivity, rescaled for the updated volumetric rates provided in~\cite{LigoRate2016} after the end of O1.\\
$^{d}$ See~\cite{Foucart2012} for a more quantitative division of the black hole-neutron star parameter space.
\end{table}
\normalsize

\subsubsection{Outflow mechanisms}
\label{Sec:NSMerger_Outflows}
The connection between binary neutron star mergers and $r$-process nucleosynthesis provides us with a source of information about the production mechanism and the properties of neutron-rich elements that can complement experiments at RIB facilities.
To understand that connection, and place it in the context of current and future observations of neutron star mergers, it is useful to review the various mechanisms by which matter is expected to be ejected in these systems.

Numerical simulations of mergers have revealed that both binary neutron star and black hole-neutron star mergers can eject a significant amount of rapidly expanding neutron-rich material, providing conditions favorable to the production of $r$-process elements. The properties of these ejecta vary significantly with the parameters of the binary (component masses and spins, eccentricity) and the unknown equation of state of dense neutron-rich matter. Additionally, within a single merger event, numerical simulations have identified various components of the ejecta, with different thermodynamical properties and composition. 

The tidal disruption of a neutron star by a black hole companion can eject cold and very neutron-rich material (e.g.~\cite{Rosswog:2005,Foucart:2013a,Kawaguchi:2015,Foucart:2017}). The merger of two neutron stars can produce similar cold tidal ejecta, complemented by hotter, less neutron-rich material originating from the contact region between the two neutron stars (e.g.~\cite{hotokezaka:13,Sekiguchi:2015,Radice:2016dwd,Dietrich:2017}).
We will call those two components, which are ejected within $\sim 1\,{\rm ms}$ of the merger, the dynamical ejecta.
For both types of binaries, more matter can then be unbound from the post-merger remnant -- either a massive neutron star or a black hole surrounded by a hot, strongly magnetized accretion disk. These post-merger outflows are ejected over much longer timescales, $\sim 0.01-10\,s$ (e.g.~\cite{Fernandez:2013,Just+15,Siegel:2017,Fujibayagshi:2017}).

The outcome of nucleosynthesis in the ejecta from a compact binary merger mostly depends on the composition, entropy, and expansion timescale of the ejected matter. In particular, many $r$-process nucleosynthesis calculations (e.g.~\cite{Meyer+92,Korobkin:2012uy,lippuner15,Martin:2015hxa}) have found that there is a critical threshold of $Y_e \sim 0.25$. For $Y_e \lesssim 0.25$, the full range of heavy $r$-process elements is produced (beyond the third peak, up to uranium), while for $Y_e \gtrsim 0.25$ only a weak $r$-process (up to $A \sim 120$) is possible. Accurate determination of the mass and composition of the ejecta in
numerical simulations is thus an important component in understanding the impact of binary mergers on $r$-process nucleosynthesis. Nuclear physics inputs (e.g. masses and half-lives of neutron-rich nuclides, nuclear reaction rates far away from $\beta$-stability, and fission fragment distributions) are also crucial for $r$-process nucleosynthesis calculations (see Sec. ~\ref{section_NSS}).

In recent years, numerical simulations have made significant progress towards taking into account the many physical effects important to the determination of the ejected mass and its composition. General relativistic and Newtonian simulations of black hole-neutron star and binary neutron star mergers can now use nuclear-theory based, finite temperature, composition dependent equations of state. Neutrino-matter interactions, which play a critical role in the cooling of the post-merger remnant and the evolution of the composition of the ejected material, can be approximately taken into account (see e.g.~\cite{Sekiguchi:2015,Foucart:2015b,Just+15,Perego:2014fma} for algorithms including both neutrino cooling and absorption). The full Boltzmann equations for neutrino transport remain, however, too costly to be directly included in global 3D simulations of mergers. The improvement of approximate transport methods, and the determination of the systematic errors in the composition of the ejecta due to approximate weak reaction rates and neutrino transport, are important objectives for future simulations. Magnetic fields also play an important role in the evolution of the post-merger remnant (see e.g.~\cite{Rezzolla:2011,Neilsen:2014,Kiuchi:2015,Siegel:2017,Paschalidis:2017}). The growth of magnetic fields due to small scale instabilities and the effects of magnetically driven turbulence in the merger remnant are generally under-resolved in existing simulations~\cite{Kiuchi:2015}. Quantitative estimates of the effects of magnetic fields thus remain challenging. The long-term ($\sim 10\,s$) evolution of the merger remnant, which is necessary to determine the properties of the post-merger outflows, also requires additional approximations. Recently, a 3D, general relativistic magnetohydrodynamics simulation of an accretion disk~\cite{Siegel:2017} has been performed. However, most simulations of the post-merger remnant are generally performed in 2D (e.g.~\cite{Fernandez:2013,Just+15}). Despite these limitations, existing merger and post-merger simulations of compact binaries can already predict a number of qualitative features of the ejecta.

\subsubsection{Dynamical ejecta}

The mass and composition of the dynamical ejecta obtained in numerical simulations of black hole-neutron star (BHNS) and binary neutron star mergers vary with the type of binary under consideration. In a black hole-neutron star merger, the disruption of the neutron star by the black hole can lead to the ejection of a large amount of cold, neutron-rich material at velocities $v\sim 0.1-0.3\,c$. These are favorable conditions for a robust strong $r$-process, including fission cycling. We should however note that whether any mass is ejected by the system is sensitive to the binary parameters (black hole mass and spin, neutron star equation of state~\cite{Foucart2012}). For example, for a black hole of mass $M_{\rm BH}=7M_\odot$ and a neutron star of mass $M_{\rm NS}=1.3M_\odot$ (typical masses of galactic compact objects), a neutron star of radius $R_{\rm NS}=11$~km would not be disrupted unless the component of the dimensionless black hole spin aligned with the orbital angular momentum of the binary satisfies $a_{\rm BH}\geq 0.7$. Here the dimensionless spin is the angular momentum in units of the maximum possible value $GM_{\rm BH}^2/c$.  The production of $r$-process elements in black hole-neutron star mergers will thus strongly depend on the distribution of black hole masses and spins, and on the neutron star equation of state.
On the other hand, currently the dynamical ejecta of black hole-neutron star mergers is the only ejecta component for which the nucleosynthesis can be reliably predicted: it robustly produces heavy $r$-process elements. Reasonable estimates of the ejected mass and its velocity have also been derived from simulations~\cite{Kawaguchi:2016,Foucart:2017}.

While a similar cold tidal component is observed in binary neutron star mergers, numerical simulations indicate that the dominant source of ejecta in most of these systems originate from the shocked contact region between the two stars. That ejecta is hotter than the tidal ejecta, and generally less neutron rich due to weak reactions~\cite{Wanajo+14,Sekiguchi:2015,Radice:2016dwd}. Future simulations need to clarify the precise impact of weak reactions on the properties of the shocked ejecta~\cite{Foucart:2016b} and, through nuclear network calculations, on the final abundance pattern in the ejected material. The total amount of ejected mass is generally lower than in black hole-neutron star mergers, and estimates from numerical simulations are currently more uncertain~\cite{Dietrich:2017}. Yet, except for very massive binary neutron star systems in which the remnant promptly collapses to a black hole, most binary neutron star mergers are expected to eject {\it some} neutron-rich material, with a broad range of composition likely to include matter with $Y_e>0.25$.

For both black hole-neutron star and binary neutron star mergers, the amount of mass ejected by the merger is sensitive to the equation of state of neutron stars: larger neutron stars generally eject more mass in black hole-neutron star mergers, and less in binary neutron star mergers. This dependence of the ejected mass on the properties of neutron-rich nuclear matter exemplifies the need for tighter constraints on the properties of high-density matter. Indeed, knowledge of the equation of state is necessary in order to determine the yield of heavy elements for a given compact binary population. 

\subsubsection{Post-merger outflows}
Depending on the binary parameters the central object of the post-merger configuration is either a black hole or a massive neutron star, which eventually can collapse to a black hole on a secular time scale. Except for black hole-neutron star mergers with a high black hole mass, for which the entire neutron star plunges into the black hole, that central object is surrounded by a differentially rotating disk (or torus) consisting of neutron star debris. Current numerical models of post-merger remnants suggest the existence of several agents able to contribute to matter ejection, which we briefly outline in the following.

The hot and dense remnant releases neutrinos at high rates that are comparable to core-collapse supernovae. Neutrino heating is therefore sufficiently powerful to gravitationally unbind the surface layers of the remnant in the form of
a neutrino-driven wind. However, in a black-hole torus remnant the luminosities, and therefore the wind power, quickly decline as torus matter is swallowed by the central black hole, resulting in negligibly small wind masses \cite{Fernandez:2013, Just+15}. In contrast, in a neutron-star torus remnant the central object represents another large energy reservoir for neutrinos. The luminosities decline on much longer timescales and allow for higher wind masses \cite{Metzger:2014, Perego:2014fma, Martin:2015hxa}. Since the neutrino emission in the equatorial direction is obscured by the surrounding torus, most of the neutrino-driven wind is expelled in the polar directions. The electron fraction in the wind is mainly determined by neutrino captures and strongly depends on the lifetime of the NS-torus system. Its general distribution turns out to be slightly higher, $Y_e\sim 0.2-0.5$, than in the case of other post-merger outflows (see below).

The differentially rotating disk around the central object is subject to the magnetorotational instability, which leads to turbulence on small scales and to viscous angular momentum transport and heating on macroscopic scales. While
the disk matter dilutes due to expansion and accretion onto the central object, neutrino cooling eventually becomes inefficient in balancing viscous heating. The increasing thermal pressure then leads to a rapid expansion of the torus
and ultimately to the expulsion of a significant fraction of its original mass mainly, but not exclusively, around the equatorial direction. Since neutrino irradiation is inefficient in the expanding disk, the electron fraction in this
viscous outflow is mainly determined by electron/positron captures and typically freezes out at $Y_e\sim 0.1-0.3$~\cite{Fernandez:2013,Just+15,Metzger&Fernandez14,Fujibayagshi:2017}.

Recent general relativistic simulations also indicate that magnetohydrodynamic turbulence alone can give rise to massive winds early in the evolution of the disk~\cite{Siegel:2017}, and that rapid redistribution of angular momentum within a differentially rotating massive neutron star remnant can unbind $\sim 0.01M_\odot$ of material in the $\sim 10\,{\rm ms}$ following merger~\cite{Fujibayagshi:2017}. 

Although all existing simulations of post-merger remnants still contain considerable simplifications in one way or another, they already indicate that post-merger outflows can be the source of a significant amount of $r$-process elements, essentially within the entire $r$-process range but probably dominated by intermediate mass ($\sim$2nd peak) elements. In addition to the variety of outflow properties that can be encountered in a single merger event, the post-merger outflows also vary with the initial binary parameters and the nuclear equation of state. For example, the amount of neutrino-driven outflows steeply increases with the lifetime of the central neutron star before its collapse to a black hole, and the amount of disk outflows is roughly proportional to the initial disk mass. Future improvements in modeling the merger and in constraining the equation of state will therefore directly translate into better predictions for the properties of post-merger ejecta. 

\subsubsection{Future prospects}
Models of the optical and infrared emission from the ejecta of GW170817 currently favor a two-component ejecta, with $\sim 0.01M_\odot$ of fast ($v\sim 0.3c$), neutron-rich ejecta and $\sim 0.04M_\odot$ of slower ($v\sim 0.1c$), less neutron-rich ejecta~\cite{Cowperthwaite:2017}. These two components could plausibly be associated with, respectively, a shocked dynamical ejecta and a viscously-driven disk outflow. However, at this point, modeling uncertainties place significant limits on our ability to make robust claims for the properties of the ejecta. An upcoming challenge will be to combine neutrino transport, magnetohydrodynamics and general relativity to obtain better predictions for the mass and composition of the ejecta, and to infer its properties and those of the merging objects from such observations. 

Improved nuclear physics models, aided by the insights gained from future nuclear experiments, will play an important role in this process. They will constrain the nucleosynthesis yields of post-merger ejecta, which impact the duration, color, and magnitude of the associated electromagnetic signal~\cite{Barnes&Kasen13}, and are required to assess the role of neutron star mergers in galactic chemical evolution.  

\subsection{Magneto-rotational supernovae}
\label{Sec:MHD_SN}

Some very energetic supernova explosions are observed and cannot be
explained by the standard neutrino-driven mechanism. These rare supernovae are
thought to be driven by a magneto-rotation mechanism \cite{LeBlancWilson70} and may be the explanation
for long gamma-ray bursts. Moreover, these explosions have been also suggested as a potential
$r$-process site \cite{Cameron2003, Nishimura.etal:2006}. During
collapse and post-bounce, the magnetic field of the stellar progenitor
can be amplified by the magnetorotational instability (MRI)
\cite{Obergaulinger.etal:2009} or by the dynamo instability
\cite{Moesta.etal:2015}. The strong magnetic field leads to a
collimation of matter and thus to a jet-like explosion. Moreover,
matter can be promptly ejected by the magnetic field without long exposure
to neutrinos thus maintaining in the outflow the neutron
richness of the proto-neutron star. The resulting relatively neutron-rich outflows are suitable sites for a full $r$-process. 

Magneto hydrodynamic simulations of core-collapse supernovae present 
computational and physics challenges. Investigating these
explosions and their nucleosynthesis requires three dimensional simulations
including detailed neutrino transport and enough resolution to
resolve the MRI. Several efforts have been reported towards this
direction but the results are not yet conclusive. Based on 2D
simulations with parametric neutrino treatment, Nishimura and
collaborators \cite{Nishimura:2015nca, Nishimura.etal:2017} have
compared the effect of neutrinos and magnetic field. They conclude that
for explosions with strong magnetic fields the $r$-process can produce elements 
up to the third peak. However, according to these studies, when neutrinos become the dominant mechanism for ejection of matter, the third $r$-process peak is underproduced. A strong $r$-process up to the third peak has been also found in 3D
simulations that include neutrino leakage and very strong  magnetic fields
($\sim 10^{13}$~G) \cite{Winteler+12, Moesta.etal:2018}. However, such strong
magnetic fields are not very likely to occur. For more realistic
magnetic fields ($\sim 10^{12}$~G), the third peak abundances are strongly reduced or even negligible
\cite{Moesta.etal:2018}. Recent efforts to include realistic magnetic
fields, rotation, and accurate neutrino transport have been reported in
2D simulations \cite{Obergaulinger.etal:2017}.

If a strong $r$-process occurs in magneto-rotational supernovae,
then they contribute to galactic nucleosynthesis with $10^{-3} - 10^{-2}\,M_\odot$ per event
\cite{Winteler+12, Moesta.etal:2018}. These are rare events and are unlikely to account for the majority of $r$-process nuclei in the Galaxy. However, they may dominate shortly after the Big Bang until neutron star mergers, which only occur with a significant time delay, begin to contribute. This may explain  
the early occurrence of $r$-process elements in the Galaxy inferred from the observation of $r$-process abundance signatures in extremely metal-poor stars (see Sec.~\ref{subsection_GCE_low_Z_MW}). The heavy element abundances observed in very weakly $r$-process enriched metal-poor stars may be signatures of these events 
\cite{Nishimura.etal:2017} (see also Sec. \ref{Sec:lowlevelr}).  MHD-supernovae require extremely rapidly rotating progenitor stellar cores, which are likely to be very rare in Nature.  However, the combined effect of a strong magnetic field and even less extreme rotation can impact nucleosynthesis in the proto-neutron star wind phase as a result of magnetic acceleration of the wind material through the seed forming region.  In particular, magnetars born with rotation periods of $\lesssim 10$ ms can produce a successful 2nd peak r-process for the otherwise identical conditions in which a normal (non-rotating, unmagnetized) proto-neutron star wind would fail \cite{0004-637X-659-1-561,Vlasov2017}.



\section{Nuclear sensitivity studies}
\label{section_NSS}

Sensitivity studies explore the detailed dependence of $r$-process observables on specific nuclear properties. Typically, these observables have been the abundances of the produced elements, but other observables such as heating and radiation transport in kilonovae also need to be considered. Such sensitivity studies play a key role in nuclear astrophysics. They guide nuclear experimental and theoretical research, including the development of new facilities, and they provide astrophysicists and astronomers with information about model uncertainties that ultimately limit the information that can be extracted and the conclusions that can be drawn by comparing models with observations. Sensitivity studies provide the intellectual connection between astrophysics and nuclear physics, and are essential for achieving a complete understanding of how the elements were formed in nature. 

Modern sensitivity studies can be categorized into two groups depending on their primary goal: 

(1) {\em Studies that aim at quantitatively propagating nuclear physics uncertainties to astrophysical observables.} These studies provide the nuclear error bars, for example, for the predicted $r$-process abundances. They are critical for enabling meaningful comparisons of $r$-process models with observations. For example, only when discrepancies with observations exceed the nuclear error bars, limitations of the astrophysical model or site can be revealed. The nuclear error bars are also essential to determine the confidence limits for inferring unknown astrophysical parameters from observed abundances, such as temperatures, densities, neutron-to-seed ratios, or neutrino fluxes. The method of choice for these types of studies are Monte Carlo/Bayesian calculations that sample appropriate probability/posterior distributions for all nuclear input parameters, run a particular model for each sample, and thus determine the corresponding uncertainties of predicted observables. Common limitations are unknown input probability/prior distributions, computational cost, and the strong correlations between the various nuclear physics uncertainties in the case of nuclear theory predictions on which most $r$-process calculations rely heavily. 

(2) {\em Studies that aim at identifying the nuclear uncertainties that most strongly affect observables.} These studies focus on providing an appropriate relative ranking of importance of nuclear physics inputs that can guide and focus future work in nuclear theory and experiment. While correlations in Monte Carlo/Bayesian studies can be used to identify at least some of these critical nuclear physics quantities, another common approach is a 
variation of individual nuclear physics properties. The key for these types of studies is the definition of a figure of merit that quantifies the importance of a particular variation in the predicted observables. This depends on the scientific question that is being explored by a particular comparison between model and observations. Example questions could be: (a) what are the astrophysical parameters needed to get the best match to the complete solar $r$-process abundance pattern?  (b) can the observed Eu/Ba ratio in metal-poor stars be created by the $r$-process? Clearly the figure of merit for a sensitivity study would be very different for these two questions. For (a) some global measure such as the sum of all abundance changes may be appropriate. The range of astrophysical parameters needed to achieve a new best fit for each nuclear physics variation would be an even better choice. For (b) the impact on the Eu/Ba ratio would be the best choice. Clearly for each case the resulting key nuclear physics parameters, and therefore the most important nuclear physics needs, can be very different. 

It should also be emphasized that sensitivity studies depend on the astrophysical model used. While there are certain nuclear properties that are needed in a broad range of models, there are many others that are only relevant for certain, or even just a single model. This is not a limitation. It just means that nuclear physics work is not necessarily ``important for the $r$-process", but provides the data needed to use and validate one particular model. As long as this astrophysical model is a reasonable choice, the work will be important to make progress. This underlines the importance of close collaboration between nuclear physicists and astrophysicists. 

In this context, the recent observations of the neutron star merger GW170817 via gravitational waves \cite{GW170817} and the associated kilonova \cite{EM170817} play an important role for nuclear physics. While the observation does not eliminate the need to study a broader range of alternative $r$-process sites, it provides a fresh impetus for studying neutron star mergers as a key site. This should serve as a strong justification to redirect significant effort in astrophysical modeling, associated sensitivity studies, and nuclear physics to the neutron star merger scenario. This is very important, as it is not feasible for the field to explore the large number of proposed $r$-process sites with equal effort, and because the neutron star merger site, especially because of its inherent lack of spherical symmetry and the range of astrophysical conditions, is a particularly challenging problem that requires significant resources and therefore a strong motivation. 

\subsection{Overview of important nuclear physics in the $r$-process} \label{sec:sens-input}

From its inception the study of $r$-process nucleosynthesis has been inextricably linked to the details of the structure of neutron-rich nuclei and found to depend on the nuclear uncertainties of far-from-stability isotopes. In fact, the existence of an $r$-process in nature is inferred from the prior understanding of $N=50$ and $N=82$ shell effects in neutron-rich nuclei \cite{BBFH, Cameron+57}. Because of these nuclear structure effects, an $r$-process provides the most natural explanation for the observed peaks in the cosmic abundance distribution of the elements. 

It was recognized from an early stage that masses and $\beta$-decay rates play the most dominant role in the determination of the abundances \cite{Seeger+65,Cwn1991,kratz1993}. This is because for the hot and neutron-rich conditions expected for most $r$ process sites, an equilibrium is established at least for some period of time, between neutron capture and photodissociation; while $(n,\gamma)$-$(\gamma,n)$ equilibrium persists, relative abundances within individual isotopic chains are determined by a Saha equation that depends primarily on neutron separation energies and nuclear partition functions (the spins of ground and low lying $E_x < kT \approx 100$~keV states). The total summed abundances in each isotopic chain are set by the $\beta$-decay lifetimes that connect them. $\beta$-decay half-lives are therefore another important nuclear ingredient that determine the speed of the $r$-process and the relative abundances in each isotopic chain. $\beta$-delayed emission of neutrons also becomes important once equilibrium fails, in determining decay paths back to stability and in providing additional neutrons for late time captures \cite{Mumpower2016b}. 

Modern reaction network calculations of the $r$-process now include theoretical predictions of additional nuclear properties, e.g., neutron capture rates as well as neutron induced, $\beta$-delayed and spontaneous fission, and associated fission fragment distributions. Individual neutron capture rates are important after $(n,\gamma)$-$(\gamma,n)$ equilibrium fails, typically towards the end of the $r$-process when the neutron abundance drops and neutron capture lifetimes increase. They are also critical in $r$ processes that occur in cold conditions where equilibrium is established only briefly if at all, and the $r$-process path is instead determined by competition between neutron capture and $\beta$-decay rates. Fission rates and product distributions are crucial for very neutron-rich $r$ processes characterized by fission recycling. 

The importance of the nuclear physics ingredients discussed so far on $r$-process models has mostly been determined by comparing model results obtained with different choices for nuclear theory predictions, such as mass models, $\beta$-decay models, or fission models or by simply repeating an $r$-process model calculation after a particular nuclear physics quantity has been measured in the laboratory. Examples for more systematic attempts to link certain isolated abundance features or discrepancies with certain specific nuclear structure features include the detailed analysis of 
Kratz \textit{et al.} \cite{kratz1993}; the classical $r$-process model study of Chen \textit{et al.} \citep{Chen1995}, who studied the impact of quenching of the spherical shell gaps far from stability on abundances just below the $A=130$ and $A=195$ peaks; the study of Schatz \textit{et al.} \cite{schatz02}, who explored the link between a possible $N=184$ shell closure and the $r$-process production of uranium and thorium; the work of Baruah \textit{et al.} \cite{Baruah2008}, who performed a quantitative analysis of the impact of mass uncertainties around $^{80}$Zn on the astrophysical model conditions needed for the synthesis of a $A=80$ abundance peak; and the recent analysis of Lippuner \textit{et al.} \cite{lippuner15}, who identified the individual contributions of longer-lived isotopes to the heating of kilonovae. 

\subsection{Results from large scale sensitivity studies for the main $r$-process}
\begin{figure}[!htb]
	\centering
		\includegraphics[width=1.0\textwidth]{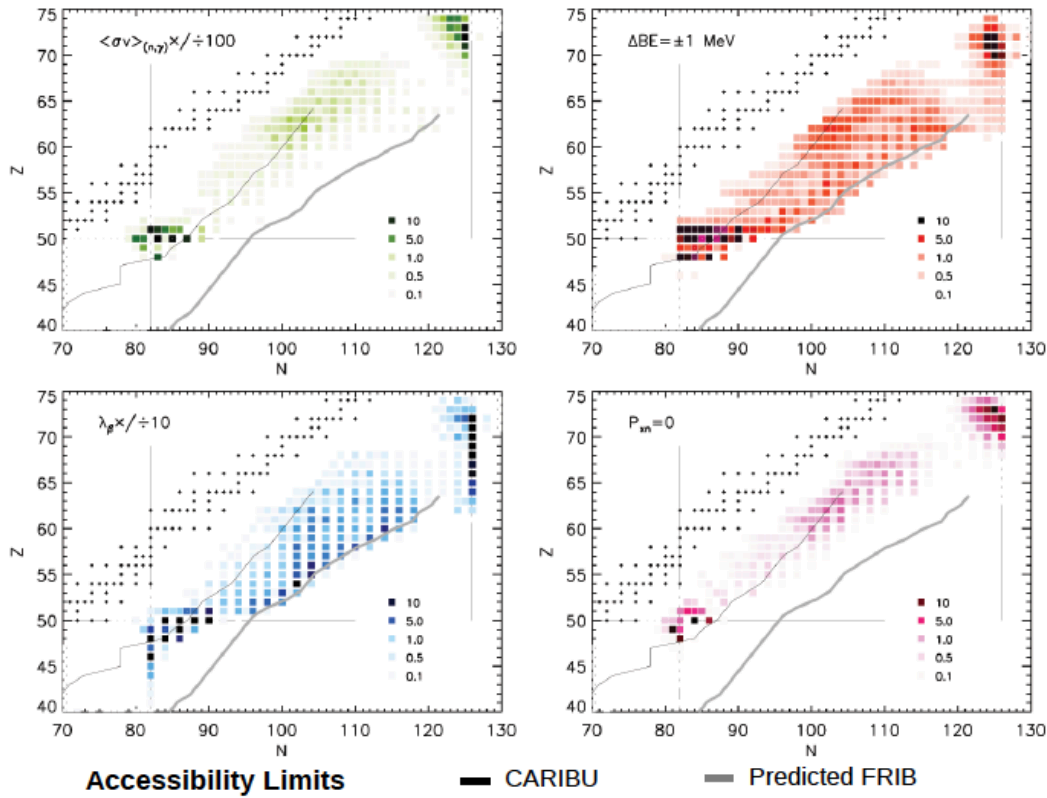}
	\caption{Nuclear properties that have the strongest global effect on the isotopic abundances produced in a $\nu$-wind $r$-process model. Shown are the most important neutron capture rates (upper left), masses (upper right), $\beta$-decay rates (lower left), and $\beta$-delayed neutron emission branchings (lower right). These nuclei should be a priority for nuclear experiment and theory to best constrain the global $r$-process abundance pattern, in particular the main abundance peaks. The reach of existing facilities (example CARIBU at ANL) and of next generation facilities (example FRIB) is indicated by the thin and thick gray lines, respectively. Modern neutron star merger models predict similar conditions.
	\label{fig:sensitivity}}
\end{figure}

Recently more systematic large scale sensitivity studies that pinpoint important individual nuclear properties in $r$-process models, and first complete $r$-process Monte Carlo studies that propagate nuclear errors to abundance observables have been carried out. The recent advances are summarized in the recent review of Mumpower\textit{ et al.} \cite{Mumpower2016}. The studies focus on the main $r$-process, i.e., the synthesis of $A \ge 120$ nuclei, and examine model sensitivities to atomic masses, $\beta$-decay half-lives, $\beta$-delayed neutron emission probabilities, and neutron capture rates (Fig.~\ref{fig:sensitivity}).

For these global sensitivity studies, a baseline abundance pattern is generated with chosen nuclear physics inputs for a fixed astrophysical trajectory. Next, individual nuclear properties are varied and compared to the baseline. The impact of the variation is measured using certain metrics, e.g., the sum of the absolute final abundance changes relative to the baseline. The study is repeated for each of a set of five $r$-process trajectories that, while not necessarily realistic, are thought to sample a wide range of conditions encountered in various possible $r$-process models. Typical results are shown in Fig.~\ref{fig:sensitivity}. The important masses are spread over a broad range of neutron-rich nuclei, with some concentration near the $N=82$ and $N=126$ spherical shell closures (the models focus on the main $r$-process, which excludes the $N=50$ region) and the rare earth region. For $\beta$-decay half-lives, the most important nuclei are more concentrated. They are located at the spherical shell closures at $N=82$, $Z < 50$ and $N=126, Z<73$ and along the path of the main $r$-process, especially within 8 neutrons of $N=82$ and in the rare earth region. The most important $\beta$-delayed neutron emitters are located at $N=84-88$ in the neutron-rich tin region. The important neutron capture rates depend strongly on the $r$-process trajectory but are in general closer to stability than the important masses and $\beta$-decay-properties, just below $N=82$ and $N=126$. This reflects the fact that neutron captures only become important towards the end of the $r$-process during freeze-out when neutron densities drop and the $r$-process reaction path is moving closer to stability. 

The fact that sensitivity studies identify as most important the properties of nuclei near closed shells and the rare earth region is no surprise. These nuclei tend to slow the reaction sequence and create the well known $A=130$, rare earth ($A\approx165$), and $A=195$ peaks in the final abundance distribution. Any changes that affect the $r$-process speed through these regions will lead to global changes in the final abundance pattern, which is the chosen figure of merit of these sensitivity studies. 

The recent review \cite{Mumpower2016} also summarizes first Monte Carlo studies that propagate uncertainties (see Sec.~\ref{Sec:NucTheory_Current}) in nuclear masses, $\beta$-decay rates and neutron capture rates to the final abundances for some of the same $r$-process model trajectories. These calculations led to a number of important conclusions. First, the $r$-process model trajectories chosen do not reproduce very well the solar $r$-process abundance distribution in all mass regions above $A\sim120$ with discrepancies ranging from factors of 2-3 to an order of magnitude in some places. The Monte Carlo studies demonstrate that only the mass uncertainties come close in possibly explaining these discrepancies, and only barely, implying that there are likely additional non-nuclear physics issues in the astrophysical models. This illustrates the importance of error propagation studies in judging the quality of an $r$-process model. The finding that, compared to $\beta$-decay rates and neutron capture rates, mass uncertainties lead to by far the largest uncertainties in the final abundances is another important result in itself. This result was obtained assuming uncorrelated mass uncertainties of 500~keV, which is a comparable value to what mass models achieve for experimentally known nuclei. However, uncertainties for predicting unknown masses are likely higher (see Sec.~\ref{Sec:NucTheory_Current}) and they are certainly correlated. In addition, these first Monte Carlo mass sensitivity studies do not account for the full range of mass-related model changes in other quantities such as $\beta$-decay properties. An example are changes in the statistical decay from final states. For all these reasons it is likely that uncertainties in the predictions of masses of $r$-process nuclei are underestimated. 

In contrast, $\beta$-decay half-lives and neutron capture rates lead to significant smaller uncertainties though they can reach about an order of magnitude. Clearly better nuclear physics is mandatory for any meaningful comparison of $r$-process model calculations with observations. These first studies suggest that in order for details of the abundance patterns to stand out over nuclear uncertainties, a mass accuracy of better than 100 keV is needed and $\beta$-decay and neutron capture rates should be known to within a factor of two \cite{Shafer2016, Liddick2016}.

\subsection{The special case of the rare earth peak}
The rare-earth peak at $A$$\approx$165 and $Z$=58$-$71 is a distinct signature of $r$-process nucleosynthesis. Its formation is subject to ongoing research (see e.g. \cite{Surman+97, Mumpower+12a, Mumpower+12b, Mumpower2016c, Mumpower2017}). In contrast to the $A$=130 and 195 abundance peaks, which arise from the accumulation of material at the $N$=82 and 126 shell closures, the mini-peaks at $A$$\approx$100 and 165 may originate from the decay of isotopes through regions of double sub-shell closures or deformed single particle energy gaps (e.g. at $Z$=40, $N$+64 and $Z$=64, $N$=106). In the case of the rare earth peak a so called `dynamical' formation mechanism during the decay back to stability has been proposed. The isotopes in the relevant mass region are strongly deformed (for the even-even nuclei this corresponds to a large quadrupole deformation $\beta_2$) which may lead to a localized enhancement in stability that causes the rare earth peak to form. Another possible formation mechanism is strongly asymmetric fission of neutron-rich actinides. 

The significance of these two formation mechanisms is that they are intimately coupled to the astrophysical conditions. While the dynamical mechanism can potentially operate in both hot and cold freezeout conditions, the fission formation mechanism requires more extreme conditions where fission recycling can occur, such as the tidal ejecta of neutron stars. Further, the dynamical mechanism formation can be studied in the laboratory at RIB facilities offering a path forward in ruling out this possibility (e.g. in the case that no feature is found in nuclear structure) and in understanding the late-time $r$-process conditions. In either case, the properties of the involved nuclei play an important role for understanding the $r$-process.

During extremely neutron-rich conditions, rare-earth nuclei with $S_\mathrm{n} \sim 2-3$~MeV will set the $r$-process path. In this phase, the nuclear properties shape the peaks and troughs in the abundance pattern \cite{Martin:2016}. During freeze-out, the radioactive progenitor nuclei will decay to stability and form the final $r$-process abundance distribution. 
As $\beta$-decay drives the abundances towards less neutron-rich nuclei, the shapes of the relevant nuclei may change. This induces changes in trends for nuclear masses and neutron capture rates that affect the final abundances. However, the location of these shape transitions on the chart of nuclides are predicted differently by various theoretical models. 

\begin{figure}[!htb]
	\centering
		\includegraphics[width=0.75\textwidth]{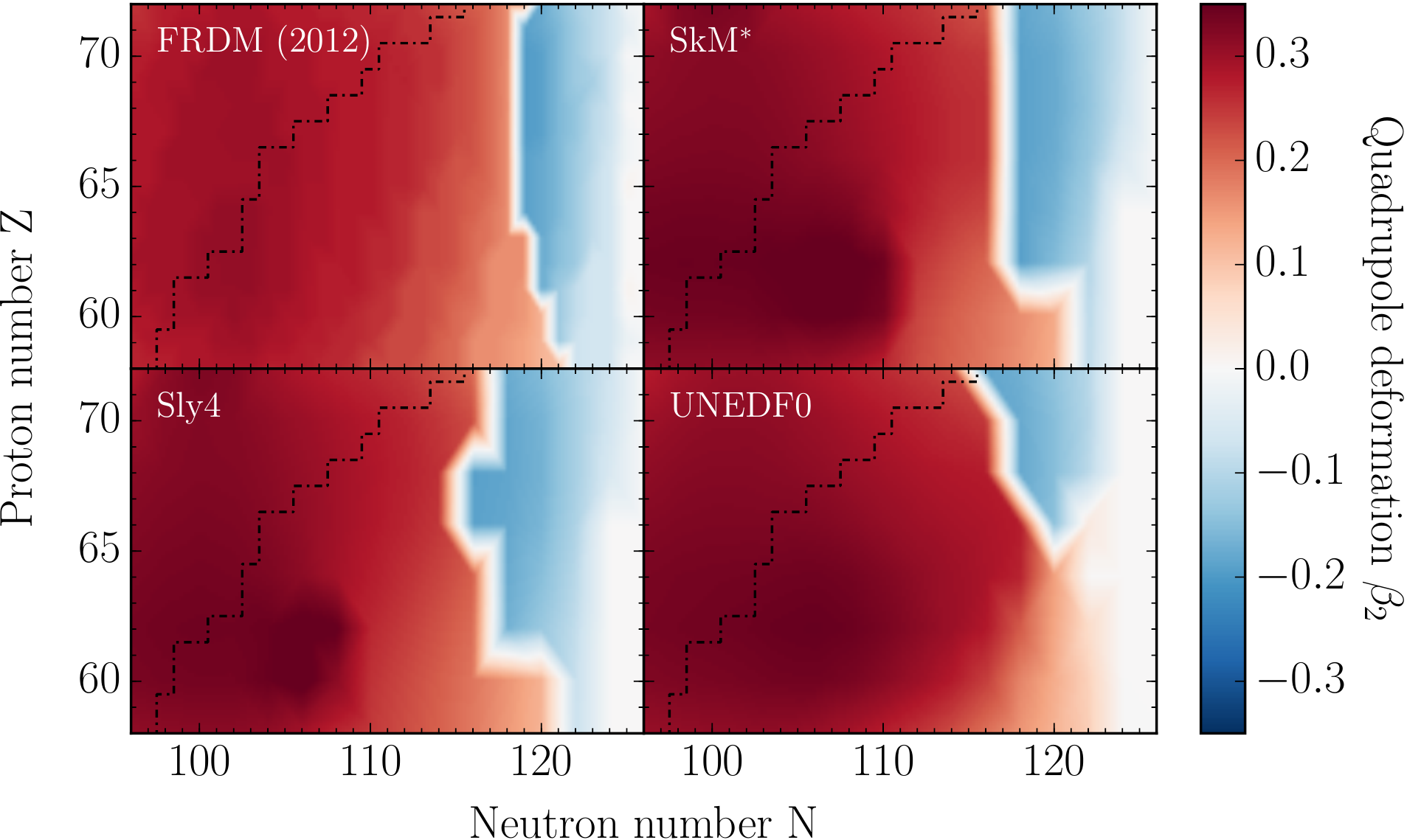}
	\caption{Quadrupole deformation $\beta_2$ as predicted by the mass models FRDM \cite{Moller2016}, SkM$^*$ \cite{Bar82}, SLy4 \cite{Chabanat1998} and UNEDF0 \cite{UNEDF0}. Note that the quadrupole deformations of odd-A and odd-odd nuclei are interpolated from the predicted values for even-even nuclei in the last three models. The dotted-dashed line marks the limit of known (neutron-rich) nuclei (as given on the NuDat website, \url{http://www.nndc.bnl.gov/nudat2/}).}
	\label{fig:quadrupole-deformation}
\end{figure} 

Figure~\ref{fig:quadrupole-deformation} shows the quadrupole deformation $\beta_2$ for even-even nuclei in the region $N=100-126$ for the four mass models FRDM 2012 \cite{Moller2016}, SkM$^*$ \cite{Bar82}, SLy4 \cite{Chabanat1998} and UNEDF0 \cite{UNEDF0}. The limit of currently known nuclei is marked by a dotted-dashed line. The region of largest deformation around the mid-shell closure is visible as well as the region for a sudden shape transition towards the $N$=126 shell closure.

For example, already the isotopes of Yb ($Z=70$) are predicted to show a rapid shape transition. The presently heaviest known Yb isotope is $^{180}$Yb ($N$=110), and the shape transition is predicted to be between $N$=116 and $N$=118 in SkM$^*$, SLy4, and UNEDF0, whereas for the FRDM(2012) it is shifted to $N$=118-120. 
Pushing further neutron rich into this part of the present "Terra Incognita" will be possible with the new fragmentation facilities FRIB and FAIR. Knowledge about the properties of rare-earth isotopes in this region will help to constrain and improve nuclear models as well as understand this important piece of $r$-process nucleosynthesis.

\subsection{Reverse engineering nuclear properties from $r$-process abundances}
\label{sec:rep}
In the traditional sensitivity studies discussed above, nuclear physics properties are used as input in $r$-process model calculations to determine their impact on the predicted $r$-process abundances. Features in the nuclear structure, such as shell closures, translate then into features in the calculated abundances. It has recently been shown that working in the reverse direction is also a powerful tool to illuminate the interdependence between nuclear physics and astrophysical observables \cite{Mumpower2016c}. In this approach one takes local features of the observed $r$-process abundance pattern together with conditions suggested by astrophysical simulations and predicts the trends in the nuclear structure necessary to produce the observed features. 

The method has recently been demonstrated in Refs.~\cite{Mumpower2016c, Mumpower2017}, who reconstructed the mass surface required to reproduce the rare earth abundance peak for various astrophysical conditions. A Markov-chain Monte Carlo approach has been used, where  neutron capture and $\beta$-decay rates have been consistently updated as the nuclear mass surface is varied. The variation of the mass surface uses a relatively featureless baseline mass model, and parametrizes deviations from this baseline trend. The result of these calculations are predictions for nuclear masses with quantitative uncertainties. For a given astrophysical model, masses must lie in this range to reproduce the solar isotopic $r$-process abundances in the rare earth peak, {\em provided that other nuclear structure input is correct}. Different results are obtained for different astrophysical conditions, making it possible one day to distinguish among the range of possibilities using new measurements at RIB facilities. 

\begin{figure*}[htb] \centering
\includegraphics[width=0.9\textwidth]{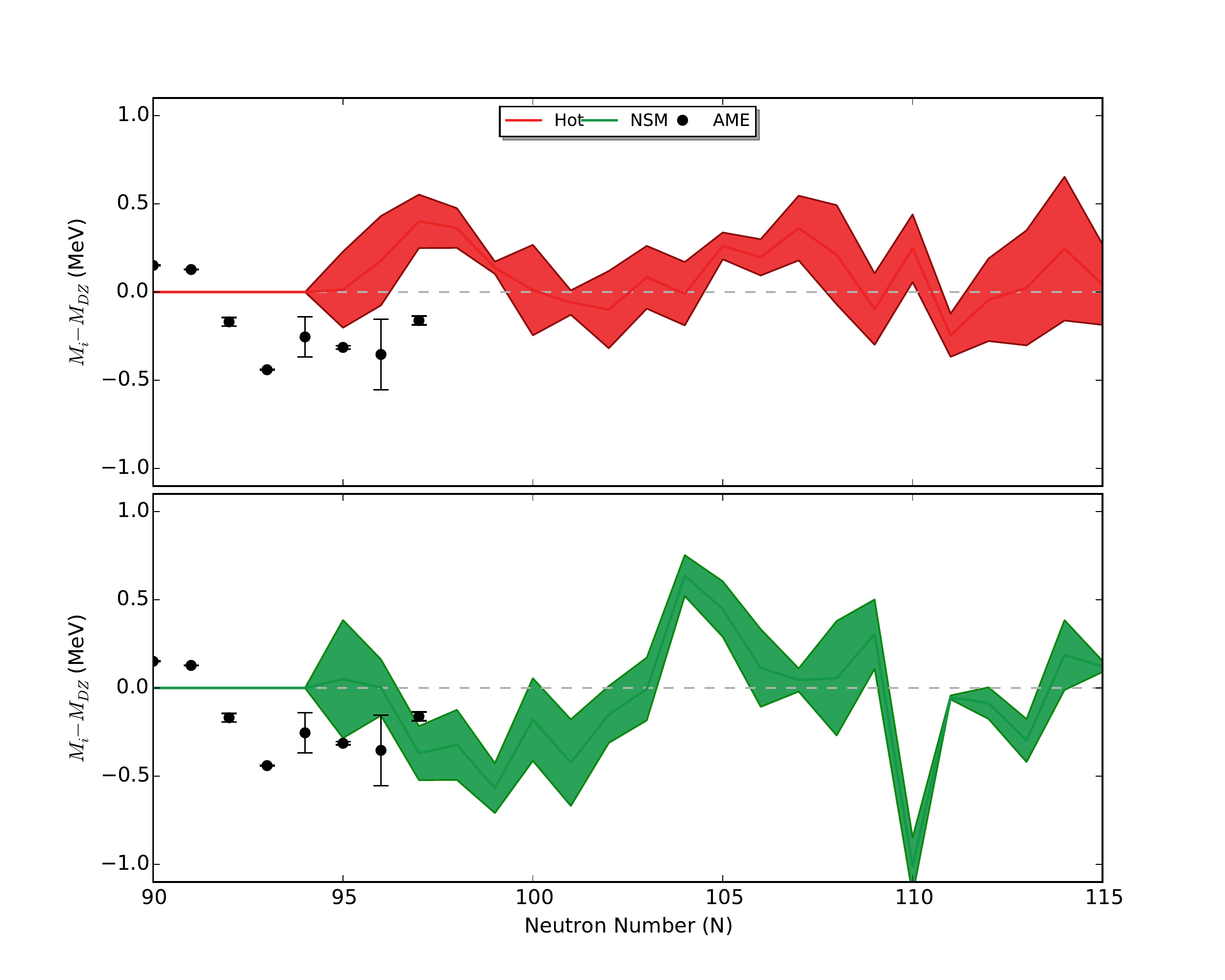}
\caption{Differences in mass datasets from Duflo-Zuker along the Z = 60 (Nd) isotopic chain. The shaded regions show the predicted change to the Duflo-Zuker mass surface using a Markov chain Monte Carlo technique for a hot (red) and very neutron-rich cold (green) $r$-process. Points show experimental data from the Atomic Mass Evaluation Audi et al. (2012). Note peak formation depends on trends in the mass surface, not the absolute values of the masses. Fig. from \cite{Mumpower2016c}}
\label{fig:inversion_example}
\end{figure*}

An example is shown in Fig. \ref{fig:inversion_example}, which shows that this method can be a very powerful tool to to guide future experimental and theoretical efforts, in particular in terms of how future experiments can discriminate between different sets of astrophysical models and conditions. We therefore discuss this method in the context of sensitivity studies. Compared to schematic sensitivity studies that vary individual masses independently and assume no inter-nuclei correlations exist, this reverse engineering approach has a number of advantages. One key advantage is that the method takes into account correlations in nuclear uncertainties that lead to systematic changes in masses along isotopic and isotonic chains.

In addition, the reverse engineering approach provides more meaningful guidance for experiments. The method identifies the key quantities that need to be measured: the masses predicted with the smallest uncertainties and the strongest deviations from the baseline trend. Moreover, the prediction of mass trends allows experimenters to  draw meaningful conclusions from early, incomplete data, and offers opportunities to test astrophysical models by exploring mass surface trends closer to stability or in nearby, easier accessible, mass regions.

Another important feature of the reverse engineering approach is that the differences in the required masses for different astrophysical models or sites provide a direct measure of the astrophysical model discrimination power of a particular mass measurement. These differences directly indicate the required mass accuracy for a meaningful experiment. If the differences are large, already a lower accuracy mass measurement may provide important insights into astrophysical $r$-process sites as long as the right nuclei are measured. An example is shown in Fig. \ref{fig:inversion_example}. For very cold neutron-rich conditions, e.g., tidal tails of neutron star mergers, the predicted mass surface is different compared to hot conditions, e.g., winds from massive neutron stars, accretion disks, or the proto-neutron stars of core collapse supernovae. Clearly, mass measurements of $^{163-166}$Nd with less than 100~keV uncertainty would be most important to calibrate models and to test the cold $r$-process hypothesis. These results clearly demonstrate that nuclear measurements are key in discriminating between different possible $r$-process models and sites. 

Looking to the future, this method can be improved on a number of fronts. First, experimental nuclear data should be included in the study to provide another anchor for the limits of model variation. Second, additional uncertainties from other relevant quantities such as $\beta$-decay and neutron capture rates should be included simultaneously with mass uncertainties. More robust nuclear models for the description of fissioning nuclei may also impact the predicted trends depending on the astrophysical conditions studied. 

The approach is general enough to expand focus on other parts of the $r$-process abundance pattern. Taking into account multiple abundance features simultaneously would further increase model discrimination power. In the end this would provide an excellent method to directly test a given $r$-process model: if the reverse engineering fails to find a set of nuclear properties that is compatible with experimental nuclear data, within uncertainties, the model can be rejected. As better nuclear data become available, the method will become more and more powerful.

\subsection{($\alpha$,n)-rate sensitivities in the weak $r$-process}
Fewer nuclear sensitivity studies focus on models producing the lighter $r$-process elements between Se and Sn, the so called weak $r$-process. Surman \textit{et al.} \cite{Surman2014} carr ied out extensive studies focusing on the neutron capture rate sensitivity, using about 90 different trajectories from a broad range of models. In about 55 of these trajectories a significant dependence on neutron capture reaction rate uncertainties is found. The important reactions are located in the neutron-rich $Z=26-34$ region, about 6-12 isotopes away from stability. 

A recent study pointed out the importance of the rates of ($\alpha$,n) reactions on slightly neutron-rich nuclei between Fe and Rh for the nucleosynthesis of elements in the Sr-Ag range \citep{bliss16}. In the weak $r$-process, the seed nuclei created by charged particle reactions such as ($\alpha$,n) are only moderately processed by the subsequent weak $r$-process. The final abundances therefore retain signatures of the charged particle reaction sequence. 

The identification and reduction of nuclear physics uncertainties is particularly important for addressing the open question of the role of the weak $r$-process in the origin of the elements. Accurate nuclear physics is mandatory for 
disentangling the weak $r$-process contribution from possible other sources of light ``$r$-process" nuclei such as $\nu$p-, charged particle, or $i$-processes.

\subsection{Outlook for nuclear sensitivity studies}

The studies discussed in the preceding sections mark just the beginning of our understanding of the detailed connection between individual nuclear properties and $r$-process model predictions. An important aspect for providing meaningful guidance to nuclear physics is the choice of the $r$-process model to be investigated. With neutron star mergers now observed as a possible site of $r$-process nucleosynthesis, future sensitivity studies should explore their various major ejecta components in more detail (see Sec.~\ref{Sec:NSMerger_Outflows}). The first Monte Carlo studies \citep{Mumpower2016} revealed statistically significant discrepancies between predicted and inferred solar abundances. As a next step it will therefore be important to refine the choice of models that agree with both the final $r$-process abundances and the most recent observational constraints from kilonovae. Owing to the studies carried out so far we are now in a position to quantify what ``better'' means. 

In addition, the range of nuclear physics inputs that are considered in sensitivity studies needs to be broadened. So far, studies have mostly focused on masses, $\beta$-decay properties, neutron capture rates, and ($\alpha$,n) rates. In the future, it will be important to include the sensitivity to uncertainties in nuclear fission properties (e.g.,  lifetimes and  yield properties) which almost certainly play a key role in many $r$-process models \citep{Goriely2015,Eichler2015,Giuliani2017} as well as other charged particle reaction rates. 

Another important question is the choice of metrics that measures the sensitivities. The most common choice is the sum of all absolute abundance residuals. This global measure is a good choice for identifying nuclear physics properties that affect the overall $r$-process conditions required for a successful $r$-process, but overemphasizes the most abundant $r$-process nuclei in the main $r$-process peaks. There is a broad range of open questions related to the $r$-process, and each requires tailored metrics to determine the key nuclear physics uncertainties. For example, when investigating the conditions required for more local abundance features such as the rare earth peak, the use of relative abundance changes as sensitivity metrics is  a better choice \cite{Mumpower2015}.

\begin{figure*}
 \begin{center}
  \centerline{\includegraphics[width=5.25in]{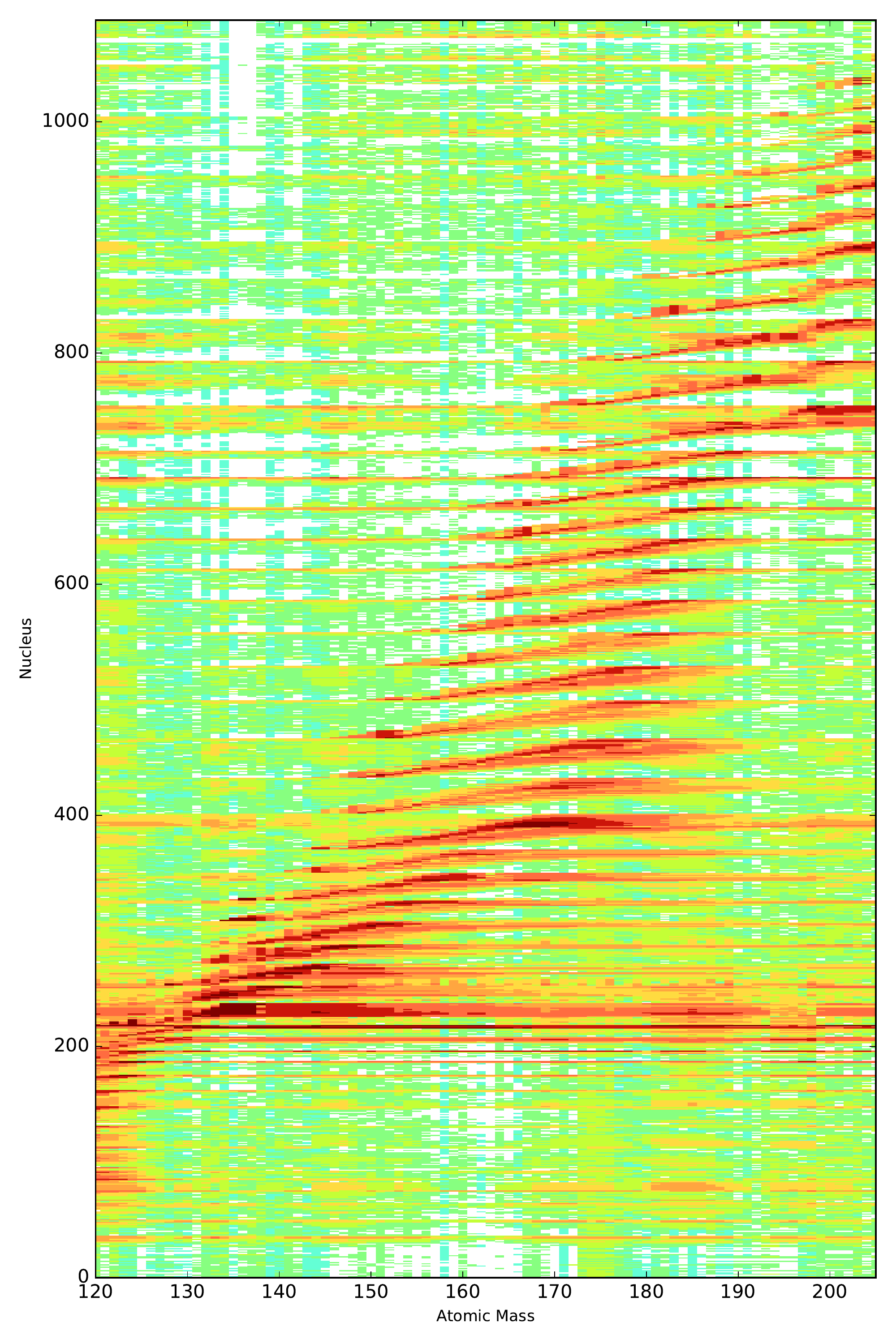}}
  \caption{\label{fig:nss-heat-map} A `heat map' showing the sensitivity of final abundances to nuclear masses in a neutron star merger $r$-process. Darker shades of red denote a larger influence on predicted abundances while green and white indicates smaller and no influence respectively. Only a few nuclei have a global impact on abundances by shifting material across a large range in atomic mass, which is indicated by the extent of the dark red shading across the x-axis \cite{Mumpower2016}. }
 \end{center}
\end{figure*}

The challenge is that there is a very large number of abundance observables one may be interested in. Sensitivity studies with global metrics do not provide information on which nuclear property affects which abundances. A possible solution to this problem is a `heat map' style analysis. An example is shown in Fig. \ref{fig:nss-heat-map}, which shows the sensitivity of all final abundances to all nuclear masses in a neutron star merger $r$-process. In this study individual nuclear masses were varied by $500$ keV and all of the nuclear properties that depend on this mass were propagated self-consistently \cite{Mumpower2015}. The color scale is defined so that dark shades of red denote a larger influence on predicted abundances while shades of green denote smaller influence. White indicates that the particular nucleus had no influence on the respective final abundance. A long dark red shading along the x-axis direction indicates broad global impact of the mass uncertainty, while small red areas indicate a more local impact on the final abundances. The nuclear masses are sorted by mass number. The diagonal correlation visible in the figure reflects the fact that the masses of heavier nuclei tend to also affect heavier final abundances. The low sensitivity in the upper left part of the plot indicates that, as expected, the mass of a nucleus mostly affects the abundances further along the $r$-process path. 

Realistic input uncertainties also are a prerequisite for realistic uncertainty predictions, especially for Monte Carlo studies. It is well known that discrepancies in predictions of masses \citep{Mumpower2016} and neutron capture rates \citep{Liddick2016} far from stability can exceed the deviations between models and experimental data closer to stability (cf. Sec.~\ref{Sec:NucTheory_Current}). The average deviation of theory and experiment, which is often used in sensitivity studies as a measure of uncertainty, therefore can be a poor predictor of theoretical uncertainties far from stability. The development of nuclear theory approaches that determine self-consistent uncertainties is therefore of particular importance for understanding the $r$-process. A recent example of this approach is the density functional theory (DFT) based mass predictions that provide statistical and systematic uncertainties \citep{Erl12a} (see also Sec.~\ref{Sec:NucTheory_New}). 

Another important issue are correlations among uncertainties. Early schematic sensitivity studies have assumed uncertainties be independent, which is certainly an incorrect assumption for theoretically predicted nuclear properties \cite{Rei10,Dob14}. One solution would be to use the nuclear ingredients that enter nuclear theoretical models as input parameters, and propagate their uncertainties through the nuclear models, and then through the astrophysical model to the final abundances. This approach has recently been attempted by \citep{Bertolli2013} for the intermediate neutron capture process ($i$-process). However, there are drawbacks to this approach: first of all, the uncertainties of nuclear structure ingredients such as level densities or nuclear potentials are even more difficult to characterize far from stability as experimental data are much more limited. Second, once the sensitivity is established it is difficult to use the information to guide specific experiments. Third, the results will be strongly nuclear model-dependent. Nevertheless this may be an interesting approach, especially for nuclear models with few free parameters, or to provide guidance for experiments that specifically target nuclear structure ingredients for nuclear theory, such as the $\beta$-Oslo method (see Sec.~\ref{sec:exptechniques:ncap}). An alternative approach to correlated uncertainties has recently been used by \citep{Martin:2016}, who use a well defined set of different energy density functionals to create a range of mass predictions for $r$-process studies. The resulting $r$-process predictions form an abundance uncertainty band that takes into account the correlated uncertainties of the model, for example in predicting shape transitions and systematic trends in neutron separation energies near closed spherical shells. 

\section{Experimental methods}
\label{sec_exp}


Nuclear physics is critical for understanding the $r$-process and addressing current open questions as it links astrophysics to $r$-process observables such as kilonova light curves and abundances. Nuclear physics also enables the prediction of the contributions from individual components in the many cases where mixtures of processes are observed. For example, abundance observations in metal-poor stars provide a compositional snapshot of the material the star is formed of, and even at low metallicities multiple events may have contributed. Similarly, observations of kilonovae from neutron star mergers probably show mixtures of distinct processes within the merger event, such as dynamical ejecta, jets, and winds (see Sec.~\ref{Sec:NSMerger_Outflows}). With reliable nuclear physics, the relative contributions from different processes can be disentangled, a pre-requisite for using observations to infer the respective astrophysical conditions. 

In general, current nuclear theory predictions are not sufficiently accurate (see Sec.~\ref{Sec:NucTheory_Current}). Experimental data are therefore required for critical quantities in $r$-process models, and to guide theory towards more reliable predictions of the properties of exotic nuclei that remain out of experimental reach.

The nuclear physics quantities needed for $r$-process studies depend somewhat on the particular $r$-process model. Data need to be provided for each reasonable model, so as to enable the testing of this model against observations. For the breadth of models available today the following types of experimental data are needed: $\beta$-decay properties, including decay rates and branchings for $\beta$-delayed neutron emission, masses, and partition functions (spin and parities of the low lying $\lessapprox 1$~MeV excited states). Neutron capture rates are essential, but there is currently no technical solution for carrying out direct measurements. Therefore, indirect measurements need to provide information for reaction and structure theory to better predict neutron capture rates. Measurements that can contribute to reducing uncertainties in predicted neutron capture rates include transfer cross sections, breakup cross sections, level densities, and $\gamma$-strength functions. Another essential nuclear physics quantity are rates for neutron induced and $\beta$-delayed fission, as well as the corresponding fission fragment distributions. 

Because $r$-process calculations will rely for the foreseeable future on theoretical nuclear physics data to complement experimental information, $r$-process nucleosynthesis models benefit strongly from experimental information on neutron-rich nuclei that improves our understanding of their structure. Nuclear structure effects such as (sub)shell closures, or changes in deformation can have a strong impact on $r$-process calculations \citep{kratz1993,Chen1995,Mumpower2016c}. This creates a close connection between nuclear physics questions, and nuclear astrophysics questions. 

The experimental requirements for the various quantities needed vary widely - some can be obtained with very limited beam intensity, some require higher beam intensities, for some the required accuracy can be obtained easily, while for others sophisticated analysis methods and reaction theory are needed. Table \ref{tab:exp_yields} provides an overview over the different requirements for various types of measurements and techniques. Generally, decay properties can be studied with the lowest beam intensities and therefore for the most neutron-rich nuclei accessible, while masses require somewhat higher beam intensities, and reaction studies are only possible closer to stability where beam intensities are still higher. In the following we discuss various experimental approaches in more detail. 

\begin{sidewaystable}
\centering
\begin{tabular}{lccc}
\br
  & Min. intensity & Achievable & Comments \\
  & (pps) & accuracy &  \\
\mr
\multicolumn{4}{l}{\textbf{Mass measurements}}\\
TOF-Ion Cyclotron Resonance & 10$^{-2}$ & $\approx$1$-$50~keV & Depends on t$_{1/2}$ and background\\
Phase Imaging- Ion Cycl. Res. & 10$^{-3}$ & $\approx$1$-$20~keV & Depends on t$_{1/2}$ and background\\
Fourier Transform- Ion Cycl. Res. & $<$10$^{-3}$ & $\approx$1$-$ 20~keV & Depends on t$_{1/2}$ and background \\
Multi Reflection TOF & $<10^{-3}$& $\approx$20$-$150~keV & Depends on t$_{1/2}$ and background \\
Schottky Mass Spectrometry & $<$10$^{-4}$ & 1-50~keV & (Storage rings) Needs $>$1~s cooling time\\
Isochronous Mass Spectrometry & $<$10$^{-4}$ & $\approx$10$-$200~keV & (Storage rings) t$_{1/2}$$>$10~$\mu$s \\
TOF-B$\rho$ (magnetic rigidity) & $<$10$^{-4}$ & $\approx$300$-$500~keV & TOF at fragment facilities, shortest t$_{1/2}$\\
\mr
\multicolumn{4}{l}{\textbf{Decay measurements}}\\
t$_{1/2}$ (Implantation detector) & 10$^{-5}$ & 50\% & Depends on background\\
t$_{1/2}$ ($\gamma$ spectroscopy) & 1 & 5$-$10\% & Depends on background and known $\gamma$-rays\\
t$_{1/2}$ ($\beta$n emitters) & 10$^{-5}$$-$10$^{-4}$ & 50\% & Depends on background of known $\beta$n-emitters\\
Nuclear structure ($\gamma$ spectr.) & 10$^{-1}$ & $-$ & First inform. about exc. states and $\beta$-feeding\\
Total absorption spectrometers & 1 & $-$ & Strength distribution\\
P$_{xn}$, $E_n$ ( neutron TOF detectors) & \multicolumn{2}{c}{$>$10$^{5}$ total events} & Strength distribution E$>$S$_n$\\
P$_{xn}$ ($\gamma$ spectroscopy) & 10 & 50\% & Needs abs. intensities and g.s./g.s. feeding\\ 
P$_{xn}$ ($^{3}$He long counters) & 10$^{-4}$$-$10$^{-5}$ & 50\% & Depends on background\\ 
P$_{1n}$, $E_n$ (Paul traps) & 0.1 & 5\% & P$_{1n}$ only and neutron spectrum\\
\mr
\multicolumn{4}{l}{\textbf{Neutron captures}}\\
$\beta$-Oslo method & 10  &  & Only compound nucleus\\
$(d,p)$ reactions & 10$^4$ &  & Compound nucleus and direct capture\\
\br
\end{tabular}\\
\caption{\label{tab:exp_yields} Lowest beam intensities required for each measurement type at the respective device. The ``Achievable accuracy" is the accuracy and precision that can be reached with the given minimum intensity.}
\end{sidewaystable}

\subsection{Masses}
\begin{figure}[!htb]
	\centering
\includegraphics[width=0.9\textwidth]{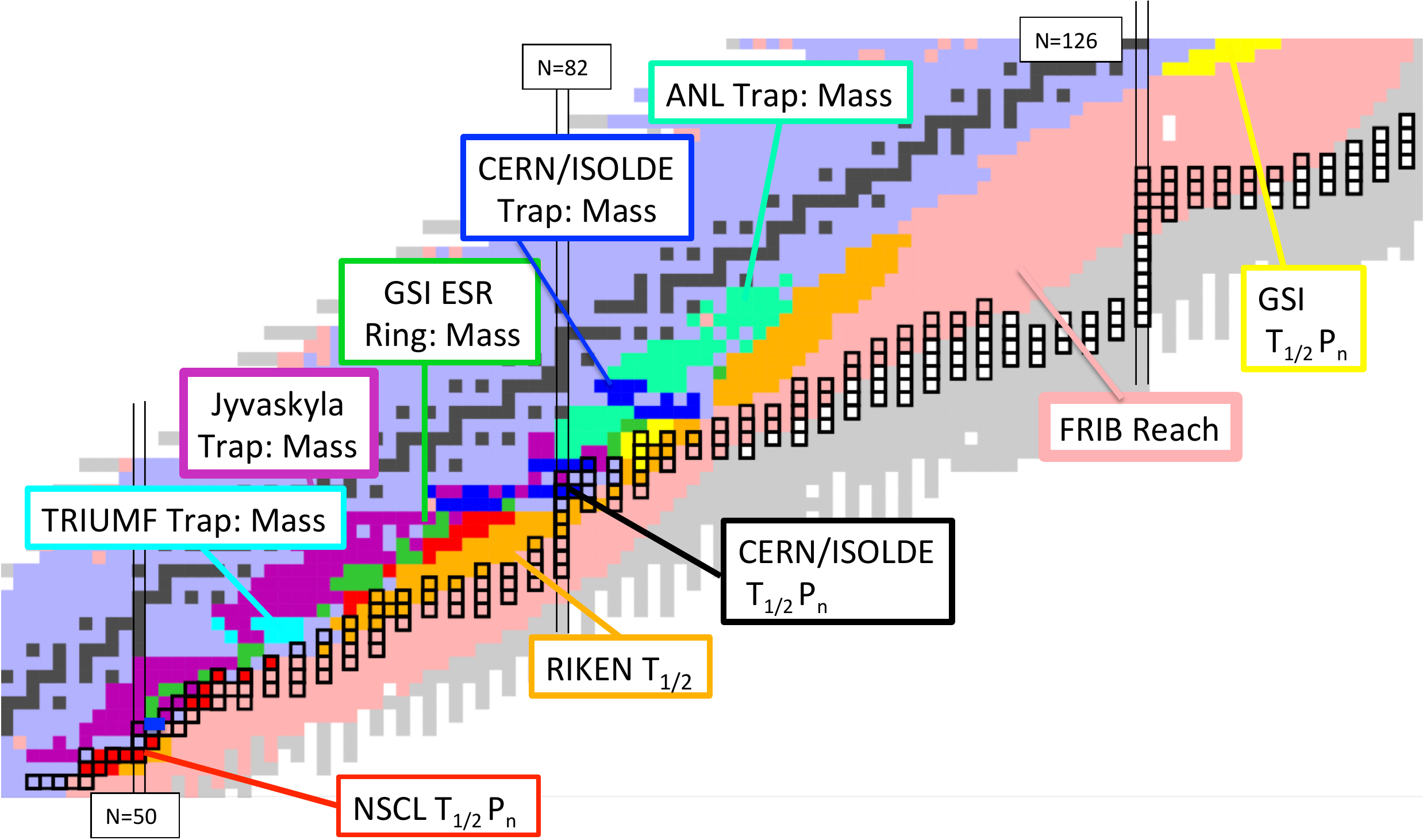}
\caption{Recent $r$-process motivated experiments measuring masses or $\beta$-decay half-lives $T_{1/2}$ at various radioactive beam facilities. The colors of the legend boxes match the colors of the chart and denote a specific facility or experimental collaboration. The pink area denotes the reach of the future FRIB facility. \label{fig:exp_reach} }
\end{figure}

There are many methods to determine binding energies of nuclei. In the past decade a large number of mass measurements of neutron-rich nuclei have been performed, approaching, and in some places reaching, the path of the $r$-process (Fig.~\ref{fig:exp_reach}). Until recently, mass measurements of nuclides in the $r$-process path have been rare, and measurements lag behind decay studies that have reached much more neutron-rich nuclei. This is about to change as new facilities are coming online and developments of experimental devices for mass measurements of exotic nuclei are completed. New facilities that are already operating and will provide a large number of $r$-process masses in the very near future include CARIBU at ANL and RIBF at RIKEN. 

Experimental mass values are not only needed as input for $r$-process models, but are also essential for validating theoretical mass models since some of the $r$-process nuclei are not experimentally reachable today and thus the simulations have to rely on theoretical mass predictions.
As discussed below in Secs.~\ref{sec:masses1} and \ref{Sec:new-masses}, current energy density functionals used in DFT calculations of nuclear masses ere deficient  near the shell closures.
One  example of is the strongly decreased odd-even mass difference for Sn isotopes as observed via the odd-even mass staggering  \cite{Hakala2012}. Whereas DFT models could not explain the observed anomalous trend in the odd-even staggering for the $N=83$ isotonic chain \cite{Hakala2012}, more detailed shell-model calculations were able to reproduce the trend \cite{Coraggio2013}. In the future, more advanced mass models should be explored to take into account these effects observed near closed shells, see Sec.~\ref{Sec:new-masses} for more discussion on this point.

In the following we provide an overview of the experimental techniques that are being used to measure masses, recent $r$-process motivated results obtained with these techniques, and an outlook on future developments (see also Fig.~\ref{fig:exp_reach} for an overview). The various techniques have different advantages and drawbacks (see Table~\ref{tab:exp_yields}). Penning traps provide the highest accuracy but measurements are only performed on a single nuclide at a time, lifetimes have to be sufficiently long, and beam intensities have to be high enough to enable transmission into the trap of a sufficiently large number of nuclei. Storage ring techniques and Spectrometer time of flight (TOF) techniques enable mass measurements of 10's or even 100's of nuclei simultaneously within a very short time (100's of ns to $\mu$s). 
Because of the longer flight path, storage rings can provide higher accuracy compared to spectrometers, but beam intensities need to be higher to compensate for the losses by transmission into a storage ring. Refs.~\cite{Meisel2013,Blaum2013,Weber2008} provide a recent overview of various techniques. Major technical developments are on the way for all these techniques. For example, the relatively new MR-TOF technique based on a multi-reflection time-of-flight spectrometer has been developed to address some of the drawbacks of the various other approaches. In the following subsections we discuss the various techniques in more detail. 

\subsubsection{Penning trap technique}
\label{sec:exptechniques:trap}
During the last decade, several hundreds of high-precision mass measurements have been performed at Penning-trap facilities worldwide using the time-of-flight ion cyclotron resonance (TOF-ICR) technique \cite{Konig1995}. Ions at very low energies are trapped in magnetic and electrical fields, and the cyclotron motion in a strong magnetic field is used to determine the mass. With TOF-ICR, typical precision achieved is on the order of a few keV for most of the neutron-rich isotopes. For the most exotic cases, the reached precision may be 10's of keV. Several reviews on the topic have been published recently. Measurements on fission fragments are summarized in Ref.~\cite{Kankainen2012}, a comprehensive review is given in Ref.~\cite{Blaum2013a}, interesting details about the history of Penning-trap spectrometry can be found from Ref.~\cite{Kluge2013}, and an insight to current and new techniques is given in Ref.~\cite{Block2016}.

Penning trap mass measurements performed in the $^{132}$Sn region are among recent highlights related to mass measurements for the $r$-process. This mass region is governed by the formation of the 2nd $r$-process abundance peak at A$\approx$130 and has a strong effect on the final $r$-process abundances in sensitivity studies \cite{Mumpower2015,Mumpower2016}. Neutron-rich Ag \cite{Breitenfeldt2010}, Cd \cite{Breitenfeldt2010,Atanasov2015} and Sn \cite{Dworschak2008} isotopes have been studied at ISOLTRAP up to $^{124}$Ag, $^{131}$Cd and $^{134}$Sn, respectively. JYFLTRAP has also measured neutron-rich Cd up to $^{128}$Cd and $^{130-135}$Sn as well as $^{129,131}$In, $^{131-136}$Sb, and $^{132-140}$Te \cite{Hakala2012}. TITAN at TRIUMF has complemented these measurements with precise determinations of masses of ground states and isomers of $^{125-127}$Cd \cite{Lascar2017} and $^{125-130}$In \cite{Babcock2018}, taking advantage of the unique capability of an electron beam ion trap (EBIT) to increase the charge state and thus the precision of the mass measurements. At CPT several isotopes in the region have been measured ($^{130-131}$In, $^{130-135}$Sn, $^{131-137}$Sb, $^{133,135-140}$Te, $^{133-135,139-141}$I, $^{142-146}$Cs) \cite{VanSchelt2012,VanSchelt2013}. The overall agreement between different Penning-trap measurements in the region is very good. However, there are a few cases, such as $^{133}$Te and $^{140}$Te, where re-measurements are anticipated to verify the mass value. This mass region shows that measurements at different facilities are essential for achieving not only precise but also accurate mass values. 

Another important mass region where Penning trap mass measurements have reached the $r$-process path are nuclides around $^{80}$Zn. This mass region governs the formation of the first $r$-process abundance peak and affects the synthesis of the lighter $r$-process elements. Here it is critical to measure masses up to $N=52$ for isotopic chains with Z$\le$30. This will enable to determine experimental two-neutron shell gap energies for N=50. The shell-gap energies are essential to test current theoretical mass models used in r-process calculations and have a strong impact on the calculated abundances. At ISOLTRAP the masses of the Zn isotopes were measured precisely out to $^{81}$Zn \cite{Baruah2008}. The authors demonstrated that with an extrapolated mass value for $^{82}$Zn, this enables a reasonable determination of the astrophysical conditions required for the $^{80}$Zn waiting point and a sufficient production of $A=80$ elements in the $r$-process. JYFLTRAP also performed mass measurements out to $^{80}$Zn in the same year \citep{Hakala2008}. Later, ISOLTRAP achieved a measurement of the $^{82}$Zn mass directly \citep{wolf2013}, making $^{80}$Zn the first major $r$-process waiting point that is completely characterized by high precision Penning-trap mass measurements. More recently ISOLTRAP mass measurements also reached the $N=50$ shell closure for the Cu isotopes \citep{welker2017}, with $^{79}$Cu being another important $r$-process nucleus. The TITAN Penning trap at TRIUMF has performed mass measurements in the path of the $r$-process of Rb and Sr isotopes out to $^{98,99}$Rb and $^{98-100}$Sr \citep{Simon2012,Klawitter2016} and demonstrated that these measurements reduce uncertainties in the prediction of the synthesis of A$\approx$90 elements in neutrino-driven wind scenarios. 

However, the TOF-ICR technique requires a measurement time of around 0.5~s to perform the required RF excitations in the trap, which sets a limit for the half-life for measurable isotopes to $T_{1/2}\gtrsim 100$~ms. The shortest-lived isotope measured with this method so far is $^{11}$Li with $T_{1/2}$= 8.75~ms at the TITAN facility \cite{Smith2008}. In addition, a lot of time is spent in collecting statistics at several frequency points to fit a TOF-ICR curve. Recently, a new method called phase imaging ion cyclotron resonance (PI-ICR)  \cite{Comisarow1974,Marshall1998} has been commissioned \cite{Eliseev2013,Eliseev2014}. The method is much faster than the conventional TOF-ICR as it does not require long excitation times in the measurement trap. It is sufficient to determine the angle between the position of the ions in a two-dimensional micro-channel plate detector, thus every ion counts and several isotopes can in principle be measured at the same time. Measurements at the SHIPTRAP facility at GSI have shown that the PI-ICR can be about 25x faster, give about 40x better mass resolving power and 5x better precision compared to TOF-ICR \cite{Block2016}. At the moment, almost all major Penning-trap facilities are either already using the PI-ICR technique or are commissioning it. 

\subsubsection{Storage ring techniques} \label{sec:exp-storage}

Storage-ring mass measurements allow masses of short-lived nuclei to be measured at  relativistic energies \cite{Franzke08}. Masses are determined from the revolution frequency of the ions in the storage ring. 
While the precision is lower compared to Penning trap measurements (100's of keV), the approach has the advantage that in principle a large number of masses can be determined simultaneously. The pioneering works have been made with the Schottky Mass Spectrometry (SMS) method with the experimental storage ring (ESR) \cite{Franzke87} at GSI Darmstadt, Germany. A mass resolving power of $\Delta m/m \sim 10^{-6}$ has been achieved with the method. However, since SMS requires cooling prior to the mass measurement, which usually takes a few seconds, it cannot be applied to mass measurements of most neutron-rich $r$-process nuclei whose half-lives tend to be much shorter than 1~s.
  
The Isochronous Mass Spectrometry (IMS) method is an alternative to measuring the masses of short-lived $r$-process nuclei in storage rings down to 10's of $\mu$s \cite{Hausmann01}. In this method the ion-optics of the storage ring is tuned so that the revolution frequency of the particle of interest (to first order) does not depend on its velocity and thus cooling is not needed. Consequently the measurement time can be decreased to less than 1~ms. IMS at the ESR has successfully been used to measure masses of $r$-process nuclei, e.g. of $^{129,130,131}$Cd \cite{Knobel2016}. The technique is also used at the HIRFL/CSRe \cite{xia2002,Xu13,zhou2016} of the Institute of Modern Physics, in Lanzhou, China. 
By applying the IMS at CSRe in Lanzhou, masses of short-lived nuclei in the A$<$100 neutron-deficient region have been measured. The best mass precision achieved to date is 8$\times$10$^{-8}$ \cite{ZHANG201720,tu2011a}. A new development is the Rare RI Ring~\cite{Ozawa12,Yamaguchi13} at RIBF, which will take advantage of the presently superior production capabilities of the most neutron-rich $r$-process nuclei at RIKEN Nishina Center.

\subsubsection{Time-of-flight with spectrometers}
\label{sec:exptechniques:tof}
Spectrometers have long been used to measure masses of exotic nuclei \cite{Wouters1987,Sarazin2000,Matos2012}. The technique relies on a simultaneous measurement of the magnetic rigidity ($B \rho$) and the time of flight (TOF) through the spectrometer beam line system. It is therefore often referred to as TOF-$B \rho$ technique. A more recent implementation of the technique uses the NSCL Coupled Cyclotron Facility at Michigan State University with the S800 spectrometer \citep{Matos2012}. A mass resolution of 1.8$\times 10^{-4}$ has been achieved and mass accuracies on the order of $10^{-5}$ \cite{Estrade2011,Meisel2015,Meisel2015a,Meisel2016}. Measurements have so far been limited to neutron-rich nuclei below iron, but it is planned to extend the method to heavier elements in the $r$-process using NSCL and, in the future FRIB. The advantage of the technique is the ability to measure a large number of masses simultaneously within a few 100 ns and with close to full transmission from the fragment separator producing the rare isotopes. However, accurate calibration masses are essential for the technique. This highlights the need for coordination among the different technical approaches to minimize systematic uncertainties. More recently the TOF--$B \rho$ technique has been implemented at RIKEN-RIBF, and first measurements have been performed with both the SHARAQ and BigRIPS spectrometers. There, another advantage of the TOF--$B \rho$  technique was exploited. Unlike other techniques it does not consume the ions measured, rather, the measurements are performed along a beam line and spectrometers where the particles enter on one side and leave on the other. This opens up the opportunity to perform mass measurements simultaneously with other experiments that can be performed downstream of the TOF--$B \rho$ section, such as decay or reaction studies. This provides a much more efficient use of expensive rare isotope facility beam time, which is especially beneficial for $r$-process research as a large number of measurements are required for which beam time is not always readily available.

\subsubsection{Multi-reflection time-of-flight spectrometer} \label{sec:MR_TOF}
In the last few years, high-resolution multi-reflection time-of-flight (MR-TOF) devices were introduced to ion-beam facilities for mass measurements \cite{wollnik90,Piechaczek2008,wolf12,schury13,Jesch2015}, bridging the gap between Penning trap and Schottky Mass Spectrometry on one hand and TOF-B$\rho$ on the other. In a MR-TOF, a trapped ion bunch is allowed to bounce back-and-forth between two electrostatic mirrors for a large number of cycles ($>$100), considerably increasing its resolving power as compared to regular TOF devices. After a suitable amount of time, allowing the temporal separation of the various species contained in the original bunch, the particles are released and detected with a micro-channel plate detector. The mass of an ion of interest is then determined from its total time-of-flight in the device compared to the time-of-flight of a calibration ion of well-known mass. MR-ToFs are also used as a high-resolution isobar separator for Penning traps and other experimental equipment requiring purified ion bunches.
The MR-TOF device has a relatively compact design (size $\approx$1-2~m) and offers non-scanning operation reaching high mass resolving power of $m/\Delta m = 10^{5}$ in just a few milliseconds. Devices of this type are already implemented or under development at almost all ion-beam facilities. Several MR-TOF systems are operational now, for example at the ISOLTRAP mass spectrometer (ISOLDE/CERN in Switzerland), at the SLOWRI setup (RIKEN in Japan) \citep{Ito2013}, at the FRS ion-catcher (GSI in Germany), at the CARIBU facility in ANL \citep{Hirsh2016}, and since 2017 also at TITAN (TRIUMF/ISAC) \cite{Jesch2015,Leistenschneider2018}. Furthermore, many new devices are under commissioning or development at other facilities and experiments, such as at the PILGRIM setup (SPIRAL facility at GANIL in France), at the University of Notre Dame \cite{SCHULTZ2016251} for the future ANL N = 126 beam factory, at the NSCL facility at Michigan State University in the USA, at the IGISOL facility at the University of Jyv\"askyl\"a in Finland, at the CAS in Lanzhou/China, and at the RISP/RAON facility in Daejeon/South Korea. 

\subsubsection{Atomic mass evaluation}\label{sec:exp-AME}
The Atomic Mass Evaluation (AME) is the most reliable source for comprehensive information related to the atomic masses. It provides the best values for the atomic masses and their associated uncertainties by evaluating all available experimental mass data. During the AME development process, all the available experimental data related to atomic masses are collected and carefully examined. It is the policy to use the experimental information as much as possible, rather than simply adopting the final mass values from the publications. The mass of one nuclidic species obtained through different methods often leads to multiple relationships between masses, thus establishing a network. In many cases the mass values are overdetermined because of a large number of measurements. A least-squares method is employed in order to unify all individual measurements. After the publication of AME2012 \cite{AME12}, the atomic mass data center (AMDC) was officially transferred from CSNSM-Orsay to IMP-Lanzhou to continue AME. 

The latest atomic mass evaluation, AME2016, was published in March 2017 in two papers. The first article provided complete information on the experimental input data and details on the evaluation procedures \cite{AME16a}.  The second one presented a table with the recommended values of the atomic masses, as well as tables and graphs of derived quantities, together with a list of all references used \cite{AME16}. Similarly to the previous distributions, the AME2016 is accompanied by the NUBASE2016 evaluation, which provides ground state information, including decay data, and thus a consistent interpretation of the individual states involved in the mass evaluation \cite{NUBASE16}. The NUBASE evaluation includes masses, excitation energies of isomers, half-lives, spins, parities, decay modes, and their intensities, for all known nuclei in their ground and excited isomeric states that have half lives longer than 100 ns. In AME2016, 13035 experimental data extracted from the available literature were accumulated and studied. 5675 of them are used as valid input data following the AME policy. In the AME2016 mass table, 2497 experimental masses for nuclides in their ground state are listed.  

For some nuclides their existence has been demonstrated experimentally, but their masses are still not known. These nuclides are typically two to three nuclides away from the known mass region along an isobaric chain. To guide further research, their masses are estimated based on the trends in the mass surface in the neighborhood and listed in the AME mass table. In AME2016, estimated mass values are given for 938 nuclides. 

\subsection{Decay half-lives}\label{exp:decay-HL}
\label{sec:exptechnique:decay}
$\beta$-decay half-lives are another important physical quantity to be measured for a better understanding of $r$-process nucleosynthesis (see Sec.~\ref{section_NSS}). While masses determine the route of the (hot) $r$ process under given astrophysical conditions, the $\beta$-decay half-lives of waiting-point nuclei dictate how fast the process can proceed towards heavy nuclei and how much material is accumulated in a given isotopic chain. In addition, in cold $r$-process scenarios such as some of the ejecta in neutron star mergers, the $r$-process path is determined by the competition between neutron capture rates and $\beta$-decay rates. 

In the last few years scientists at RIBF in RIKEN Nishina Center in Japan have pushed out the limits for neutron-rich isotopes more than any other facility. 
These measurements show the power of the new generation of RIB facilities in combination with highly-efficient state-of-the-art detection setups. As a result more than 200 $\beta$-decay half-lives were measured for neutron-rich isotopes between $^{72}$Co up to $^{163}$Pm in the rare earth region \cite{nishimura11,xu14,lorusso15,wu17}, many of them for the first time. On Jan. 15, 2018 RIKEN announced that they have discovered 73 new isotopes ranging from Mn to Er at the BigRIPS in-flight separator, totaling the number of newly discovered isotopes in 10 years of operation at RIBF to 132. About 62 more new isotopes from recent campaigns will be announced soon.  Half-life and neutron-branching measurements are expected for many of these isotopes from the BRIKEN collaboration.

\begin{figure}[!htb]
	\centering
\includegraphics[width=1.0\textwidth]{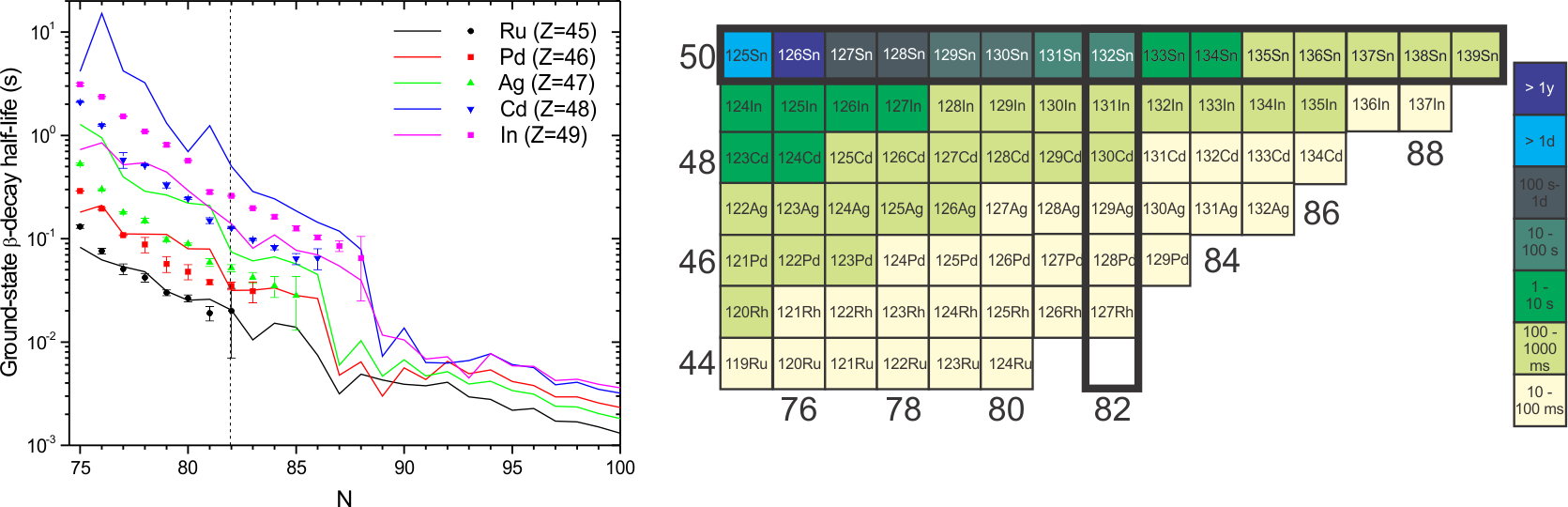}
\caption{Left: QRPA predictions for (ground-state) $\beta$-decay half-lives \cite{} of isotopes with $Z$=45-49 up to $N$=100 (lines) in comparison with experimental data (symbols). 
Right: Color-coded chart of nuclides for the experimental half-lives in this region.}
\label{fig:FRDM-plateau}
\end{figure}

The large-scale $\beta$-decay half-life measuring campaigns at RIKEN revealed some surprises. First, they emphasize the necessity of cross-checking previous values whenever possible. A striking example here is the half-life of the crucial $N$=82 isotope $^{130}$Cd, which is mainly responsible for the second $r$-process abundance peak. Previous results from ISOLDE using the time-dependence of the decay measured via $\beta$-delayed neutrons resulted in values of $t_{1/2}$= 195 (35)~ms \cite{kratz86} and later in $t_{1/2}$= 162(7)~ms \cite{hannawald01}. In the recent EURICA campaign a 20\% lower value of $t_{1/2}$= 127 (2)~ms was derived \cite{lorusso15}, which was confirmed by the GRIFFIN collaboration at TRIUMF ($t_{1/2}$= 126 (4)~ms, \cite{dunlop16}). As discussed in Ref.~\cite{lorusso15} the systematic overestimate of the half-lives for the $N$=82 isotones can be traced to the scaling of the Gamow-Teller (GT) quenching to the previously reported longer half-life for $^{130}$Cd \cite{hannawald01}. Increasing the GT quenching factor from $q$ = 0.66 to 0.75 in order to reproduce the shorter half-life resolves this discrepancy and new predictions will yield shorter half-lives for the yet unmeasured $N$=82 isotones with $Z$$<$44. Another interesting outcome from the half-life measurements of Lorusso et al.~\cite{lorusso15} is that there is no sudden drop in half-life when crossing the $N$=82 shell closure in this region (see Fig.~\ref{fig:FRDM-plateau}). 

Most of these recent $\beta$-decay half-life measurements were focused on very neutron-rich isotopes with $A$$<$170. 
A recent measurement at the Fragment Separator (FRS) at the GSI Helmholtz Center for Heavy Ion Research in Darmstadt, Germany extended the sparse knowledge in the region ``south-east" of $^{208}$Pb \cite{caballero16,caballero17}. With these new data, new tests of theoretical predictions become possible. (The likely too long value for the half-life of $^{202}$Pt ($N$=124, $t_{1/2}$=44(15)~h, \cite{shi92}) calls for a remeasurement.)

The FRDM+QRPA \cite{Moller2003} and the Gross Theory (GT2) model plus HFB21 masses \cite{GT2-privcomm} overpredict the half-lives for $N$$\leq$126 by a factor of $\approx$20-30. This is likely due to the increasing contributions from first-forbidden transitions towards higher $Z$ values that are not self consistently included in the theory. Gamow-Teller transitions become progressively Pauli-blocked by the filling of the $\pi$h$_{11/2}$ proton orbital, which reduces contributions from allowed $\nu$h$_{9/2}$-$\pi$h$_{11/2}$ transitions. This trend is inverted for $N$$>$126 which is an indication of the predominance of allowed $\nu$i$_{11/2}$-$\pi$i$_{13/2}$ Gamow-Teller-transitions ``south-east" of $^{208}$Pb \cite{caballero16,caballero17}.

The newer RHB+RQRPA \cite{Marketin2016} seems to do a slightly better job in the ``south-west" quadrant but underpredicts most of these half-lives. As soon as one crosses the shell closure, the trend drastically worsens and half-lives are underpredicted by a factor of 30 and more. The interpretation given in Ref.~\cite{caballero16,caballero17} is \textit{``This feature may be ascribed to a specific neutron-excess dependent ansatz for the strength of the T=0 proton-neutron (dynamic) pairing assumed. This free parameter has a strong influence on the $\beta$-decay strength function. This parameter was fitted to the available decay rates, which did not include the isotopes with very high (N-Z) values studied in the present work."}. 

These examples illustrate the challenges with using theoretical approaches that rely on empirically determined parameters for extrapolation into unknown mass regions, and the need for nuclear data to address these issues. 
In the following paragraphs we discuss the different methods that have been used for measurements of half-lives. The three different methods in Table~\ref{tab:exp_yields} are explained, including their strengths and weaknesses. As can be seen, short of proof of existence, half-life determinations have the highest sensitivity of all experiments and can be performed down to beam intensities of 10$^{-5}$ pps for a few weeks of beamtime. 


\subsubsection{Decay half-lives from the implant-correlation method}
\label{sec:implant-correlation}

The best suited method for measuring the half-lives of the most neutron-rich and shortest-lived (t$_{1/2}$$<$10~s) isotopes is the implant-decay correlation technique (see Table~\ref{tab:exp_yields}). The radioactive beam is implanted into a detector system, that detects the implantation event as well as the decay event, most commonly via detection of the emitted $\beta$-particle. The implantation setup can additionally be surrounded by a moderated neutron detection setup. From the time differences between an implantation and the subsequent decay the half-life can be determined. For background to be manageable the method requires the time between subsequent implantations to be larger than a few multiples of the half-lives to be measured. This is commonly achieved by segmentation of the detector system. In a segmented system, only the rate per segment must be sufficiently low. For reasonable segmentations (1000's of pixels) the method is still suited to low implantation rates and short half-lives, which is often fulfilled for $r$-process related measurements. Typical detector systems can handle a few kHz implantation rates for typical $r$-process half-lives of less than a second. 

The method is most commonly applied for relatively fast radioactive beams produced by in-flight fragmentation or fission. Such beams are implanted deep into a detector layer, enabling high efficiency for the detection of the emitted $\beta$-particle. In addition the method can then be combined with event-by-event particle identification. As a result the technique becomes extremely selective, and even with mixed beams the observation of just a few (less than 10) ions enables half-life determinations with errors of approximately factors of two to three.  

For example, the astrophysically interesting $^{78}$Ni was first measured with the technique based on the identification of only 7 ions resulting in a half-life of 110($^{+100}_{-60})$~ms \citep{Hosmer2005}. Subsequent measurements with much higher statistics have since confirmed the half-life to be 122.2 (51)~ms \citep{xu14}. A significant number of new half-lives have been determined recently using the technique that are important for $r$-process scenarios and suggest new trends relative to theoretical predictions \cite{nishimura11,xu14,lorusso15,wu17}. Additional $\approx$ 50 new half-lives are expected from the 2017 run of the BRIKEN project (see Sec~\ref{sec:exp:3He}). 

In its simplest approach the emitted $\beta^-$ particles are detected and their correlation in time with the implantation of the preceding ion is used to determine the half-life. However, if beam intensities are sufficiently high, the method can be significantly improved by detecting $\gamma$-rays or neutrons in coincidence. The most precise and least error-prone method for the measurement of a half-life is via the time-dependence of known $\gamma$-transitions in the decay. Ideally, the half-life is deduced as error- and intensity-weighted average from several transitions of the same state (ground state or isomers). This method has its limitations if the half-life is ``contaminated" by an isomer or long-lived state in the decay daughter, or if the half-lives of the ground-state and isomeric parent isotope are too similar (example: $^{129}$Cd, see \cite{dunlop16}). A drawback of this method is the requirement of reasonably good statistics ($>$1000 events) for several $\gamma$-lines. The method is therefore not applicable to the most neutron-rich nuclides that can be reached.

Alternatively, the time-dependent decrease of a $\beta$-delayed neutron emitter can be a very sensitive method and has been widely used in the past for the first-time determination of half-lives. It is particularly advantageous if the isotope of interest is the only $\beta$-delayed neutron emitter in the experiment, i.e. if daughters and beam contaminants are not strong neutron emitters. 


A number of implant-decay correlation systems exist or are being planned for current and future fragmentation facilities including the beta-counting system at NSCL and eventually FRIB \citep{Prisciandaro2003,Larson2013};  WAS3Abi \citep{Nishimura2012} and CAITEN \citep{Famiano2003} at RIKEN; SIMBA at GSI \citep{caballero17}; the active stopper for the RISING campaign \citep{Kumar2009}; DESPEC AIDA at FAIR and at RIKEN \cite{Griffin16}; and a system at Lanzhou \citep{He2014}.  The technique has also been implemented with gaseous detectors at a variety of facilities (see Ref. \cite{Janiak2017} for a recent example). 

The implant-decay correlation technique becomes progressively more difficult to apply as the half-lives become longer and the ion production rates increase.  If the secondary radioactive ion beam is pure enough then half-life information can still be obtained by pulsing the accelerator with the disadvantage of the reduction in duty cycle \citep{Huyan2016}.  There have been recent developments to apply the implant-decay correlation technique to longer lifetimes \citep{Kurtukian-Nieto2008,caballero17}.  The technique is promising but appears to require a larger number of ions compared to the standard implant-decay correlations.

\subsubsection{Decay half-lives from the moving-tape method}
For longer half-lives of the order of a few seconds and more, the implantation of a low-energy beam onto a moving tape made out of mylar is the method of choice. The measurement is performed in ``tape cycles" usually starting with a background measurement of a few seconds, then the beam is implanted on the tape for a certain time (depending on the half-life of interest). After interruption of the  beam the decay is followed for several half-lives, before the ``contaminated" tape is moved away from the implantation point behind a lead wall. Ideally, the length of the tape and tape cycles allows a "fresh" tape to be used throughout the experiment, so long-lived contaminations on the tape can be discarded afterwards. Special care has to be taken if volatile elements are implanted or produced as decay products like gases or iodine. Experiments have shown that losses due to diffusion appear if the beam is not implanted deep enough (with enough energy) into the tape. Aluminized mylar tapes are used for these cases. 
These tape stations can then be surrounded by highly-efficient detection setups, e.g. $\gamma$-spectrometers, $\beta$-decay stations, or even neutron detectors (in combination with a $\beta$-detector).

This method is commonly used at ISOL facilities with low energy beams ($<$60~keV). An example for such a setup is the newly commissioned GRIFFIN spectrometer at TRIUMF with its auxillary detectors \cite{svensson14}. Recent results from this setup include new half-life measurements of $^{128-130}$Cd for $r$-process studies \cite{dunlop16}. At FRIB it will be possible to stop the high-energy beam in a gas catcher, extract it at 40~keV energy, and then implant it on a moving tape system.   

\subsubsection{$\beta$-decay half-lives from stored ions} 

Time-resolved Schottky mass spectrometry is a non-destructive method based on Schottky-noise spectroscopy in circular accelerators and storage rings. In the ESR at GSI Darmstadt \cite{Litvinov11} the ions have typical revolution frequencies of about 2~MHz and induce in each turn mirror charges on two electrostatic pick-up electrodes. The revolution frequency measured at these pick-up electrodes is related to the mass-to-charge ratio of the ion and can be used to uniquely identify it. If after a decay the daughter nucleus continues to circulate in the storage ring, the decay can be detected by the frequency change. Hence, time dependent measurements of frequency and intensity of the pick-up electrode signals can be used to determine the decay half-life. Due to the restricted acceptance of the ESR only half-lives from electron capture (EC) and isomeric decays can be measured with this method, especially those of highly-charged ions (see e.g. Ref.~\cite{irnich95,litvinov03,akber15}). A similar approach can be applied to nuclei stored in a MR-TOF spectrometer (see section \ref{sec:MR_TOF}) \citep{Wolf2016}.

The addition of multi-purpose particle detectors like CsISiPHOS \cite{najafi16} in the respective outside or inside pocket positions of storage rings can extend this method to $\beta^+$-,  $\beta^-$-, and also to $\alpha$-decay half-lives. The unique advantage of measuring half-lives in the storage ring is the opportunity to measure the influence of the charge state on the half-life. It is well-known that in highly ionized states the decay of nuclei which decay predominantly via EC or Internal Conversion (IC) is hindered \cite{irnich95,litvinov03,akber15}, whereas the bound-state $\beta$-decay in fully ionized nuclei accelerates the decay compared to the terrestrial decay half-life \cite{jung92}. However, the ultra-high vacuum conditions in the storage ring are not identical to real ``stellar" conditions, where the nuclei are immersed in a plasma with finite electron densities. 

\subsection{$\beta$-decay strength functions}
$\beta$-decay strength distributions provide a strong constraint to theoretical models that predict $\beta$-decay properties.
While a large amount of experimental data providing $\beta$-
intensities exists in the literature, only a limited fraction 
provide a reliable measure of this important quantity due to the 
well known ``Pandemonium effect'' \citep{Hardy1977}. Pandemonium 
is a phenomenon that occurs when low-efficiency detection systems 
are used to infer the $\beta$-intensity. Such measurements can 
miss significant fractions of the low-intensity/high-energy 
$\gamma$ emission and as a result, the $\beta$-intensity to low 
lying levels is artificially enhanced. The effect is even more 
pronounced when moving away from the valley of stability, where 
the $\beta$-decay $Q$-value increases and many more states can be populated. The technique of Total 
Absorption Spectroscopy (TAS) was developed as a means to 
overcome the Pandemonium effect (see \cite{Rubio2017} for a recent example, and references therein for a history). 
The TAS technique relies on the 
use of a large volume, high efficiency $\gamma$-ray calorimeter, 
which can detect the full energy emitted in a $\gamma$ cascade, 
and therefore identify the excitation energy that was fed in the 
$\beta$ decay. Using the TAS technique, the $\beta$-decay 
intensity can be accurately measured and compared to theoretical 
calculations. 

Several total absorption spectrometers have been developed for current and next generation radioactive beam facilities. Newer examples of such detectors are the Modular Total Absorption Spectrometer (MTAS) at Oak Ridge National Laboratory \citep{MTAS2014}, the Summing NaI (SuN) detector at the NSCL and FRIB \citep{SUN2013}, the Decay Total Absorption Spectrometer (DTAS) \citep{DTAS2016} for FAIR, and others. These TAS detectors are typically coupled with an appropriate $\beta$-decay implantation setup, as described in section \ref{sec:implant-correlation}, either for fast beam implantation-$\beta$ correlation measurements or for stopped-beam moving-tape experiments. Depending on the setup, beam impurities, half-lives, and other parameters, the required beam rate for such experiments can vary, however a typical minimum requirement for extracting the $\beta$-decay intensity is of the order of 1~pps. Many nuclei are accessible for TAS studies at current facilities and many experimental campaigns are ongoing. In addition, experiments are planned at next generation facilities, especially going to heavier nuclei (around $N$=82 and the rare-earth region), in regions that are currently inaccessible. 


\subsection{$\beta$-delayed neutron emission}
$\beta$-delayed neutron emission is a common decay mode of neutron-rich nuclei, and can become the most dominant decay process for very neutron-rich isotopes in the $r$-process. If the neutron separation energy becomes smaller than the $\beta$-decay energy window ($Q_\beta$ value), the emission of neutrons after $\beta$-decay is possible. These “$\beta$-delayed neutron” ($\beta$n) emitters play a crucial role in nuclear structure, nuclear astrophysics, and for nuclear reactor applications. 

The neutron emission branching ratio ($P_n$ value) is an important physical quantity in $r$-process models. During freeze-out, 
$\beta$-decay moves the reaction flow towards stability along isobaric chains. 
$\beta$n-emitters in these decay chains can influence the final abundance distribution in two ways: 
(1) the emission of neutrons leads to a transfer of material into $\beta$-decay chains with lower mass number, thereby changing the final element created in the $r$-process, and (2) the emitted neutron is thermalized and increases the abundance of late time neutrons that may be recaptured by the decaying material. Thus an accurate knowledge of the neutron-branching ratio and half-lives of $\beta$n-emitters is needed for $r$-process models (see Sec.~\ref{section_NSS}).

Neutron spectra are less important for astrophysical applications because emitted neutrons are rapidly thermalized before they undergo a capture reaction. However, measured in the laboratory, they serve as important diagnostics of the theoretical models used to predict decay properties. 

In general two types methods can be distinguished for the measurement of $\beta$-delayed neutrons: the classical ones that are detecting the neutron (with or without moderation), and indirect methods that do not detect the neutron but identify parent and daughter nuclei. 

An overview of the different methods to extract half-lives and $P_{n}$ values from data is given in the recent summary reports of the IAEA Coordinated Research Project ``Development of a Reference Database for Beta-delayed neutron emission", INDC(NDS)-0599, -0643, -0683, and -0735 which can be downloaded from \url{https://www-nds.iaea.org/beta-delayed-neutron/}. These reports (INDC(NDS)-0683, p. 11ff.) lists eight methods for the determination of the neutron branching ratio: 
\begin{itemize}
\item (1) the ``$\beta$/n coincidence method", 
\item (2) the ``n-$\beta$" method (separately measuring $\beta$s and neutrons simultaneously but not in coincidence), 
\item (3) the method to count $\gamma$-rays in the daughter (``$\gamma$ $^A$Z+n"), 
\item (4) measuring relatively to a known $P_{1n}$ standard (``$P_n$ $^A$Z+n"), 
\item (5) counting the number of precursors and then the amount of $\beta$n daughters by any suitable method (``ion"), 
\item (6) measuring the number of precursors by fission yields and then the amount of $\beta$n daughters by any suitable method (``fiss."), 
\item (7) pure $\gamma$-counting techniques to determine both the number of mothers and $\beta$n grand-daughters (``$\gamma$-$\gamma$"), and 
\item (8) the ion-recoil method which includes trap measurements \cite{yee13}.
\end{itemize}

\subsubsection{Spectra and $\beta$-delayed neutron branching ratios from traps}

An example of the ``ion-recoil" method (8) are measurements in a Paul trap \cite{yee13,Scielzo2014}. Both the $\beta$ particles emitted from the trapped ions and the recoiling daughter nucleus, are detected using detectors that surround the trap. The time difference between the $\beta$s and the slower recoil ions produces a time of flight spectrum which consists broadly of two peaks: one at the longer time of flight from regular beta decay, and one from which the energy of the emitted neutron (and the fact that a neutron was emitted in addition to the $\beta$) can be reconstructed. The method is currently under further development at ANL. It can be complemented with $\gamma$-detectors. The advantage of this approach is a very clean and selective signal, the well defined large detection efficiency, as well as the ability to determine the energy spectrum of the emitted neutrons. However, this method is currently limited to $P_{1n}$ measurements, and is not feasible for multi-neutron emission.

\subsubsection{$\beta$-delayed neutron branching ratios from $^{3}$He and BF$_3$ detectors}\label{sec:exp:3He}

The extraction of $P_{xn}$ values from $^{3}$He or BF$_3$ arrays with a moderator can be accomplished with above mentioned methods (1), (2), (4), and (6). The emitted neutron is moderated in a suitable material, typically polyethylene, and the low energy neutrons can then be detected efficiently through $^{3}$He(n,p) or $^{10}$B(n,$\alpha$) reactions in gas-filled ion counters that detect the resulting charged particle. The ``fission" method (6) relies on older, less reliable fission yield measurements and is no longer recommended. The measurement relative to a known $P_{1n}$ standard (method (4)) requires that the chosen standard has a similar neutron energy spectrum compared to the isotope of interest. Only then the amount of counted neutrons can be directly related to the standard. This can also be achieved with a neutron detector that has a ``flat" neutron detection efficiency curve. Setups like NERO \cite{pereira10}, BELEN \cite{agramunt16} and the new BRIKEN array at RIKEN \citep{Tarifeno2017} have been designed with Monte Carlo simulations to achieve a flat efficiency curve up to $\approx$2 MeV. In these cases the different neutron energy spectra of calibration isotopes and isotopes of interest can be neglected as long as the underlying assumption of neutron energies below 2 MeV is valid. 

The methods of counting $\beta$s and neutrons either in coincidence or separately are the most reliable, provided the background and the $\beta$ energy-dependence of the detector array can be well characterized. The one-neutron branching ratio $P_{1n}$ can be deduced from the number of detected $\beta$ decays ($N_{\beta}$) and the number of detected $\beta$-neutron coincidences ($N_{\beta n}$) via $P_{1n}={N_{\beta n}}/({N_{\beta}\cdot \epsilon_n})$. The main requirement of this method is that the number of counted $\beta$s is free of contaminations, i.e. any background is subtracted. If the neutron efficiency curve is constant by design,  $\epsilon_n$ is a constant. However, if the neutron energy distribution (and ergo the respective $\beta$ energies) is very different from the calibrant isotope, systematic effects arise which cannot be corrected without theoretical predictions (because the neutron spectrum is not typically known) and simulations. In the coincidence method the $\beta$-n time correlation window needs to be long enough to include also high-energy neutrons with longer moderation times. Typical moderation time windows for modern neutron detector setups are a few hundred $\mu$s.

Since 2016 the BRIKEN project (“Beta-delayed neutron measurements at RIKEN for nuclear structure, astrophysics, and applications”) focusses on the most exotic $\beta$n-emitters which can presently be produced \cite{Tarifeno2017}. Several experiments were carried out in 2017 and covered 231 $\beta$n-emitter between $^{64}$Cr up to $^{151}$Cs. For many of these isotopes, $\beta$n-emission branching ratios have been measured for the first time, e.g. for the doubly-magic $N$=50 isotope $^{78}$Ni, as well as about 50 new $\beta$-decay half-lives. More experiments for $A$$>$150 and $A$$<$60 will be carried out in the upcoming 2-3 years, making this experimental campaign one of the largest systematic investigation of two important nuclear physics input parameters (half-lives and P$_{xn}$ values) for modeling the $r$-process nucleosynthesis. 


An estimated outcome of this project is shown in the last column of Table~\ref{tab:bn}. Almost all of the previously measured $\beta$n-emitters will be remeasured, and approximately 150 new $\beta$n emitters will be added to the list of 298 known $\beta$n-emitters. Also the number of measured multi-neutron emitters will be largely expanded. The inclusion of these new results in astrophysical network calculations will help to reduce the uncertainty in the calculated $r$-process abundances from this nuclear physics quantity.

\subsubsection{ $\beta$-delayed neutron branching ratios from HPGe detectors}

In method (3) the abundance of the precursor is determined via $\gamma$-counting of any $\beta$-decay daughter, followed by the detection of $\beta$-delayed neutrons. For this, absolute $\gamma$ intensities
have to be known (which are not available for many isotopes). When fragmentation reactions are used for the production of the precursor nucleus, the $\beta$n-daughter might also be produced, as well as isomers. The $\gamma$ counting then needs to be corrected to account for this.

Method (7) is a pure $\gamma$-counting technique to determine both, the number of mother and $\beta$n granddaughter nuclei. As mentioned before, absolute $\gamma$ intensities and a complete knowledge of the decay scheme are required, as well as the direct ground-state feeding that can lead to neutron emission without a $\gamma$-ray. If these requirements are fulfilled, the $P_{1n}$ value can be extracted from $\gamma$ efficiencies $\epsilon_{\gamma}$, number of detected $\gamma$'s $N_{\gamma}$, and the $\gamma$ intensities $I_{\gamma}$ via  $P_{1n}=(\epsilon_{\gamma,d} \cdot N_{\gamma,g}\cdot I_{\gamma,d})/(\epsilon_{\gamma,g} \cdot N_{\gamma,d}\cdot I_{\gamma,g}$), where ``d" stands for the decay daughter and ``g" for the granddaughter (after $\beta$n-decay).

\subsubsection{ $\beta$-delayed neutron branching ratio from ion counting, e.g. in storage rings or traps}

The ion-counting method (5) relies on counting the number of precursors and $\beta$n daughters, and deducing the $P_{xn}$ value from this via $P_{1n}=N_{\beta1n}/N_{ion}$. Such measurements can be performed in devices that store ions for a sufficiently long time so they can decay, and that allows single-ion counting. Examples are traps \cite{yee13} and storage rings \cite{evdokimov12}. This method is completely independent of the neutron detection efficiency, but the efficiency of and transmission through the ion counting device has to be carefully determined and known for both species, mother and daughter. This method also needs corrections for $\beta$n daughters already present in the beam cocktail and potential losses during the storage times, e.g. by reactions with rest gas particles or atomic interactions that change the ion's charge state. 

\subsubsection{$\beta$-delayed neutron spectra from neutron detectors without moderation}

The use of neutron detectors that are capable of detecting MeV neutrons with reasonable efficiency has several advantages over the use of  $^{3}$He or $^{10}$B based neutron detectors described above. The main advantage is that without moderation, the time-of-flight can be used to measure the energy of individual neutrons. This allows the determination of neutron spectra in addition to branching ratios. A disadvantage is the much lower efficiency and the need to correct for the considerable detection energy threshold, typically of the order of 100~keV, that prevents the detection of the lowest energy neutrons. The most common type of detector used for such measurements are plastic scintillators. Recently developed detector systems for use at rare isotope facilities include VANDLE at ORNL \cite{Madurga2016} and LENDA at NSCL \cite{perdikakis12}. Liquid scintillator-based detection systems are DESCANT at TRIUMF \cite{garrett14} and MONSTER \cite{Martinez16} for FAIR.

\subsubsection{Recent evaluation of $\beta$-delayed neutron emitters}\label{sec:exp-bn}
The latest Atomic Mass Evaluation (AME2016 \cite{AME16}) has identified 2451 isotopes, from which 621 are $\beta$-delayed neutron emitters ($Q_{\beta xn}$$>$0~keV). The present status (June 2017) is summarized in Table~\ref{tab:bn}, together with an estimate of new data from the ongoing BRIKEN campaign (see Sec.~\ref{sec:exp:3He}). 

\begin{table}[!htb]
\caption{\label{tab:bn} Number of identified $\beta$-delayed neutron emitters with $Q_{\beta xn}$$>$0~keV (within the uncertainties, from AME~2016 \cite{AME16a}) and number of isotopes where the neutron-branching ratio has been measured but not necessarily to the required precision (Status: June 2017, isomeric states not included). The last column gives the estimated number of new emitters that will be measured by the BRIKEN collaboration from 2017-19. 
}
{\centering
\footnotesize
\begin{tabular}{lccccc}
\br
  & Identified & Measured & Fraction (\%) & Isotopes &  BRIKEN (est. new)
  \\ \hline
$P_{1n}$ & 621 & 298 	& 48.0\% & $^{8}$He-$^{216}$Tl & $\approx$150 \\
$P_{2n}$ & 300 & 23 	& 7.7\% & $^{11}$Li-$^{136}$Sb & $\approx$50 	\\
$P_{3n}$ & 138 & 4 		& 2.9\% & $^{11}$Li-$^{31}$Na	 & $\approx$20 	\\
$P_{4n}$ & 58  & 1 		& 1.7\% & $^{17}$B		 & $\approx$5 	\\
\mr
\end{tabular}\\}
\end{table}

\normalsize

About half of the identified $\beta$n-emitters have a measured $P_{1n}$ value, however many of them with either large uncertainties or only as upper (or even lower) limits. The situation gets worse when going to $\beta$-delayed multi-neutron emission which will be the prevalent decay mode for the most neutron-rich isotopes that are produced in any $r$-process scenario. Only 23 $\beta$2n emitters have been measured so far (out of 300 which are accessible), and only four $\beta$3n- and a single $\beta$4n-emitter ($^{17}$B) are known so far, all outside of the $r$-process mass range in the lighter mass region up to A=31.

All $\beta$n-emitters with an experimental value for the $\beta$-decay half-life or neutron-branching ratio have been recently re-evaluated in the framework of a Coordinated Research Project (CRP) of the International Atomic Energy Agency (IAEA) which ran from 2013-2017. The new recommended values for these isotopes will be available in a reference database on the IAEA website at \url{https://www-nds.iaea.org/beta-delayed-neutron/} and have been published or will be published soon. The isotopes with Z$\leq$28 are published in Ref.~\cite{Birch2015}, and the isotopes with Z$>$28 are presently under review \cite{Liang2018}.

This evaluation of the half-lives and neutron-branching ratio was performed completely independent from the efforts in NuBase \cite{AME16a,NUBASE}. While for the half-lives both evaluations should yield almost identical results, we strongly encourage users only to use the recommended neutron-branching ratio values from the IAEA CRP in the reference database since the NuBase work for this quantity is only a compilation (collection of results including extrapolated values from theory) without the detailed evaluation work that has been done by the members of the IAEA CRP group.

Apart from the evaluated data the IAEA CRP reference database will also include values from various theoretical predictions and semi-empirical estimates for the given isotopes, and much more additional information like links to digitized neutron spectra. 

Although the CRP finished in 2017, the evaluation effort and maintenance of the reference database will be continued. The present database is an important basis for the vast amount of new data on neutron-rich isotopes that is expected to be published in the upcoming years, from running projects like the BRIKEN project at RIKEN and then from even more neutron-rich new isotopes to be produced and investigated at the new generation of RIB facilities which will come online in the next decade.

\subsection{Neutron capture rates}
\label{sec:exptechniques:ncap}
Due to the lack of a suitable neutron target, or a target of unstable isotopes, determining neutron capture rates for nuclei far from stability with short half-lives ($<$ 1 day) currently requires indirect techniques. Proposals have been made for a radioactive beam in a storage ring intersecting either a nuclear reactor \citep{Reifarth2014a} or a neutron target produced by ongoing spallation reactions \citep{Reifarth2017}, but these ideas remain a scenario for the far future, if they are feasible at all. Indirect techniques use alternative reactions to populate compound nucleus states that are important for the neutron capture reaction, and reaction data can then be used to constrain the relevant properties such as neutron and $\gamma$-strengths. A broad range of techniques is under development under the label of ``Surrogate Reactions" \citep{Escher2012}. All these approaches require a very close interplay of experimental work and reaction theory. An approach that has been used with an $r$-process motivation are neutron transfer reactions such as (d,p) (see \cite{Bardayan2016} for a recent review). Pioneering measurements have been carried out at ORNL's HRIBF facility probing neutron captures on $^{130}$Sn \cite{Kozub2012} and $^{132}$Sn \citep{Jones2010}. These reactions have so far only been used to probe bound states relevant for the direct capture component of the reaction rate. 

Resonant states are more difficult to constrain as level spacings can be small compared to level widths, and configurations can become more complex. An alternative approach that avoids the need to characterize individual states is to determine average neutron and $\gamma$-strength functions, as well as level densities, that can then be used as input in statistical model calculations of neutron capture rates using the Hauser-Feshbach approach. Statistical model codes commonly used for $r$-process calculations are TALYS \citep{talys1.6} and NON-SMOKER \citep{nonsmoker} (or its more recent incarnation SMARAGD). Statistical model predictions have been shown to be accurate to about a factor of 2 for neutron capture rates on stable nuclei where data from direct measurements exist \cite{Beard2014}. However, there are indications from theoretical calculations with a range of input models for level densities and $\gamma$-strength functions that uncertainties blow up considerably to orders of magnitude, just a few mass units away from stability \cite{Liddick2016}. The likely reason is the fact that the global parametrizations that are used for strength functions and level densities have been fitted to data near stability. A promising approach is therefore to experimentally constrain these input quantities for neutron-rich nuclei to enable statistical model calculations to at least achieve the same level of accuracy as for stable nuclei. 

A technique that has been recently developed with this goal in mind is the $\beta$-Oslo method \citep{Spyrou2014,Liddick2016}. In this method, a radioactive isotope is placed into a segmented total absorption spectrometer (such as SuN at the NSCL \citep{SUN2013}) and the $\beta$-delayed $\gamma$-ray emission is monitored.  The total excitation energy of a state populated by $\beta$-decay is obtained from the sum of all $\gamma$-ray energies deposited in the detector from the deexcitation cascade. Individual $\gamma$-rays originating from any of these excited states can be identifed from the energy deposited in individual segments of the detector. From the combination of these experimental data, the nuclear level density and the $\gamma$-ray strength function can be extracted \citep{Guttormsen1996,Schiller2000,Larsen2011}, which can then be used to constrain the neutron capture cross section within a statistical model.  The technique has been applied to the neutron capture of $^{75}$Ge and $^{69}$Ni, and has resulted in a reduction of the neutron capture uncertainties to factors of a few.  The technique can be applied to nuclei with production rates of the order of ten particles per second.  The completion of FRIB should enable the study of a significant number of the most sensitive neutron capture rates related to the formation of the the $N$=82 $r$-process abundance peak.  Further it would allow checks on the ratio between direct and statistical neutron capture \citep{Xu2014}.

\subsection{Experimental studies of fission \label{section:FissionExperiment}}
 Although, as discussed in Section~\ref{section:FissionTheory}, information on fission rates and fragment distributions of neutron-rich heavy nuclei is crucial to understand $r$-process nucleosynthesis, little is known experimentally as well as theoretically. Pioneering work to investigate fission fragment distributions in proton-rich nuclei~\cite{Schmidt01} has demonstrated the feasibility of fission studies with radioactive ion beams, but the information obtained so far is still too limited to help theorists in establishing more reliable theoretical predictions. 
 
Experimental approaches to study fission have recently been summarized in a review article \cite{Andreyev2018}. New experiments are proposed at RIKEN Nishina Center to determine fission barrier heights and fragment distributions in the region of $Z$=82--85. The experiments employ the $(p,2p)$ knockout reaction in inverse kinematics. Precise measurements of scattering angles and energies of the two protons allow one to determine the excitation energy in the residual nucleus using the missing mass technique. The expected resolution for the excitation energy is 2~MeV. The residual nucleus, whether or not fission takes places, is identified with the SAMURAI spectrometer and its focal plane detector setup. 
 
 A new aspect in the experiments is that both $Z$ and $A$ of the fission fragments are identified event-by-event and their distributions are determined as a function of excitation energies. This provides nearly complete experimental information of nuclear fission and will be an essential input both to $r$-process network calculations and to theories of fission.

\subsection{($\alpha$,n) reactions}
\label{sec:exptechniques:an}
Most current $r$-process models predict that the $r$-process site creates its own seed from a mix of protons, neutrons, and $\alpha$-particles at high temperatures. Charged particle reactions such as 3$\alpha$, $\alpha \alpha n$ and ($\alpha$,n) reactions on heavier nuclei create seed nuclei, typically in the $A\approx 80-90$ mass region as the increasing Coulomb barrier makes such reactions on heavier nuclei inefficient. These seed nuclei serve then as starting points for the rapid neutron captures and $\beta$-decays that constitute the $r$-process. 

While it is typically assumed that the seed production is a hierarchical freeze-out sequence from equilibrium and therefore relatively insensitive to the individual reaction rates (except for major bottle necks such as the $\alpha \alpha n$ reaction, or rather its slower subsequent $^{9}$Be($\alpha$,n)$^{12}$C reaction), it has recently been shown that in a weak $r$-process producing predominantly elements in the Sr, Y, and Zr range, ($\alpha$,n) reaction rates do affect the nucleosynthesis significantly  \cite{Hansen.etal:2014}. An example is the fast-expanding ($\sim$few ms) neutrino-driven winds following core collapse supernovae \cite{Arcones.Thielemann}. Under these conditions, the neutron-to-seed ratio is small, and the $r$-process ends in the $A$=80-90 region with the final abundances produced by a mix of ($\alpha$,n) reactions, neutron captures, and $\beta$-decays. Recent studies \cite{bliss16} show that $(\alpha,n)$ reactions can have a strong impact on the synthesis of Sr-Zr elements at temperatures between $\sim$2$-$5~GK (see also Sec.~\ref{section_NSS}). 
The most important $(\alpha,n)$ reactions involve neutron-rich nuclei about $\sim 2-10$ neutrons away from stability in the region between Ga and Sr. Unfortunately, none of the important $(\alpha,n)$ reaction rates are experimentally known at $T$$\sim$2$-$5~GK \cite{bliss16}, and must consequently be calculated with reaction codes like e.g. TALYS \cite{talys1.6} and NON-SMOKER \cite{nonsmoker}. Recent studies have shown that the theoretical uncertainty of calculated rates for these reactions can reach factors up to 10 in the range of temperatures relevant for the $\alpha$ process \cite{pereira16} (see Sec.~\ref{Sec:Theory:Rates}). These results emphasize the need to experimentally study the relevant $(\alpha,n)$ reactions in detail. 

As the importance of ($\alpha$,n) reactions in the weak $r$-process has only been discovered recently, experimental work has only just begun. The required neutron-rich low energy beams are not too far from stability, and some are available with sufficient intensity at existing facilities. A first program to measure these reactions has started at MSU's NSCL ReA3 facility that provides reaccelerated radioactive beams at low astrophysical energies. A first experiment using a He gas cell and the newly developed HABANERO neutron detector has been  carried out and is currently under analysis. HABANERO is a $^{3}$He- and BF$_3$-based long counter with a polyethylene moderator matrix, similar to NERO \cite{pereira10}. The main difference is a redesign optimized for higher energy neutrons from reactions in inverse kinematics that provides a relatively constant efficiency up to 20 MeV. 

In the future, there are also opportunities for measurements at other facilities such as TRIUMF/ISAC or ANL/ATLAS. Alternative detection methods to be explored may involve active targets such as the NSCL Active-target Time Projection Chamber (AT-TPC) \cite{suzuki12} or the ATLAS MUlti-Sampling Ionization Chamber (MUSIC) \cite{carnelli15}, possibly in conjunction with neutron detectors \cite{ayyad16}. 

\section{Theoretical methods}
\label{sec_theory}

The majority of data used in $r$-process calculations have to be obtained from theoretical models. 
For the ``hot" $r$-process, experimental $\beta$-decay half-life measurements  have reached the $r$-process path only up to around $A$$<$140. For masses and $\beta$-delayed neutron emission branching ratios a much smaller number of nuclei have been reached. There is no experimental information available on neutron capture rates, fission rates, or fission fragment distributions in the $r$-process. Even when selected observables such as half-lives or masses are measured, they need to be corrected for astrophysical conditions. For the ``cold" scenario, the $r$-process path runs near the neutron dripline completely outside of current experimental reach (though as the $r$-process progresses, the path shifts closer to stability were some data are available). With continued experimental efforts at RIKEN, and in later years with the new generation of radioactive beam facilities now under construction, the amount of experimental data available for $r$-process studies will increase dramatically. For example, more than 200 $\beta$-decay half-lives were measured and remeasured for neutron-rich isotopes between $^{72}$Co up to $^{163}$Pm by the EURICA collaboration at RIKEN (see Sec.~\ref{sec:exptechnique:decay}). The ongoing experimental campaigns of the BRIKEN collaboration (see Sec.~\ref{sec:exp:3He}) 
are expected to complement this dataset with more half-life measurements of neutron-rich isotopes, as well as a (re)measurement of $\beta$-delayed neutron branching ratios for nuclei with $A$=30$-$210. In 2017 alone the BRIKEN experiments have covered 231 $\beta$n-emitters between $^{64}$Cr up to $^{151}$Cs, many of them for the first time, as well as about 50 new $\beta$-decay half-lives. 

Nevertheless, depending on the astrophysical models, some sections of the $r$-process path, such as  the heavy-element fission region, are going to remain out of experimental reach. In addition, theoretical corrections for the astrophysical environment will always be required. Nuclear theory will thus always be an integral part of $r$-process studies. Of key importance will be to take full advantage of experimental capabilities to improve theoretical predictions, and the reliable characterization of the uncertainties of theoretical predictions - also guided by data. With new experimental data on neutron-rich nuclei, and advanced uncertainty quantification techniques, astrophysical information can be extracted from observations in a fully quantitative way. 



\subsection{Theoretical data in current $r$-process models and their limitations}
\label{Sec:NucTheory_Current}

Theoretical predictions of masses, decays, neutron capture rates, and fission-related observables  have made great strides. However, astrophysical $r$-process model calculations do not typically employ the most cutting-edge nuclear theory. There are two chief reasons for this: 

(1) Progress in nuclear theory often does not translate into improved predictions for global data sets that include all nuclei. Such consistent global data sets are required for astrophysical calculations. The challenge is that advanced theoretical approaches are often only possible in limited regions of the nuclear chart, or they are computationally so expensive that global calculations are not feasible. Nevertheless, the generation of new large-scale datasets should be accelerated. In addition, opportunities to benchmark and then improve global theoretical surveys using computationally more demanding smaller-scale simulations should be exploited in order to ensure that progress in nuclear theory has its full impact on nuclear astrophysics. 

(2) Predicted observables are often strongly correlated. This requires a consistent recalculation of all datasets once improvements in a global model are made. For example masses, through the corresponding $Q$-values, affect  predictions of decay properties, neutron capture rates, and fission properties. Many current $r$-process models still rely on the FRDM mass model from 1995 \cite{Moller:1995} or the HFB-21 mass model from 2010 \cite{Goriely:2010}, despite newer versions being available. One reason is that the available decay and reaction rate datasets are based on the older mass models. A capability to timely update  global datasets used for $r$-process calculations is needed. This would also facilitate sensitivity studies, where variations in one nuclear quantity should be propagated to other observables, see Sec.~\ref{section_NSS}. 

\subsubsection{Masses}\label{sec:masses1}

\label{sec:current_mass_models} Nuclear masses are crucial for $r$-process simulations  for any  astrophysical scenario (see Sec.~\ref{sec:sens-input}). As many of the relevant nuclei are out of experimental reach, present $r$-process model calculations must rely on global mass models. The most commonly used mass models in current $r$-process simulations are the finite range droplet model FRDM 1995 \citep{Moller:1995} (though there is a newer version FRDM 2012 \citep{Moller2016}), the Hartree-Fock Bogoliubov mass model series with the most commonly used HFB-21 version \citep{Goriely:2010} (though HFB-31 is the most recent model \cite{Goriely2016}), the Duflo-Zuker (DZ) mass formula \citep{DufloZuker:1995}, and the more recently developed WS3 mass formula \citep{WS3}. The FRDM is a microscopic-macroscopic approach and HFB-n is  rooted in a self-consistent mean-field approach with some phenomenological corrections added; in both cases 
parameters are globally optimized to experiment. The rms deviations from experimental masses typically range from 500$-$700~keV for FRDM and HFB-n to 300$-$400~keV for WS3 and DZ \citep{Mumpower2016}. These deviations are significantly larger than the $\approx$100~keV uncertainty required for astrophysical applications. 

However, as all mass models are fitted to experimental data, the rms deviation only provides a lower limit of the model uncertainty in predicting unknown masses. Rather, the predictive power of the models towards more unstable nuclei is the key criterion to assess the quality of a mass model for $r$-process studies. This predictive power is difficult to determine. One approach is to compare models to new masses that were not known at the time the mass prediction was published. The most recent study of this type has been performed in Ref.~\cite{Sobiczewski14} who compared the performance of 10 different models for the masses measured before and after the AME2003 mass evaluation. The study has some limitations: it does group nuclei by mass region, but does not look specifically at neutron-rich nuclei of relevance for the $r$-process. It also includes mass models developed after 2003, for which the new masses were used in the mass fit. Nevertheless it is clear that the mass models such as FRDM 1995 or HFB-21 show only a limited increase in rms when compared to new masses (up to about 20\% or 100 keV), while phenomenological mass formulae with their lower overall rms deviation show much stronger increase in relative and absolute errors. In fact, the paper finds an interesting anti-correlation between global rms mass deviation and extrapolation quality, with the models with the worst rms deviations compared to known masses performing the best in terms of extrapolation. This demonstrates the inadequacy of the rms mass deviation as a measure of global quality of a mass model. Indeed, if a model does not have sound microscopic foundations it cannot be trusted when it comes to huge extrapolations outside the experimentally-known regions. 

Another challenge is that progress in mass measurements towards more neutron-rich nuclei has been slow. For this reason, studies of mass model extrapolability are based on a  small number of masses of nuclei located just a few mass units away from the masses used to optimize the model. They do therefore not necessarily provide a reliable estimate for the predictive power of mass models in the $r$-process path that can be 10 or more mass units away. Mass measurements at the new generation of rare isotope facilities will dramatically change this picture. With the large number of new masses of neutron-rich nuclei expected from these facilities we will be able to  quantify better the predictive power of the global  mass models used in $r$-process simulations. 
In this respect, as discussed in Sec.~\ref{Sec:new-masses}, modern statistical tools such as hierarchical Bayesian approaches  utilizing our prior knowledge of systematic trends in mass residuals, can significantly improve quality of extrapolations based on global microscopic models.  

Another approach to assess the predictive power of current theory is to assess the spread of predictions using different  models that have demonstrated  good reproducibility of known masses. Another measure of a mass model's performance is its ability to reproduce important binding energy differences such as   neutron separation energies or $Q$-values for $\beta$-decay. For such quantities, the model performance is often  significantly better compared to masses as many systematic theoretical errors are expected to cancel out.

\begin{figure*}
 \begin{center}
  \centerline{\includegraphics[width=4.5in]{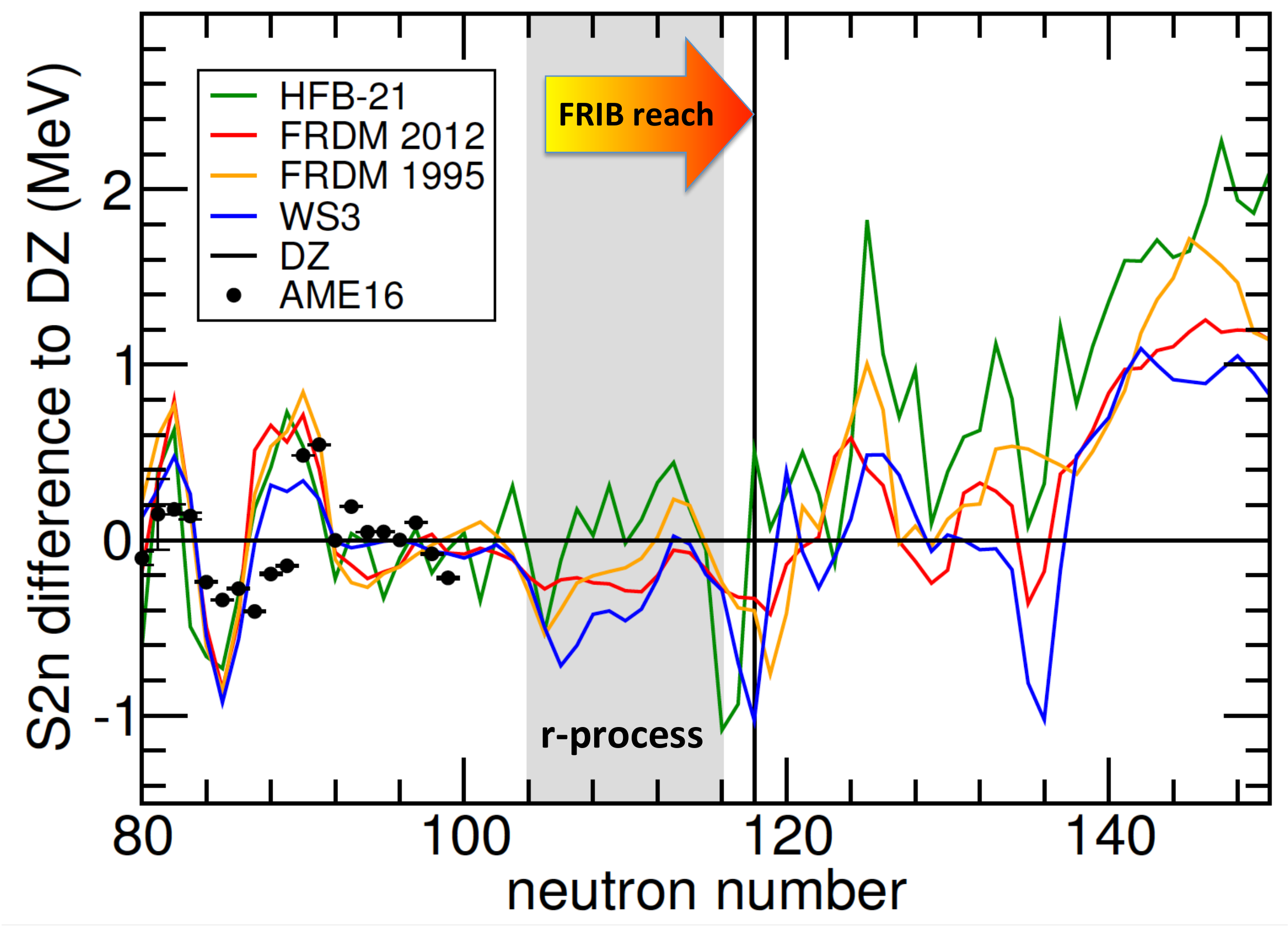}} 
  \caption{\label{fig:GdS2n} Two-neutron separation energies (S$_{2n}$) of Gd isotopes obtained in various mass models often used in $r$-process simulations shown relative to DZ mass formula predictions. The experimental values from AME2016 \cite{AME16a,AME16} are also shown. The gray band marks the important region suggested by the sensitivity studies for a typical hot $r$-process \citep{Mumpower2016}. The reach for mass measurements at FRIB (assuming an intensity limit of around 10$^{-3}$~pps) is also indicated. }
 \end{center}
\end{figure*}

Figure~\ref{fig:GdS2n} shows a typical example of the current theoretical situation, using the two-neutron separation energies $S_{\rm 2n}$ of the Gd isotopes and mass models often used for $r$-process simulations. In general, mass predictions are consistent where experimental data are available.

A few mass units past the last measured point, as one enters the region of interest for the $r$-process, differences between models increase significantly. Nevertheless, discrepancies stay within about 800~keV for about 24 mass units across the $r$-process path. Around the $N=126$ shell closure and towards the neutron dripline, differences between  $S_{\rm 2n}$ values become much larger and can reach up to 2~MeV. While this may provide some indication of the expected uncertainties, it is not necessarily a reliable measure. Uncertainties may be overestimated because of one low-performing model, or uncertainties may be underestimated because not all models shown in Fig.~\ref{fig:GdS2n} are based on similar phenomenology near stability. An indication for the latter is the good agreement between mass models near stability, despite large discrepancies with data. 
Such model dependencies introduce an appreciable nuclear physics uncertainty into calculation of $r$-process abundances, especially because  mass errors exponentially impact  the half-lives. See for example Refs.~\cite{Mumpower2016,Mendoza:2015,Martin:2016} for $r$-process model calculations based on different mass models. 

\subsubsection{$\beta$-decay half-lives}
In a hot $r$-process, $\beta$-decay half-lives determine, for a given reaction path, the speed at which the $r$-process can create heavier elements and the abundance pattern that is created along the path (with more abundance accumulating where the decays are slow). For a cold $r$-process, the competition of $\beta$-decay and neutron capture directly determines the reaction path. Sensitivity studies indicate that $\beta$-decay half-lives need to be known to at least a factor of 2 (see Sec.~\ref{section_NSS}).

There are two global sets of theoretical predictions of $\beta$-decay half-lives that are currently used in most $r$-process calculations: the QRPA model of Ref.~\cite{Moller2003} and the gross theory model of Ref.~\cite{Tachibana1990}. The half-life predictions depend strongly on the mass model used. Datasets available for $r$-process calculations include the QRPA predictions based on FRDM 1995 masses, and gross theory predictions using the HFB-21 mass model (see Sec.~\ref{sec:current_mass_models}).

The performance of the two approaches is comparable when it comes to the reproduction of measured $\beta$-decay half-lives of neutron-rich nuclei  \cite{Mumpower2016}. For short-lived nuclei with half-lives below 1\,s, deviations from experiment are within about a factor of 10.  For very short-lived nuclei below a few 100\,ms, the accuracy becomes somewhat better. Predictions for short-lived nuclei are more reliable because the larger $Q$-value window enables contributions of many transitions, and reduces the sensitivity to the excitation energy of a particular transition (which depends on (Q$_\beta$$-$E$_x$)$^5$). Clearly, theoretical accuracy is still far from the desired factor of 2. Note, however, that these deviations are due to a combination of mass uncertainty and intrinsic model uncertainty as for most experimental half-life data far from stability no mass measurements have been carried out.

While the majority of $\beta$-decay half-lives for $r$-process models needs to be predicted theoretically, the number of available experimental $\beta$-decay half-lives is steadily increasing with new rare isotope production capabilities coming online. One issue with some of the experimental half-lives is the possible existence of isomers. In this case, the measured half-life may be related to a long-lived excited state, or it may be a mixture of ground and excited state decays. It is not always possible to clarify this aspect experimentally and theory is needed to provide guidance. Even if the ground state half-life has been measured, it needs to be corrected for possible decays from excited states that can be thermally populated in the hot astrophysical environment where the $r$-process occurs. For nuclei far from stability, where $Q$-value windows are large and many allowed transitions are typically possible, such a correction for thermally populated excited states is expected to be smaller than in some cases near stability. Initial estimates indicate that the effects of excited state $\beta$-decays  can be significant enough to affect final $r$-process abundance distributions \cite{Famiano2008}.

\subsubsection{$\beta$-delayed neutron branching ratios}\label{sec:bnbranchings}
In addition to predicting the $\beta$-decay strength function, models of $\beta$-delayed neutron emission must describe the competition between the neutron emission and the $\gamma$-ray de-excitation of  excited states above the neutron separation energy. Another challenge is to understand the competition between the different neutron emission channels ($\beta$1n, $\beta$2n, $\beta$3n...), especially in light of the lack of experimental data (as discussed in Sec.~\ref{sec:exp-bn}). 

\begin{figure}[!htb]
	\centering
\includegraphics[width=0.9\textwidth]{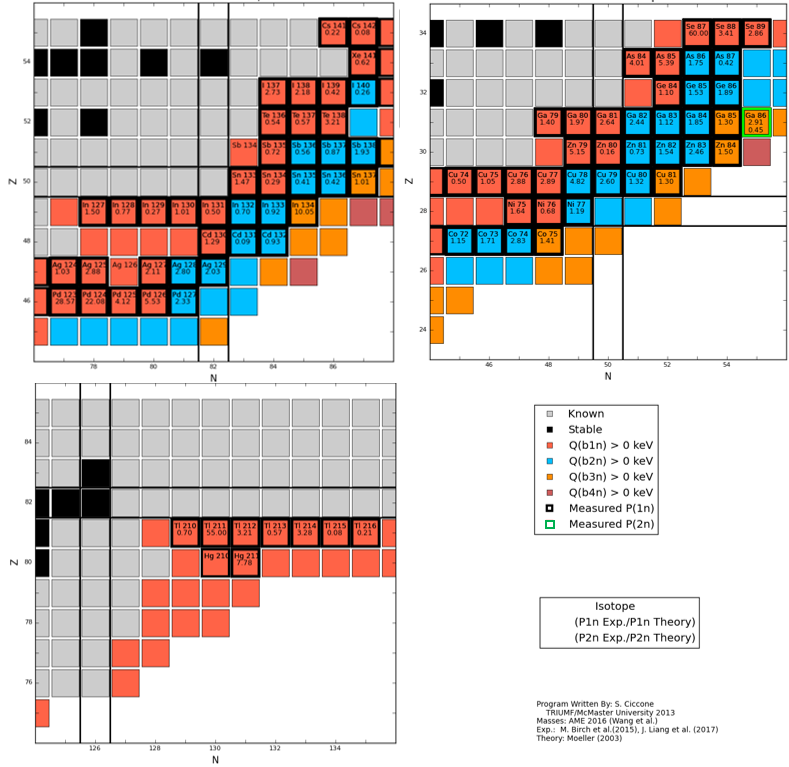}
\caption{Ratio of measured and theoretical $P_{1n}$ ground-state values from  Ref.~\cite{Moller2003} for the $N$=50, 82, and 126 regions.}
\label{fig:bn-ratio}
\end{figure}

The QRPA approach can be employed to compute the Gamow-Teller (GT) strength used to predict $\beta$-delayed neutron emission. Here, the  standard is the QRPA model \cite{Moller:1997,Moller2003} using masses from FRDM 1995. The model from 2003 \cite{Moller2003} combines GT QRPA calculations with an empirical spreading of the quasi-particle strength and the gross theory for the first-forbidden part of $\beta$-decay. This model is still widely used in the community for inferring neutron-branching ratios for $r$-process studies. In Fig.~\ref{fig:bn-ratio} the ratio between the measured P$_{1n}$ values relative to the QRPA predictions from M\"oller 2003 \cite{Moller2003} are shown for the $N$=50, 82, and 126 regions. The experimental values are generally reproduced within a factor of 5.

Current models typically assume that there is no competition between the various de-excitation channels ($\gamma$-, one- and multi-neutron emission), and allow as many neutrons as energetically possible to be emitted. Thus the predicted average number of emitted neutrons is always overestimated.

\begin{figure}[!htb]
	\centering
\includegraphics[width=1.0\textwidth]{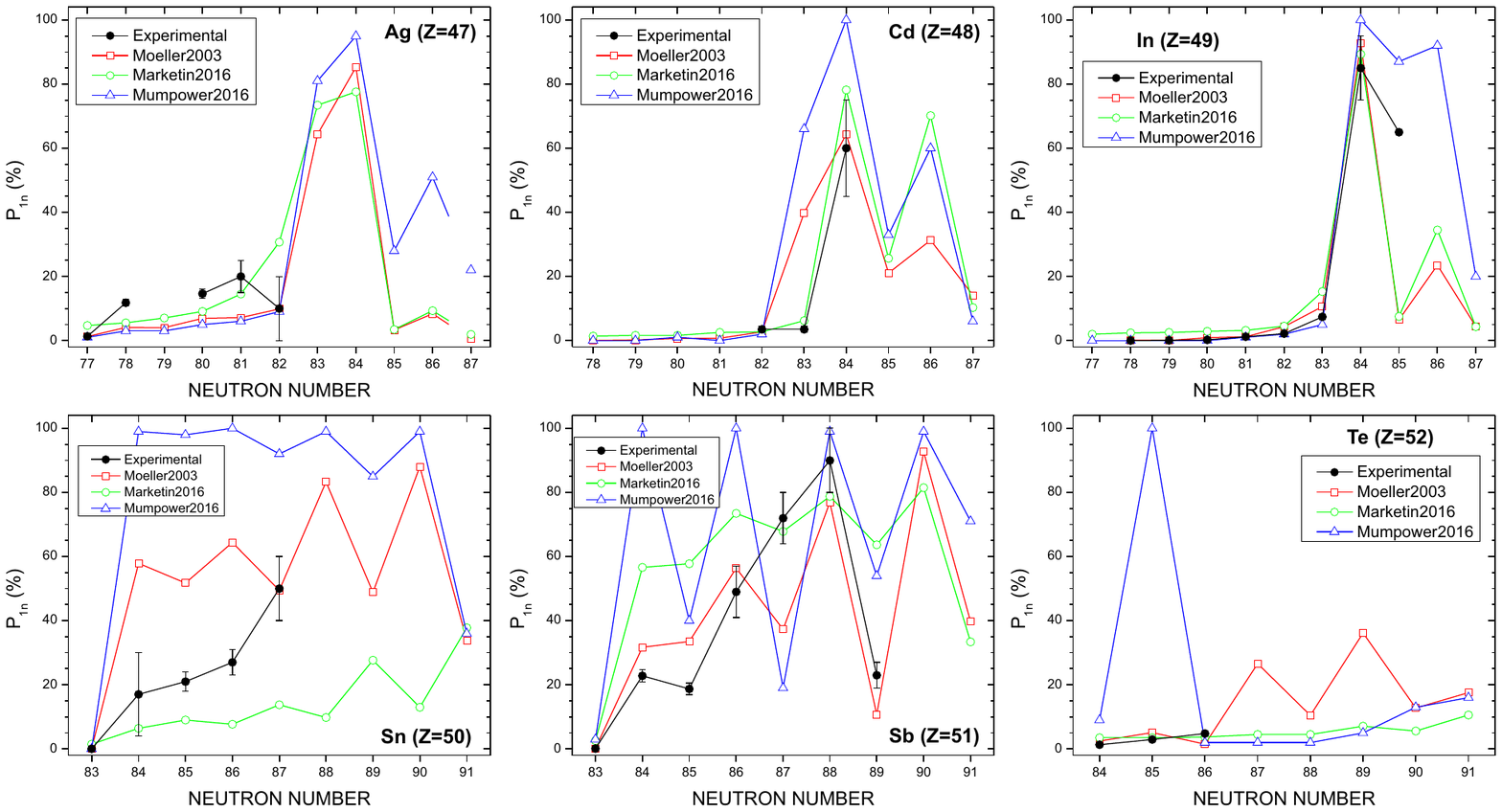}
\caption{Comparison of measured and theoretical $P_{1n}$ ground-state values from Refs.~\cite{Moller2003,Marketin2016,Mumpower2016b} for isotopes with $Z$=47-52.}
\label{fig:bn-comp}
\end{figure}

The recent approached  are these of Refs~\cite{Marketin2016,Mumpower2016b}. In both cases, excited states are obtained within the proton-neutron QRPA.
An improvement is to combine the theoretical  $\beta$ strength  with emission cross-sections provided by a statistical model to include the competition of $\gamma$-rays with neutron emission above $S_{\rm n}$. In the model of Ref.~\cite{Mumpower2016b}, the Hauser-Feshbach formalism is used to estimate $\gamma$ spectra as well as delayed particle spectra and probabilities. It predicts that on average more $\beta$-delayed neutrons are emitted for nuclei near the neutron drip line compared to models that do not consider the statistical decay. 
Figure~\ref{fig:bn-comp} compares the aforementioned three models with experimental data for isotopes around the doubly-magic $^{132}$Sn. For these six isotopic chains, very different behavior can be observed. For indium ($Z$=49) the experimental trend with the sudden jump at $N$=84 is reproduced until the $\beta$2n emission channel opens. For tin ($Z$=50) and antimony ($Z$=51) all theoretical models predict a even-odd staggering which is not reproduced by the data. This figure is just one example that shows the necessity to improve theoretical models and their predictive power towards $\beta$-delayed neutron emission in the $r$-process. 

In the absence of reliable theory, phenomenological models using systematics \cite{Miernik2013,Miernik2014,KHF1973,McCutchan2012} provide a better description of the available data and may even allow a short-range extrapolation of P$_{1n}$ values of a few mass units. However, these phenomenological models do not have the predictive power needed for $r$-process applications. The large amount of new data on $\beta$-delayed neutron emission expected from rare isotope beam facilities (see e.g. Sec.~\ref{sec:exp:3He}) will provide a unique data set needed to asses the predictive power of various approaches.

\subsubsection{$\alpha$-decay}
In principle, $r$-process models also include $\alpha$-decay, which can become relevant in the actinide and superheavy element region. If not available experimentally, these rates are often estimated using phenomenological expressions based on the Geiger-Nuttall $\alpha$-decay law~\cite{Koura2012,Qi2014}. 
$\alpha$-decay is mostly important after freeze-out when heavy elements decay back into long-lived actinides, Pb, or Bi. However, it is not clear whether the final results are sensitive to the rates. 

\subsubsection{Neutron capture and $(\alpha,n)$ rates}
\label{Sec:Theory:Rates}
Neutron capture rates in $r$-process simulations are exclusively based on the Hauser-Feshbach statistical model. The most commonly used models are TALYS \cite{talys1.6}, which employs microscopic level densities and strength functions, and NON-SMOKER \cite{nonsmoker}, which uses phenomenological descriptions of these important input parameters. Rate datasets from TALYS go beyond a compound nucleus approximation and also include direct capture and pre-equilibrium capture effects. 
For stable nuclei, where neutron capture data are available, these models can predict neutron capture rates with typical uncertainties of around a factor of 1.4, which can increase near shell closures to about a factor of 2 or more \cite{Beard2014,Mumpower2016}. These uncertainties would be acceptable for $r$-process model calculations \cite{Mumpower2016}. However, uncertainties for neutron capture rate predictions for unstable nuclei are more difficult to estimate. Comparisons between predictions of the two codes indicate differences that quickly increase with neutron number and can reach up to 3 orders of magnitude (though some of this may be due to differences in predicted masses; see also Sec.~\ref{sec:exptechniques:ncap}). 

Another approach is to vary within TALYS  various models for level density, $\gamma$-strength functions, and optical model potentials. While the predictions with the different options agree within a factor of a few for capture rates near stability, the differences quickly increase to 1-2 orders of magnitude just a few mass units away from stability \cite{Liddick2016}. Level densities and $\gamma$-strength functions seem to be particularly important. 

\begin{figure*}
 \begin{center}
  \centerline{\includegraphics[width=2.5in]{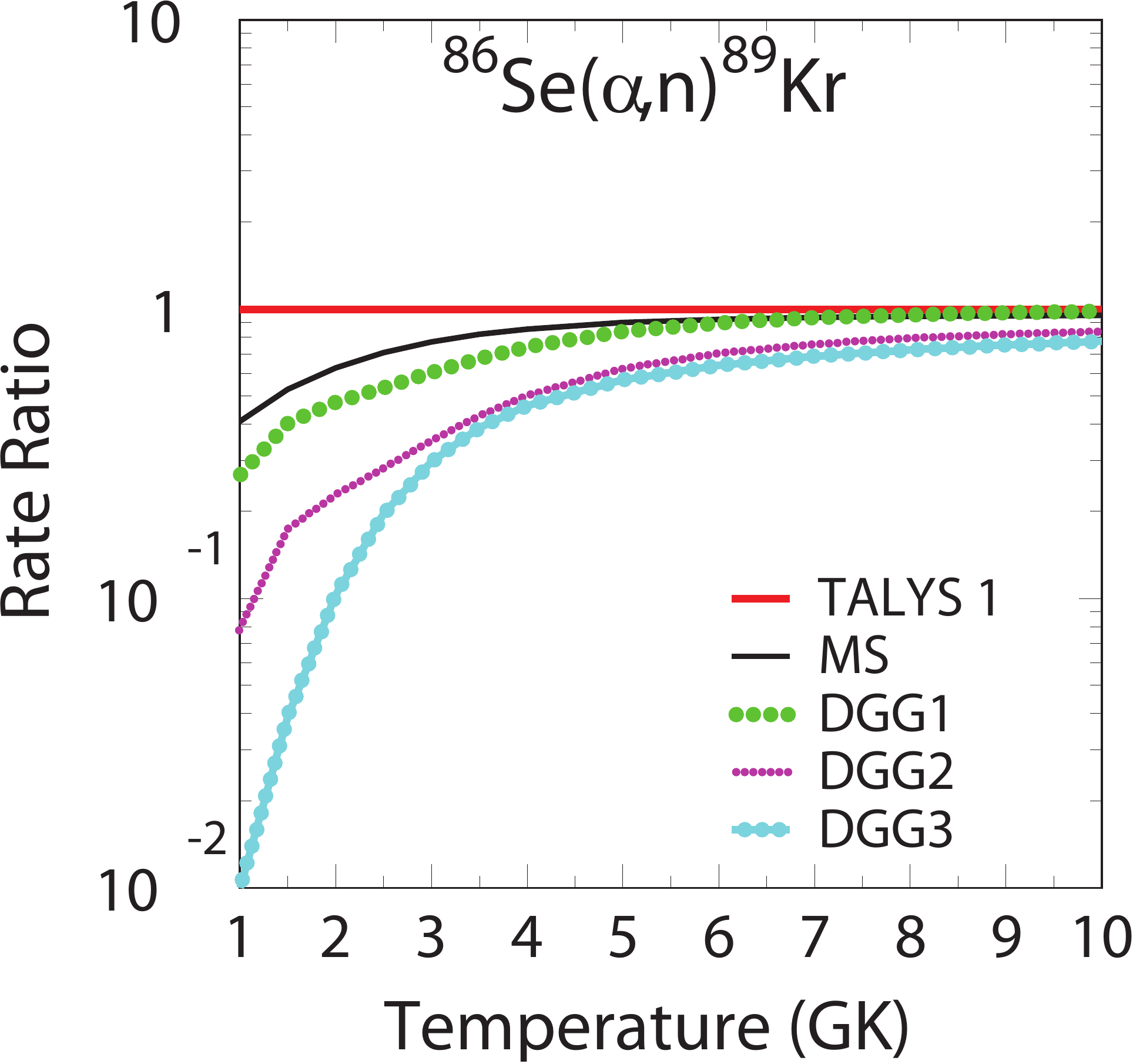}}
  \caption{\label{fig:talys-aop-se86an} TALYS rates for the reaction $^{86}$Se$(\alpha,1n)$$^{89}$Kr calculated using the alpha optical-potentials from the TALYS global model (red solid line), the McFadden and Satchler model (MS), and three versions of the Demetriou-Grama-Goriely model (DGG1, DGG2, DGG3). The rest of the nuclear inputs are taken from the packet TALYS 1 explained in Tables I and II of \cite{pereira16}. All the rates are normalized to the results obtained using the packet TALYS 1.}
 \end{center}
\end{figure*}

The situation is similar for ($\alpha$,n) reactions \cite{pereira16}. In this case, however, the most important source of uncertainty is the $\alpha$-nucleus optical potential \cite{pereira16,mohr16}. Figure \ref{fig:talys-aop-se86an} compares the TALYS-calculated reaction rates of $^{86}$Se$(\alpha,1n)^{89}$Kr at temperatures between 1\,GK and 10\,GK using different  optical-potential models. As can be seen, the differences in the reaction rates increase rapidly at low temperatures. Besides $\alpha$-nucleus optical potentials, there are technical aspects intrinsic to the reaction codes that can introduce additional uncertainties. For instance, while TALYS can calculate the exclusive reaction channels $(\alpha,\times n)$, which must be explicitly included in network calculations, the NON-SMOKER rates are inclusive \cite{pereira16}. Another important aspect that needs to be investigated is the role played by reaction mechanisms that go beyond the Hauser-Feshbach formalism, such as direct or pre-equilibrium components. 

\subsubsection{Fission rates and yields}
Fission plays a critical role in most $r$-process models. Neutron-induced fission provides a natural endpoint of the $r$-process that most models predict to lie in the $A$$\approx$260 mass region. Neutron-to-seed ratios in neutron star merger $r$-process models are predicted to be large enough to not only reach this mass region, but also to undergo so-called "fission (re)cycling" \cite{Seeger+65}, where the fission fragments serve as new seeds for the $r$-process. Fission cycling has been proposed as a possible explanation for the apparent robustness of the $r$-process indicated by observations of $r$-process signatures in metal-poor stars \cite{Beu08} (see Sec.~\ref{lightrpro}). Systematic sensitivity studies have not yet been performed for fission rates in the $r$-process. $r$-process model calculations with different fission model predictions indicate a significant impact on the final abundance distribution \cite{Gor13,Eichler2015,Gor15}. 

The most important fission mechanism is neutron-induced fission during the $r$-process. $\beta$-delayed fission plays a role when heavy nuclei decay back to stability and can for example affect the final U and Th abundances \cite{schatz02}. Both mechanisms are governed by fission from excited states. The key observables here are fission rates, and fission fragment distributions. 

Only recently have $r$-process models begun to incorporate comprehensive models for fission rates and fragment distributions. 
One commonly used set of rates for neutron induced \cite{Pan2010} and $\beta$-delayed fission \cite{Pan05} uses the statistical model code SMOKER and the same QRPA model used for the $\beta$-decay rate set of \cite{Moller2003}, respectively. A variety of 
schematic approaches utilizing computed fission saddle points obtained in various mass models are employed to generate fission rates \cite{Moller2003,Gor09,Gor13}. Fission fragment distributions are typically obtained from simple parametrizations, often guided by limited experimental data (see for example Ref.~\cite{Eichler2015} for a brief overview). 

\subsection{Developments in nuclear theory}
\label{Sec:NucTheory_New}

\subsubsection{New mass predictions}\label{Sec:new-masses}

Most of the masses of nuclei that are required for the $r$-process element
abundance calculations come from theoretical models. 
Theoretically, there has been exciting progress in global modeling of nuclear properties, greatly facilitated by high-performance computing. A microscopic tool that is well suited to provide quantified microphysics throughout the nuclear chart is nuclear DFT \cite{Ben03}. The in-medium effective interaction of DFT is modeled in terms of the energy density functional (EDF), whose parameters  have been fit to measured mass data and other global nuclear properties \cite{Erl12a,Goriely2013,Goriely2016,UNEDF2,Afanasjev15,Afanasjev16,Wang2015,Xia17}. This approach is capable of predicting a variety of observables needed, and is able to assess the uncertainties on those observables, both statistical and systematic \cite{Dob14,DFTBayes}. Such a capability is essential in the context of making extrapolations into the regions where experiments are impossible. 

A basic test of any EDF parameterization is its ability to reproduce binding energies and other basic nuclear properties across the nuclear chart. 
The overall rms deviation between theoretical and experimental masses for $A$=16-250 is 500 keV or greater \cite{Goriely2016,UNEDF2,Afanasjev16}. Currently, the best overall agreement with experimental masses obtained with the Skyrme-like EDF is 0.561 MeV (HFB-31) \cite{Goriely2016}. However, this excellent result was obtained at a price of several corrections on top of the EDF used. 
\begin{figure*}[htb] \centering
\includegraphics[width=0.5\textwidth]{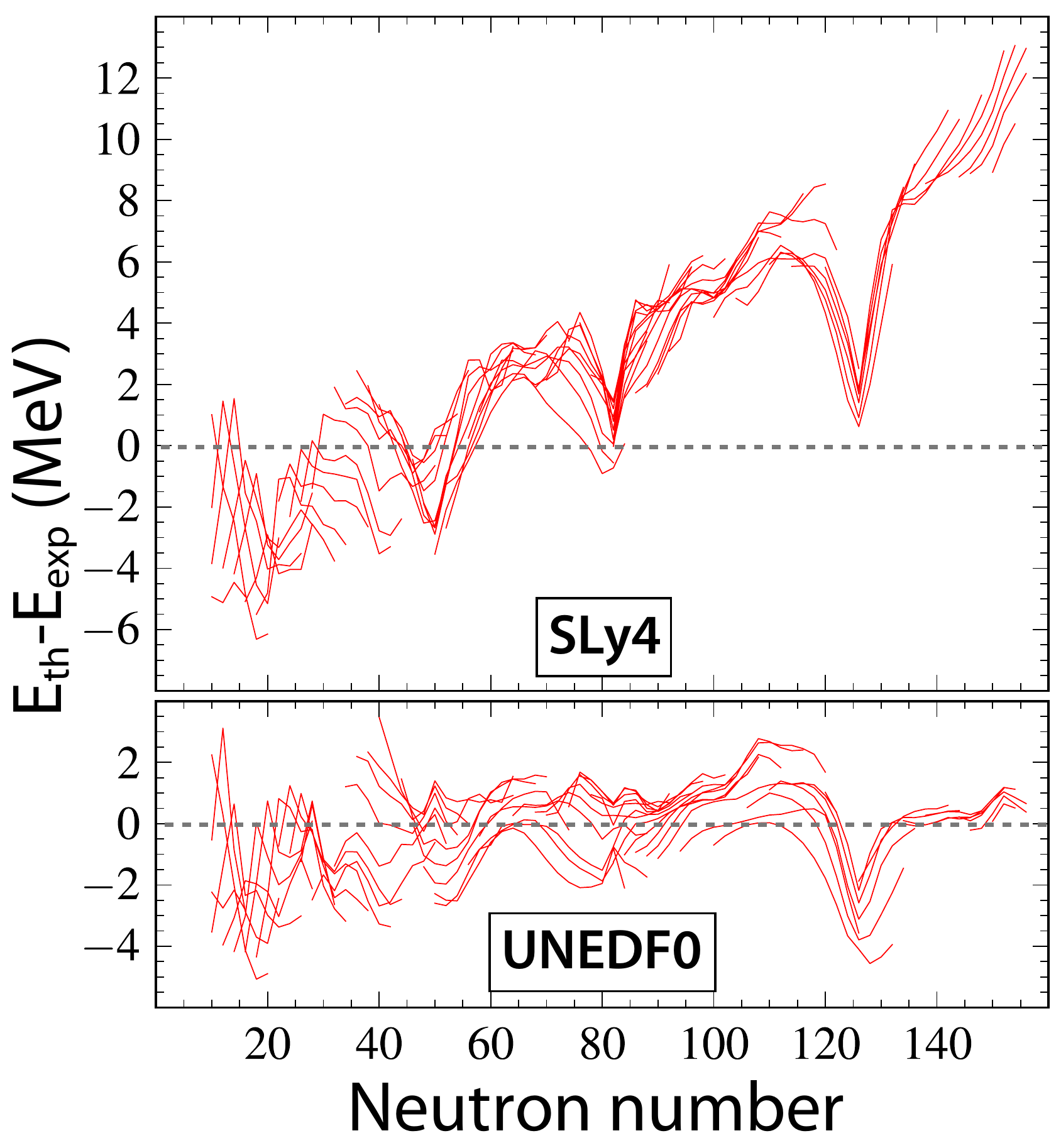}
\caption{Binding energy residuals between DFT calculations and experiment for 520 even-even nuclei \cite{UNEDF0}. The DFT results were obtained with functionals SLy4 (top) are UNEDF0 (bottom).}
\label{fig:E-residuals}
\end{figure*}

The binding energy residuals (i.e., differences between experimental and theoretical values) for the 520  even-even nuclei are shown in Fig.~\ref{fig:E-residuals} for SLy4 \cite{Chabanat1998} and UNEDF0 \cite{UNEDF0} EDFs. 
A pronounced systematic trend  is seen for  SLy4. By contrast, carefully optimized UNEDF0 shows a much flatter behavior, while simultaneously reducing the mass residuals: from rms deviation of 4.80\,MeV in SLy4, to 1.45 MeV in UNEDF0. By inspecting  Fig.~\ref{fig:E-residuals} it is apparent  that -- while the global trend of binding energy residuals in UNEDF0 has been improved --  significant systematic variations  remain. 

\begin{figure*}[htb] \centering
\includegraphics[width=0.5\textwidth]{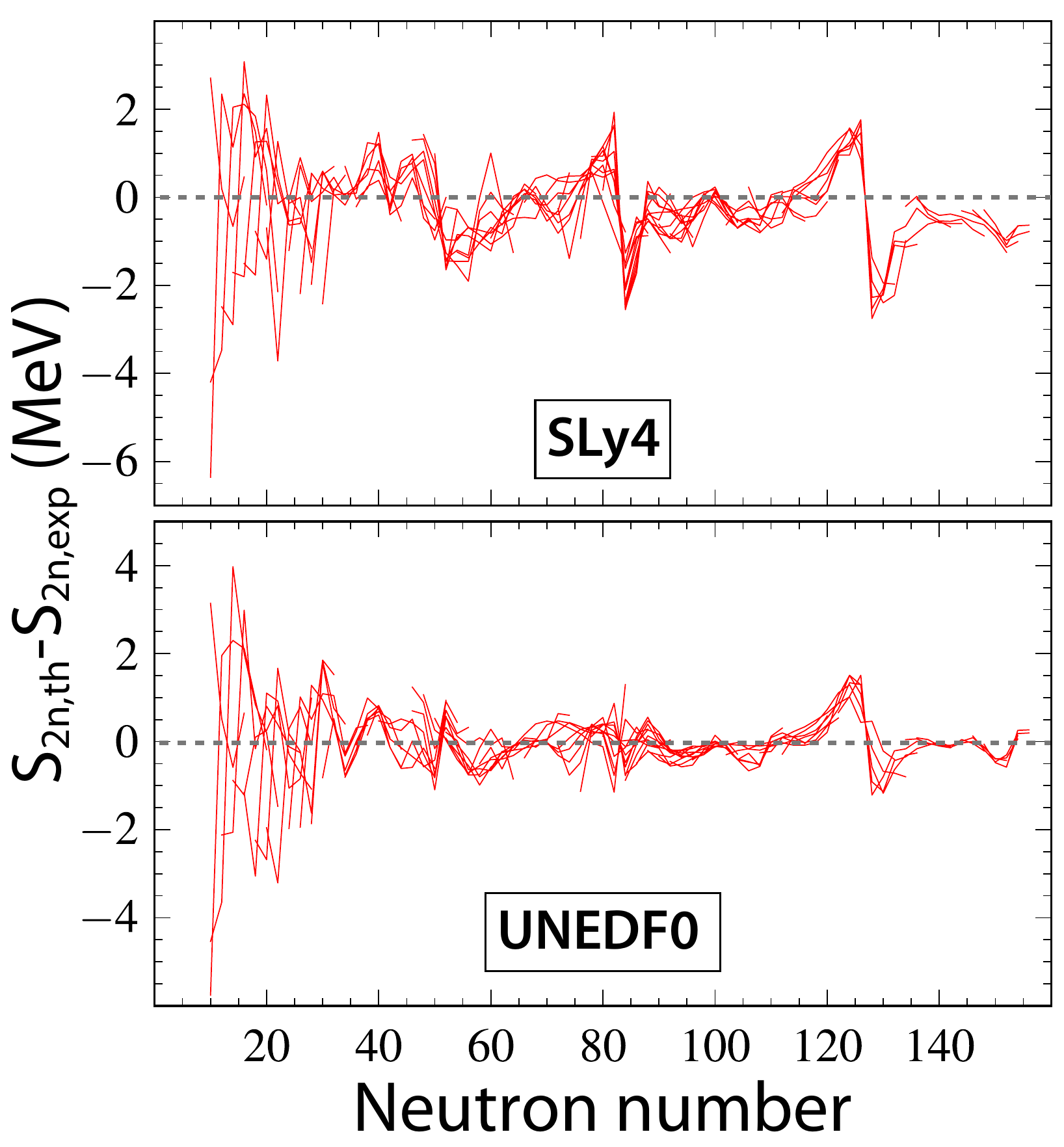}
\caption{Similar as in Fig.~\ref{fig:E-residuals} but for two-neutron separation energy residuals}
\label{fig:S2n-residuals}
\end{figure*}

To quantify this, Fig.~\ref{fig:S2n-residuals}  shows two-neutron separation energy residuals. The rms deviations of $S_{2n}$ for SLy4 and UNEDF0 are, respectively, 0.99 and 0.76\,MeV. This improvement is expected as some systematic uncertainties cancel out in binding energy differences.  For the subset of heavy nuclei ($A \geq 80$), the rms variations are 0.85\,MeV for SLy4 and 0.45\,MeV for UNEDF0. 

In the context of the $r$-process, the challenge is to carry out reliable model-based extrapolations into the neutron-rich region where experiments are not available. To improve the quality of theoretical input, advanced statistical techniques of uncertainty quantification must be used. 

If the experimental mass (or binding energy) $m_{\rm exp}(Z,N)$ is known, it can be related to the model prediction $m_{\rm th}(Z,N)$ via:
\begin{equation}\label{residual}
m_{\rm exp}(Z,N) = m_{\rm th}(Z,N,\theta) + \delta(Z,N),
\end{equation}
where $\theta$ is the vector of model parameters and  $\delta(Z,N)$ is the mass residual, which can be further split into the systematic part due to  imperfections of the nuclear physics model (incorrect model assumptions, uncontrolled simplifications, etc.) and the statistical uncertainty stemming from uncertainties on model parameters $\theta$ and experimental errors. In the following, we neglect the experimental errors on masses as those are usually well below theoretical uncertainties. In general, the problem is to calibrate the model (i.e., estimate $\theta$), predict the mass of 
isotopes  inside and outside the range of experimental data, and estimate the uncertainty
of the prediction.

The statistical uncertainty can be estimated by means of the linear regression technique \cite{Dob14,Gao2013,Kortelainen2015,Niksic15,Haverinen17} or Bayesian inference methods \cite{DFTBayes,Schunck2015e}. By propagating theoretical statistical uncertainties to unknown neutron-rich nuclei, one concludes that --
apart from the few closed-shell, waiting-point nuclei --  the statistical error on the position of the dripline is on the order of 15 to 20 nucleons
\cite{DFTBayes}.

The systematic uncertainty on masses and separation energies  can be estimated by comparing predictions of different DFT frameworks and different EDF parametrizations \cite{Erl12a,Afanasjev15,Wang2015,Xia17}. The resulting  error on the dripline position is comparable to the statistical uncertainty.   Such a strategy  has been employed in Ref.~\citep{Martin:2016} to estimate mass-related systematic uncertainties on $r$-process abundances by using  a well-defined set of different EDFs to create a range of mass predictions.

One can do much better by taking advantage of modern statistical tools. Current $r$-process simulations utilize predicted masses, $m_{\rm th}(Z,N,\theta)$, whenever experimental data are not available. A more powerful strategy is to estimate residuals  $\delta(Z,N)$  of Eq.~(\ref{residual}) by  training the corresponding emulator  on the set of known masses using  hierarchical Bayesian approaches, such as Gaussian processes, neural networks, or frequency-domain bootstrap \cite{BNN-SL,Bayram2014,Bayram2017,Yuan2016,Utama16,Utama17,Bertsch2017,Zhang2017,Niu2018}. The unknown masses (or other theoretical quantities needed such as $\beta$-decay rates \cite{Costiris2009}) are then obtained from Eq.~\ref{residual} by combing the theoretical mass prediction and estimated residual. Of course, the quality of a theoretical mass model is important when making such extrapolations. It is essential that the model 
\begin{itemize}
\item is global (i.e., can be applied throughout the nuclear chart); 
\item is capable of reproducing known global nuclear properties (such as the emergence of shell structure);   
\item is based on an effective nuclear interaction, which makes the extrapolation in isospin and mass number meaningful. 
\end{itemize}
In this respect, nuclear DFT applying  quantified EDFs is the method of choice. It is worth noting that by developing reliable emulators for $\delta(Z,N)$, which take into account correlations between masses of different nuclei, one can significantly refine  mass predictions and estimate uncertainties on predicted values \cite{Utama17,Zhang2017,Niu2018}. 

Naturally, by studying the surface of residuals $\delta(Z,N)$  one can learn a great deal about the deficiencies of the model itself. This information is crucial for developing  higher fidelity models. Meanwhile,
there are other ways of reducing the calculated mass uncertainty. For instance, it may be possible to decrease the 
mass residuals locally by fine-tuning model parameters to selected regional data. Such a strategy has been successfully employed in heavy nuclei \cite{Shi2014}. 
More systematically, new EDFs for $r$-process studies can be obtained  by assigning higher weights to observables in neutron-rich regions during the optimization process \cite{Rei10}.
In this respect, experimental masses measured at RIB facilities will greatly add to the data set
that can be used in such fits, however, see also Ref.~\cite{DFTBayes}.

A significant challenge for nuclear theory is its predictive power for odd-$A$ and odd-odd nuclei as the currently employed EDFs do not have the necessary spectroscopic quality \cite{Bonneau2007,UNEDF2,Kortelainen08,Schunck2010,Tarpanov2014,Afanasjev2015,Baldo2015}. Moreover, the odd-even staggering of binding energies is only roughly reproduced by the current pairing functionals  \cite{Bertsch2009,Bertulani09,Robledo2012}. Clearly, novel pairing functionals are called for to describe pairing in neutron-rich systems \cite{Yamagami2009,Yamagami2012,Pastore2013,Changizi,Hove13}. While the pairing EDF optimization  often employs data on odd-even binding energy  differences, it has recently been pointed out  \cite{Hinohara16} that the pairing-rotational moments of inertia, extracted from binding energy differences of even-even nuclei (hence free from ambiguities attributed to odd-mass systems), are excellent pairing indicators.

\subsubsection{New approaches to $\beta$-decay}\label{sec:bdecayth}
Modern large-scale calculations of $\beta$ decay based on the proton-neutron QRPA begin with the work of Ref.~\cite{Moller2003} within the macroscopic-microscopic model. Other approaches used include the extended Thomas-Fermi plus Strutinsky integral method \cite{Borzov2000} and nuclear DFT with Fayans  \cite{Borzov2003}, Skyrme \cite{Engel1999,Sarriguren2017}, Gogny \cite{Martini2014}, and covariant \cite{Niksic2005,Niu2013}  EDFs.  In Ref.~\cite{Marketin2016}, global $\beta$-decay rates,  including  first-forbidden transitions, have been carried out in the covariant DFT+QRPA approach, with the approximations that all nuclei are spherical and that odd-$A$ and odd-odd nuclei can be treated as even-even systems but with the expectation value of the particle number operator constrained to an odd number of protons and/or neutrons. 

Recently, Ref.~\cite{Mustonen2016} employed a newly developed finite-amplitude method (FAM) for solving proton-neutron QRPA equations \cite{Mustonen2014} for  almost all neutron-rich even-even spherical and deformed nuclei in Skyrme DFT. They used experimental $\beta$-decay rates and energies of Gamow-Teller and spin-dipole resonances  to optimize  previously unconstrained parameters in the charge-changing time-odd part of the functional that
have no effect on the ground-state properties of even-even systems.  More recently, the FAM+DFT calculations were extended \cite{Shafer2016} to odd-even and odd-odd nuclei in the ``equal filling" approximation, which includes some of the polarization of the even-even nuclear core by the valence nucleon(s) \cite{Schunck2010}. They optimized Skyrme parameters locally in the rare-earth and $A=80$ region and investigated the consequences of newly calculated rates for $r$-process simulations. 

\begin{figure}[!htb]
	\centering
\includegraphics[width=0.5\textwidth]{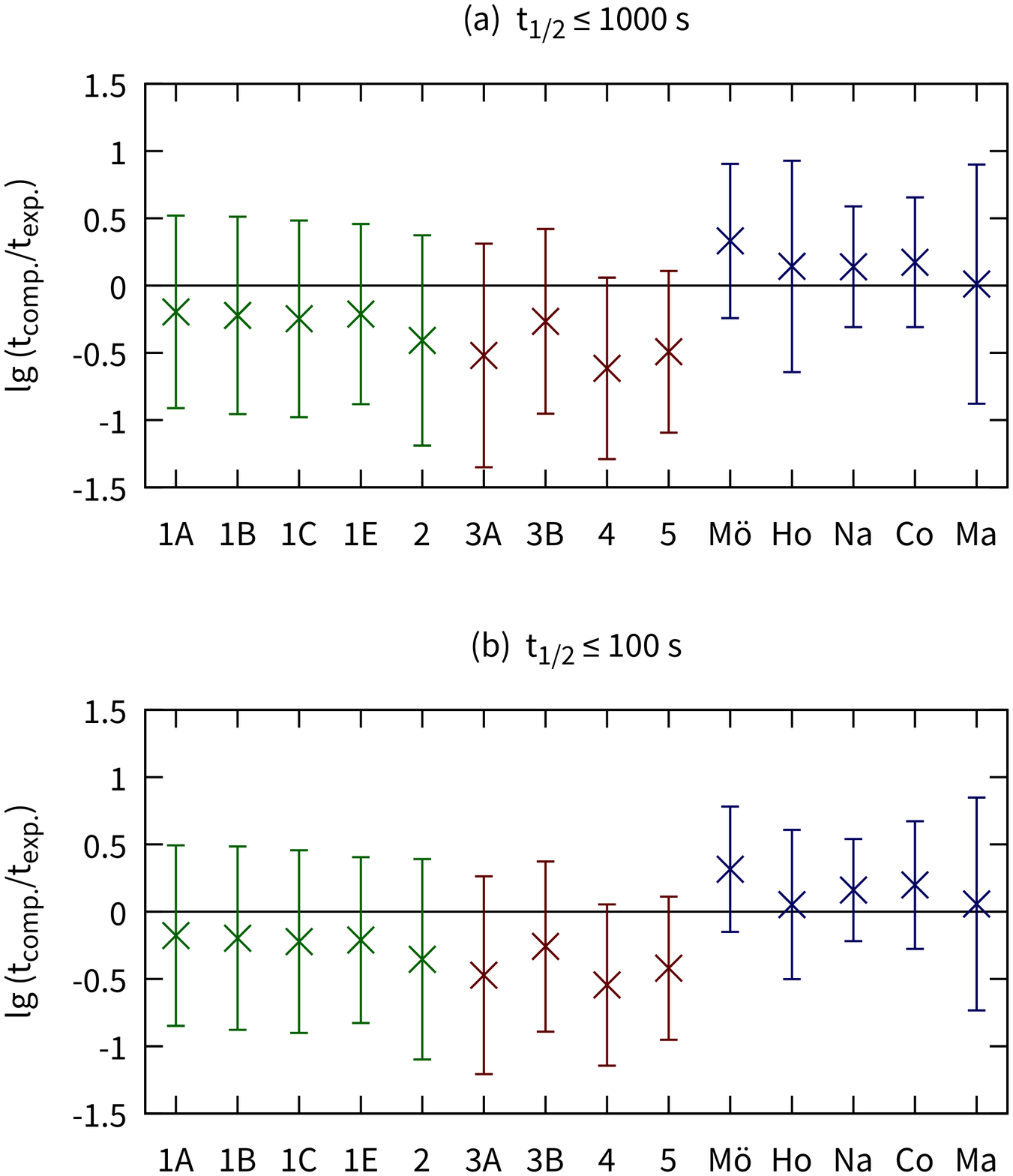}%
\includegraphics[width=0.5\textwidth]{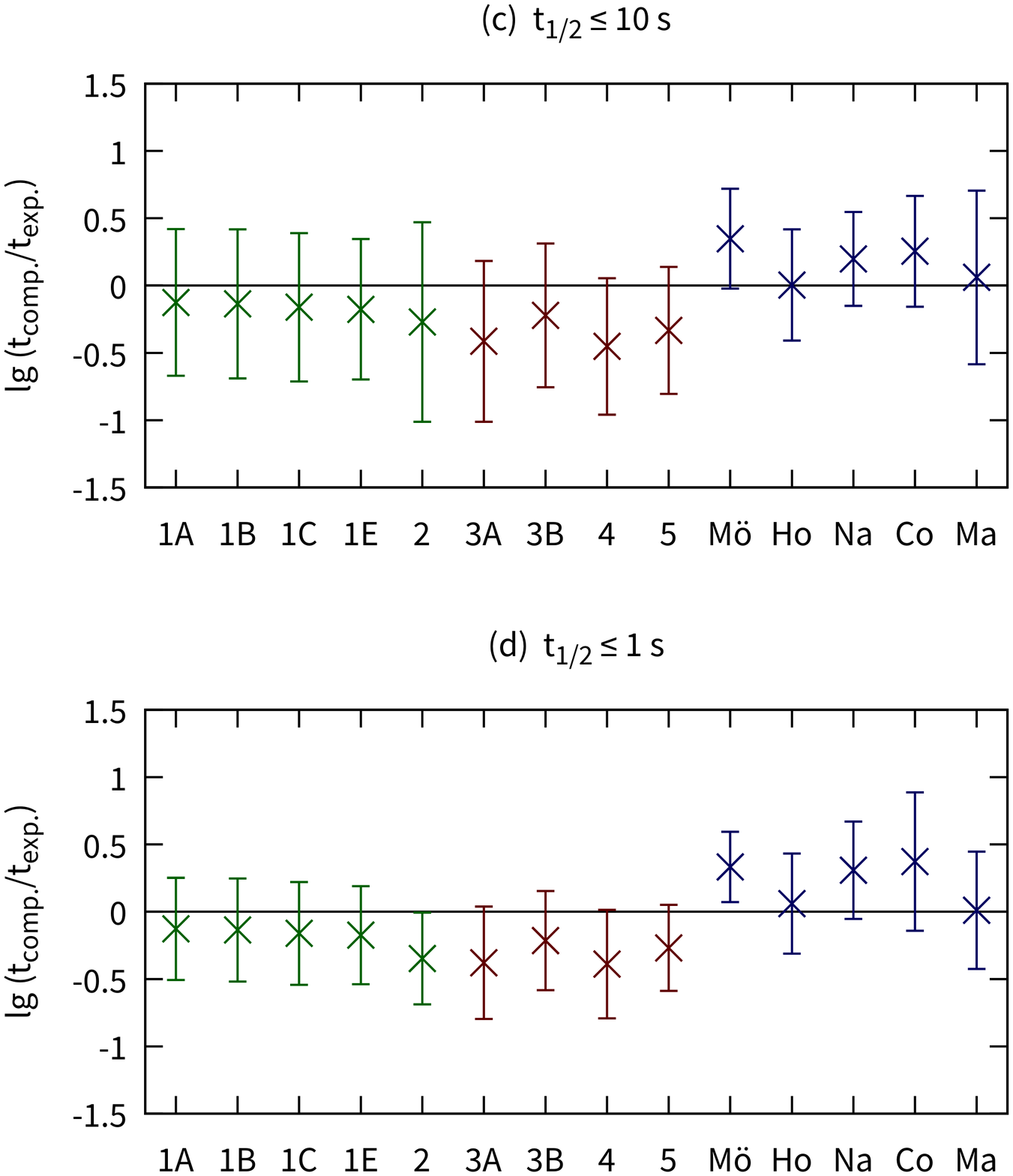}%
\caption{Comparison of the mean and standard deviation of the $\lg t$ values for several EDF models of Ref.~\cite{Mustonen2016}.  The results from prior work are contained in \cite{Moller2003} (M\"o), \cite{Homma1996} (Ho), \cite{Nakata1997} (Na), \cite{Costiris2009} (Co), and \cite{Marketin2016} (Ma).  Only even-even isotopes are considered. (From Ref.~\cite{Mustonen2016}.)}
\label{fig:decay-theory}
\end{figure}
The $\beta$-decay half-lives calculated in nuclear DFT \cite{Marketin2016,Mustonen2016,Shafer2016} are fairly similar to those obtained in other global calculations. Figure \ref{fig:decay-theory} from Ref.~\cite{Mustonen2016} compares  predictions  of half-lives grouped by half-life range for a number of different models: the first nine from Ref.~\cite{Mustonen2016} with different methods for fitting the parameters of the time-odd Skyrme EDF, and the remaining ones from prior work.  All these computations yield predictions of similar quality. Some improvements are expected by refining EDFs used in $\beta$-decay calculations, especially in the spin-isospin sector. It is to be noted, for instance, that  tensor terms were neglected in the calculations of Refs.~\cite{Mustonen2016,Shafer2016}. Another option will be to include more correlations than the ones included in the QRPA.  Although some beyond-QRPA schemes have been proposed \cite{Severyukhin2014,Konieczka2016}, they are restricted thus far to spherical nuclei.  The next few years should see the development of such schemes and their application to decays across the isotopic chart.

\subsubsection{Improved predictions of $\beta$-delayed neutron emission}

A  review of models used to describe $\beta$-delayed neutron emission can be found in the Appendix 2 of Ref.~\cite{bdelayednreport}.
The current leading global models based on QRPA and Hauser-Feshbach approach \cite{Marketin2016,Mumpower2016b} are discussed in Sec.~\ref{sec:bnbranchings}
around Fig.~\ref{fig:bn-comp}.

As indicated in Sec.~\ref{sec:bdecayth},  global  models of   $\beta$- decay, based on microscopic DFT+QRPA and EDFs that have been optimized to charge-exchange processes, are now becoming available. Importantly, some of these models include the contributions of the forbidden transitions on equal footing with allowed GT transitions.
It is anticipated that those modern  DFT+QRPA frameworks, such as 
the FAM+DFT approach~\cite{Shafer2016}, will be used to globally predict  half-lives and P$_{xn}$-values of neutron-rich nuclei.

\subsubsection{Improved predictions of capture rates}
 
The statistical Hauser-Feshbach theory \cite{Hauser52} is the commonly used tool to generate theoretical reaction rates for the $r$-process. This theoretical framework is suitable for the description of reactions involving the decay of nuclei excited at a sufficiently high energy to contain a large number of levels per MeV. When this condition is met, an energy-averaged reaction cross section in the region of highly overlapping resonances is meaningful. The main source of physics-related uncertainty in the calculated reaction rates is the modeling of the formation and decay of the excited compound nucleus through the respective transmission coefficients. Such calculations use nuclear level densities to describe the excitation of each compound nucleus, optical potentials to describe the emission or absorption of particles, and $\gamma$-ray strength functions to describe the emission of $\gamma$-rays. In the Hauser-Feshbach picture all energetically possible reaction channels are in statistical competition with a probability evaluated by dividing over the sum of transmission coefficients for all channels.
In the context of $r$-process nucleosynthesis ($\alpha$,n) and (n,$\gamma$) reactions are of particular interest. 

Neutron capture rates are sensitive to the parameters that describe the formation of the compound system and its decay by $\gamma$-ray emission. 
The calculated reaction rates strongly depend on the nuclear level density and  $\gamma$-ray strength, and much less on the choice of the neutron-nucleus optical potential \cite{Gor11,Xu2014}. 
The Hauser-Feshbach approximation is valid in the limit of high density of resonances in the compound system. If this condition is not met, 
a  procedure  based on the generation of statistical resonances can be used \cite{Rochman17}. This  technique  is more realistic than Hauser-Feshbach approach for  neutron-rich nuclei or at  low energies. 

The nuclear level density is often approximated through the back-shifted Fermi-gas formalism \cite{Rauscher1997}. Microscopically, several approaches have been proposed that  relate the level density to the actual Hamiltonian, or density functional. Within nuclear DFT,  the  combinatorial model \cite{Demetriou2001,Gor08,HilGir12}  predicts the experimental low-$\ell$  neutron resonance spacings and provides a reliable extrapolation at low energies. 
Another DFT-based approach to nuclear level density is
finite-temperature  DFT \cite{Egido2000}, see Ref.~\cite{Banerjee17} for recent applications.
The above approaches are of  particular interest if one's goal is  to provide the global nuclear structure input for $r$-process simulations based on one consistent framework (here: DFT). 

Hamiltonian-based approaches include the schematic pairing model \cite{Hung2017} and the finite-temperature shell model quantum Monte Carlo
 approach \cite{Alhassid2015a,Alhassid1999,Alhassid2007,Mocelj2007,Ozen15,Alhassid2015}, recently enhanced to circumvent the odd-particle sign problem. This method has been  applied to deformed heavy rare-earth nuclei and provides good agreement with experimental data obtained by various methods, including level counting at low energies, charged particle spectra and Oslo method data at intermediate energies, neutron and proton resonance data, and Ericson's fluctuation analysis at higher excitation energies.

(n,$\gamma$) rates in the $r$-process
strongly depend on the photon de-excitation
probability, which is related to the
$\gamma$-ray strength function. Here, large-scale DFT+QRPA calculations of
the E1 strength function for the $r$-process were carried out   with Skyrme \cite{GorKhan04,Terasaki2006,Matsuo15},  Gogny  \cite{Martini16}, and covariant \cite{Daoutidis2012} EDFs. In addition to E1, a  low-energy enhancement of the radiative M1 strength function is also anticipated \cite{Loens2012,Sieja17} in the large-scale shell model. So far, no global DFT+QRPA calculations of both, the M1 and E1 strength have  been carried out.

\subsubsection{Advances in fission theory\label{section:FissionTheory}}

Fission impacts 
the formation of heavy elements through the recycling mechanism \cite{Bel67,Cwn1991,Pan03,Pan05,Mar07,Pan13,Arn07,Beu08,Erl12b,Gor13,Giuliani2017}. 
Information on fission rates and 
fission yield properties are thus key ingredients of $r$-process reaction network  calculations.

Unfortunately, a comprehensive, microscopic explanation of nuclear fission still eludes us due to the complexity of the process \cite{schunck2016}. This fundamental nuclear decay is an example of a quantal large-amplitude collective motion. During  fission,   the nucleus evolves in a multidimensional manifold of collective coordinates, often   through a classically forbidden region. The resulting evolution can be understood in terms of the  competition between the static structure of the collective manifold and the
stochastic dynamics involving transitions between mean fields with different intrinsic symmetries.

\begin{figure*}[htb] \centering
\includegraphics[width=0.9\textwidth]{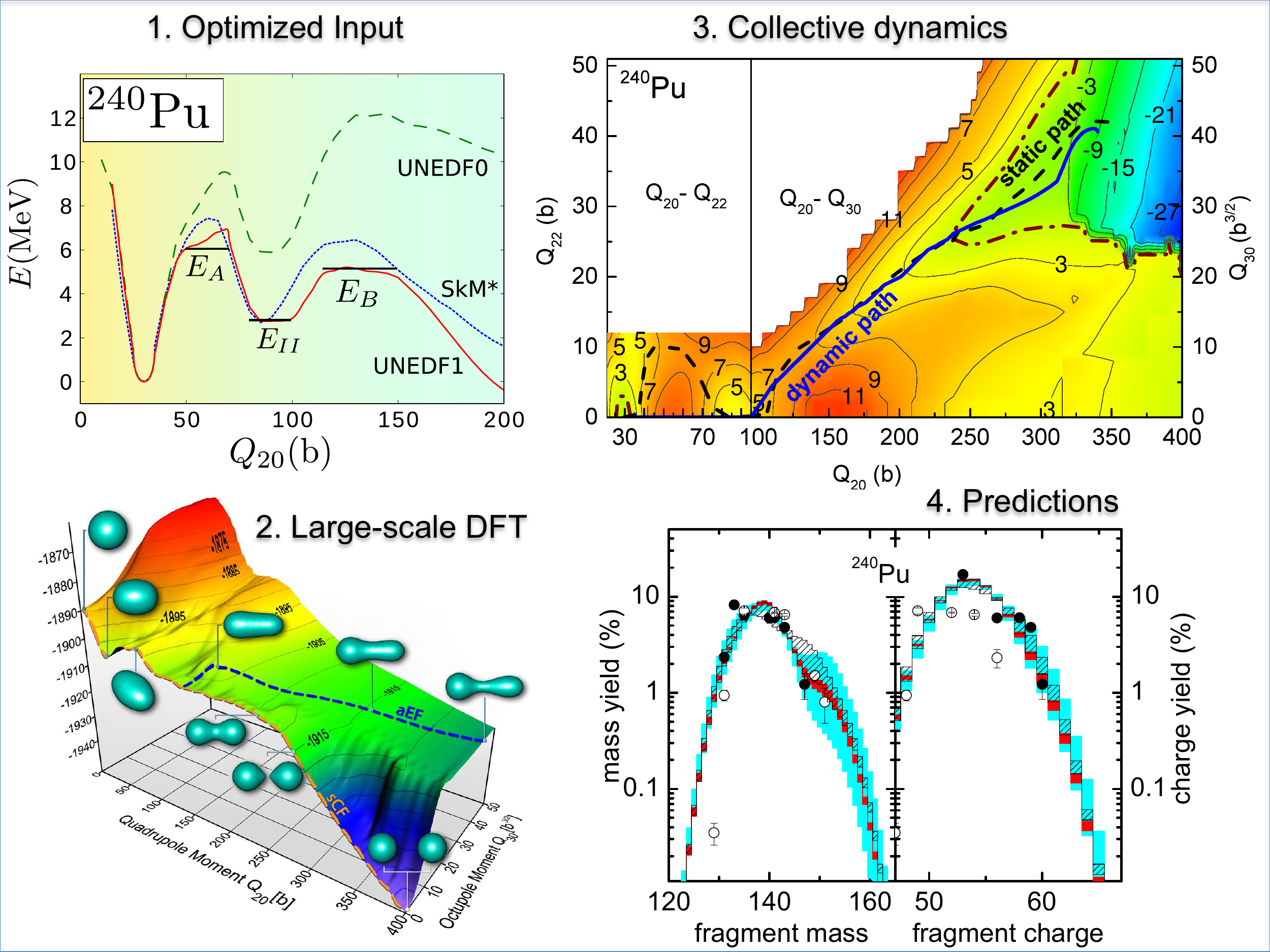}
\caption{Nuclear DFT approach to fission. Calculations are based on quantified density functionals optimized for large deformations, such as UNEDF1 \cite{UNEDF1} (upper left) and state-of-the art numerical techniques. 
Large-scale  calculations in multidimensional collective spaces are needed to produce accurate potential energy surfaces, which 
enable us to identify the multiple fission channels \cite{Sta09,Sta13} (lower left). Based on 
this information, dynamical simulations are carried  out  \cite{Sad13,Sad14} (upper right)
to calculate fission  properties 
\cite{Sad16,Sad17} (lower right).}
\label{fig:fission_strategy}
\end{figure*}

The goal of modern nuclear theory is to obtain a comprehensive understanding of the nuclear  fission process by taking advantage of state-of-the-art theoretical techniques and advanced computational tools, including the leadership-class computers \cite{exascaleage}.
Figure~\ref{fig:fission_strategy} presents key elements of the microscopic approach to fission within nuclear DFT. The  quality of a DFT calculation  relies an underlying energy density functional. 
A significant effort has been devoted to
develop validated energy functionals that produce correct physics at large 
shape elongations \cite{Ton85,Gor07,Nik11,UNEDF1},  advance numerical techniques and tools that would facilitate constrained DFT calculations, and benchmark theoretical models of fission \cite{Ber15}.

A starting point in many approaches to fission is the capability to compute accurate 
multidimensional potential energy surfaces (PES), and use them to predict 
observables such as fission half-lives or fragment distributions. 
The methodology to compute multidimensional PESs and corresponding collective mass (inertia) tensor is in place. It allows us to characterize competing fission pathways and compute the collective action \cite{Bar11} needed to predict half-lives and properties of fission fragments. 
It is worth noting that calculations of
self-consistent PESs in many-dimensional collective spaces are computer intensive; hence, massive parallel computing platforms  must be used.
In practice, one considers the nuclear collective coordinates associated with shape (elongation, triaxiality, mass asymmetry, necking) and pairing (proton and neutron pairing gaps).
The collective inertia (or mass) tensor can be obtained
from the self-consistent densities by employing the ATDHFB approximation \cite{Bar11,Giuliani2017}.
Since static
fission barriers are often both high and wide, at low energies fission  lifetimes
can obtained semiclassically by minimizing the  action
integral in the collective space \cite{Bra72,Bar81}.
To evaluate the barrier penetration probability, or a fission half-life, one has to  integrate the collective action along the optimum path.

It is important to realize that predicted fission pathways strongly depend on the choice of the collective inertia \cite{Sad13,Sad14,Giu14,Zha15,Zha16,Tao17,Giuliani2017,Rodrıguez2017}. In particular, it has been realized that pairing correlations can dramatically
alter fission trajectories. For instance,  in $^{240}$Pu pairing correlations  basically restore the axial symmetry along the dynamic fission trajectory \cite{Sad14}. This result indicates that, in the dynamical description of nuclear fission, pairing correlations should be considered  on the same footing as those associated with shape degrees of freedom. In some cases, the dynamical coupling between shape- and pairing degrees of freedom can lead to a dramatic departure from the static picture; hence,  the very notion of fission barrier, typically extracted from a saddle point of a static PES,  is very limited.

\begin{figure*}[htb] \centering
\includegraphics[width=0.8\textwidth]{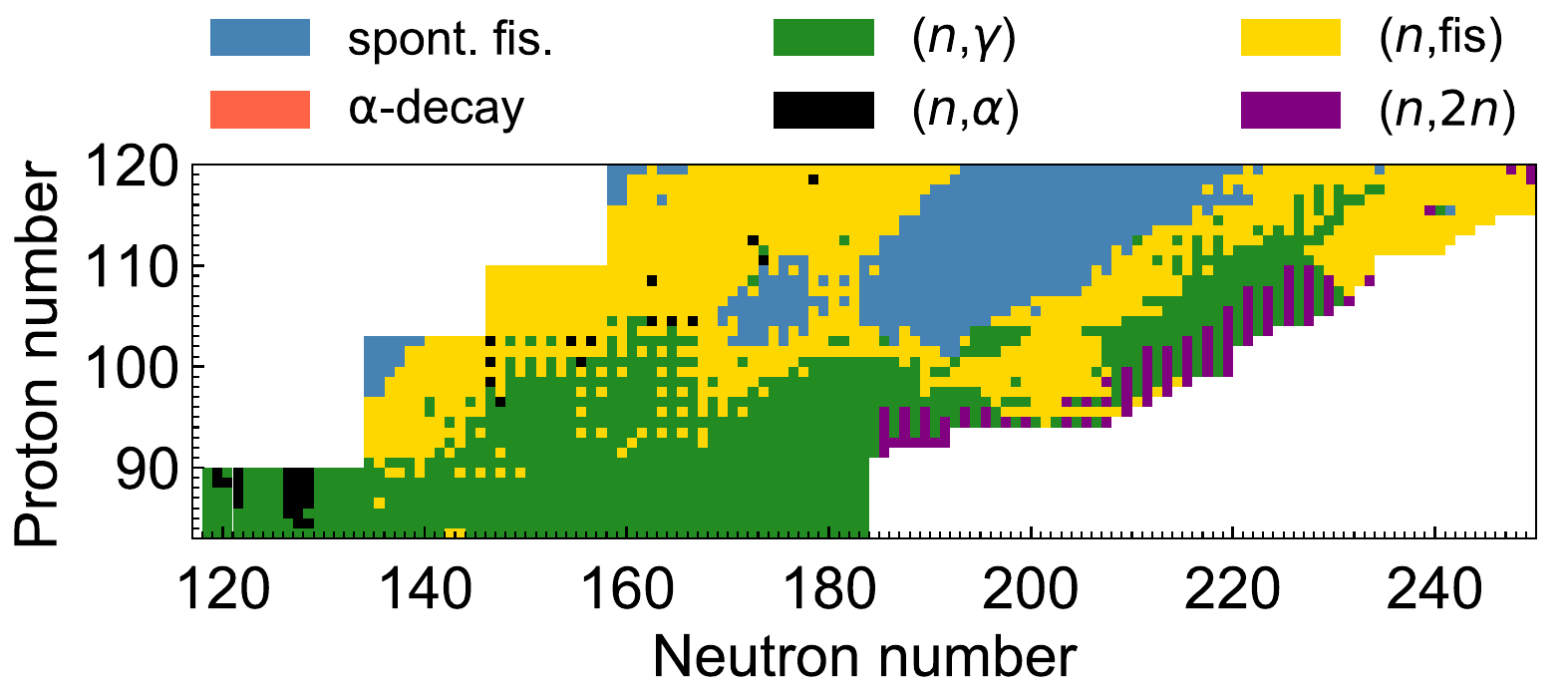}
\caption{Dominating decay channels predicted in DFT calculations of Ref.~\cite{Giuliani2017} for typical conditions of $r$-process in neutron star mergers ($T = 0.9$\,GK, $n_n = 10\times 10^{28}$cm$^{-3}$) \cite{Mendoza:2015}: spontaneous fission,
$\alpha$-decay, neutron capture, neutron-induced $\alpha$ emission, neutron-induced fission and two-neutron emission.}
\label{channels}
\end{figure*}

The fissioning $r$-process nuclei are produced by neutron capture  at some excitation energy. Therefore, in reaction network simulations a multitude of possible decays of heavy nuclei must be considered.
Figure~\ref{channels} shows the  dominating decay channel of heavy and superheavy nuclei predicted in DFT  calculations of Ref.~\cite{Giuliani2017} for typical conditions during the $r$-process in neutron star mergers.
On can see that -- according to this model -- $r$-process nucleosynthesis of nuclei with $N > 184$ is going to be strongly hindered due to the dominance of fission channels over neutron capture. Moreover, for transfermium nuclei with $N < 184$, neutron-induced fission is expected to dominate.

The neutron-induced fission rates are estimated in Ref.~\cite{Giuliani2017} within
the Hauser-Feshbach theory.
A more microscopic treatment of neutron induced fission can be carried out  using  finite-temperature  DFT \cite{Egido2000,Pei2009,McD14}, see recent Refs.~\cite{Schunck2015,Zhu16}.

The characterization of fission fragments poses additional  challenges as it involves  many-dimensional tunneling followed by a dissipative motion  from the outer turning points to scission where the nucleus splits.
In principle, time-dependent DFT (TDDFT) methods should be able to describe this latter phase of  
fission process \cite{Simenel14,Scamps15,Reg16,Bul16}. Unfortunately, each such calculation simulates only one possible  fission event: reconstructing the entire distribution of fission fragment yields,  based on different initial conditions outside the outer turning point, can become 
prohibitively expensive especially when pairing correlations are considered.

The situation becomes more complicated for induced fission from  excited states, where pairing is quenched and  dynamics becomes strongly dissipative and non-adiabatic \cite{Norenberg1983}. 
 In this regime, stochastic transport theories have been successfully applied  to describe the energy transfer between the collective and intrinsic degrees of freedom of the
fissioning nucleus \cite{Abe96,Sierk17}. Of particular interest are  dynamical approaches based on the Langevin equation and its derivatives; such methods  have been remarkably successful in reproducing properties of  fission fragments \cite{Arimoto14,Randrup15,Denisov17,Sierk17,Sad16,Sad17}. In particular, the width of the fission yield distributions 
is primarily determined by near-scission fluctuations caused by the random force  \cite{Arimoto14,Sad17}.

The rupture of the neck is a rapid and violent process. As demonstrated in Refs.~\cite{Younes11,Schunck2014,Sad17},  prefragments are strongly entangled near the scission  point and a full quantum mechanical treatment is needed to describe  the split. In this respect,  the stochastic mean-field technique coupled to TDDFT \cite{Lacroix14} offers interesting opportunities for microscopic studies \cite{Tanimura17} that go beyond the Langevin approach.

\section{Facilities}
\label{sec_facilities}
Radioactive beam facilities have come a long way since the first half-lives of $r$-process nuclei have been measured in 1986 - the $^{80}$Zn waiting point at OSIRIS in Studsvik \cite{Ekstrom1986}, and the $^{130}$Cd waiting point at ISOLDE/CERN \cite{kratz86}. Over the last decades a large number of facilities around the world have made significant progress in developing capabilities in beam production, beam purification, and experimental techniques. Experiments exploiting these new capabilities have generated an increasing body of data on $r$-process nuclei. Despite this progress, the majority of $r$-process nuclei have been out of reach, and the vast number of successful measurements have been limited to 
decay studies, primarily half-lives, but also branchings for $\beta$-delayed neutron emission (see Fig.~\ref{fig:exp_reach}) in the mass regions around A$\approx80$ and $A\approx130$, and to some extent in between. Examples include measurements at ISOLDE \cite{Hannawald2000}, NSCL \cite{Montes2006,Hosmer2010}, TRIUMF \cite{dunlop16}, GSI \cite{caballero16}, and more recently RIKEN/RIBF \cite{nishimura11,xu14,lorusso15,wu17}. Mass measurements have only recently reached the $r$-process in a few cases, for example with Penning Trap measurements at Jyv\"askyl\"a \cite{Hakala2012}, TRIUMF \cite{Simon2012}, ISOLDE \cite{Breitenfeldt2010}, and ANL \cite{VanSchelt2013}. Measurements to constrain neutron capture rates have been performed using neutron transfer reactions at ORNL on radioactive beams of $^{130}$Sn \citep{Kozub2012} and $^{132}$Sn \citep{Jones2010}, and more recently using the new $\beta$-Oslo technique at NSCL \cite{Spyrou2014,Liddick2016}. These experiments and techniques are discussed in more detail in Section \ref{sec_exp}. 

We are now at the threshold of major new capabilities for experimental $r$-process studies with the emergence of a new generation of very powerful radioactive beam facilities. These facilities promise to bring the majority of $r$-process nuclei up to $A\approx200$ within experimental reach. The broadest reach is provided by fragmentation-type radioactive beam facilities, where an intense heavy ion beam is fragmented on a thin production target at energies in the range of 50-1000 MeV/u. All the produced radioactive fragments leave the target with somewhat lower energies per nucleon. A fragment separator is therefore essential to limit the radioactive beam to the species of interest. The most powerful new facility of this type will be FRIB in the US (400 kW beam power). FRIB is currently under construction at Michigan State University and is expected to be completed in 2022. Complementary to FRIB will be the planned FAIR facility in Darmstadt, Germany, with somewhat less beam power (50 kW) but higher energy and a beam time structure that is well-suited for an extensive storage ring program. RIKEN/RIBF with 10 kW beam power is the first of the new generation of facilities to come online and has in its first 10 years of operation produced spectacular results related to $r$-process nuclei. The RISP project at the planned RAON facility in Korea will also have a fragmentation capability. 

At ISOL facilities, radioactive isotopes far from stability are produced via fragmentation, spallation and fission induced by impinging intense light particle beams onto a solid target or a gas cell. While for thick solid target facilities the resulting radioactive beam intensities are strongly dependent on the chemical properties of the produced radioactive isotopes, beam intensities in facilities employing gas cells are chemically insensitive. 
Radioactive beam intensities at ISOL facilities can be higher compared to fragmentation facilities in favorable cases, especially for low-energy high-quality beams that are needed for reaction studies. TRIUMF/ISAC, CERN/ISOLDE, and IGISOL at Jyv\"askyl\"a continue to be the leading facilities in this area. The new RISP/RAON facility will also have an ISOL capability. A new generation ISOL facility is under discussion in Europe as EURISOL \citep{cornell2009} with the goal to exceed radioactive beam intensities of facilities existing or under construction by at least a factor of 100. As an organizational step towards this goal, the EURISOL Distributed Facility (EURISOL-DF) collaboration has been formed by various European ISOL facilities.   

Gas stopping and reacceleration schemes are being developed to use beam production via fragmentation to produce ISOL-quality low energy radioactive beams for reaction studies that are also relevant for the $r$-process. This is currently pursued at the NSCL ReA3 facility, which will later become part of FRIB. The goal is to produce low energy beams that are difficult or impossible to produce with the ISOL technique. 

A third type of facility uses fission, either spontaneous or induced by photons, electrons, or light ion beams, to produce neutron-rich radioactive nuclei. The nuclei that can be produced are limited to what is included in actinide fission fragment distributions but intense beams of neutron-rich nuclei near or in the $r$-process can be produced in two localized mass regions around $A=90-110$ and $A=130-140$. The CARIBU facility at ANL uses an intense $^{252}$Cf source to successfully produce neutron-rich $r$-process nuclei. Next generation facilities under construction include ARIEL at TRIUMF/ISAC (using photofission) and SPIRAL2 at GANIL. These facilities are expected to have significant reach into the $r$-process. 

With these developments we can expect that in the coming 10-20 years $r$-process research will change dramatically. Large amounts of experimental nuclear data with well defined uncertainties will be available, and improved nuclear theory will be able to predict much better the nuclear data that remain out of experimental reach. 


\subsection{Fragmentation facilities}
A number of facilities are using in-flight fragmentation to produce radioactive beams. A heavy ion beam is accelerated to energies in the 50 MeV/u to 1000 MeV/u range and impinges on a relatively thin target. Fragmentation of the beam particles produces a broad range of stable and radioactive nuclides that emerge from the target with energies per nucleon of the same order of magnitude as the incident beam. The desired fragment is then selected with a fragment separator, typically employing magnetic dipoles. The advantages of this technique are flexibility to quickly select a broad range of beams simply by adjusting the fragment separator, radioactive beam production that is independent of the chemistry of the respective element, fast transport (typically 100s of nano seconds) of the radioactive beam to the experiment that minimizes decay losses, and relatively high beam energies, which allow the use of thick secondary targets, and enable particle-by-particle identification of beam particles with various detector systems. The latter capability can be used to run with impure beams to measure large numbers of nuclides simultaneously, and to achieve nearly 100\% selectivity enabling measurements, for example of decay half-lives, with very low beam intensities pushing measurement capabilities to the most exotic nuclides. Disadvantages are the poor beam quality that often requires tracking of beam particles for reaction studies, losses from stopping or slowing down the beam for measurements of masses or decay properties, and the difficulty to reduce the beam energy to astrophysical energies for reaction measurements - this can be overcome by employing a gas stopping and reacceleration scheme. The fast transport to the experiment also means that nuclei arriving at the experiment can be in isomeric excited states, which can cause difficulties in some experiments (while in others it may be used as an advantage). 

\subsubsection{NSCL and FRIB}
The National Superconducting Cyclotron Laboratory (NSCL) is a world class international user’s facility dedicated to the production and study of radioactive isotopes.  The research undertaken at the facility falls into a few broad categories including basic nuclear science, nuclear astrophysics, fundamental symmetries, accelerator physics, and the application of isotopes for societal benefit.  The accelerator is capable of ionizing any chemical element and delivering it directly to an experimental apparatus with an energy up to 200 MeV/nucleon with a beam power up to 1 kW.  These high-energy stable ion beams are fragmented through collisions with a target to produce a wide variety of radioactive nuclei that are subsequently separated out in flight by a high-acceptance fragment separator, the A1900, and delivered to various experimental facilities.  The secondary, radioactive nuclei are delivered to experimental facilities either at high energies (fast beams), thermal energies from a gas stopping system (stopped beams), or reaccelerated to near the Coulomb barrier. To date, more than 1000 isotopes have been produced and used at the NSCL.

The Facility for Rare Isotope Beams (FRIB) will be one of the next-generation radioactive ion beam (RIB) facilities that will allow access to the uncharted territories of the chart of nuclides. Most of the expansion in terms of the reach will be on the neutron-rich side towards the location of the $r$-process path. Hence, with FRIB coming on-line, a large array of atomic masses critical for our understanding of the $r$-process will be within our reach (see Fig.~\ref{fig:exp_reach}). The baseline design of FRIB is based on a 200 MeV/u superconducting linac with a delivered beam power of up to 400 kW for beams up to uranium.  Construction of FRIB conventional facilities began in the spring of 2014.  Project completion is expected in 2022 with management toward early completion in 2021.  The predicted rates for FRIB are provided at https://groups.nscl.msu.edu/frib/rates/fribrates.html and are expected to exceed NSCL capabilities by at least three orders of magnitude.

Much of the experimental equipment for FRIB $r$-process experiments exists already or is under development, and will be used for experiments at NSCL until FRIB turns on. A state of the art decay station to measure $\beta$-decay properties using the implant-correlation method (see section \ref{sec:implant-correlation}) is currently in the planning stage and will be of particular importance to fully exploit the capabilities of FRIB for $r$-process studies. 

\begin{figure*}[htb] \centering
\includegraphics[width=0.8\textwidth]{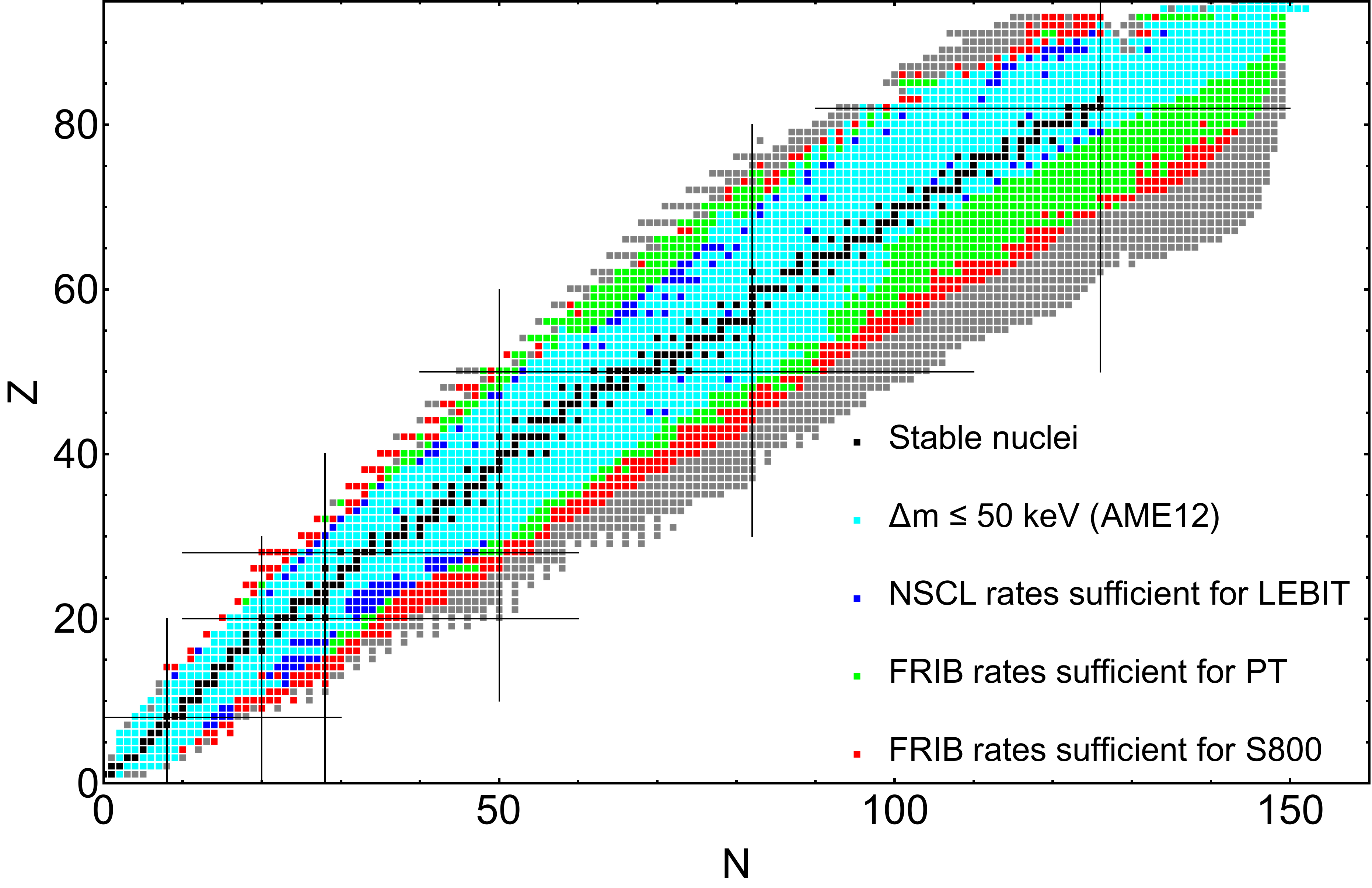}
\caption{Predicted reach of various mass measurement techniques at the NSCL and FRIB, based on LISE calculations. Shown is the reach for Penning Trap (PT) measurements (green) and the time-of-flight technique using the S800 or, for the most exotic isotopes, the HRS spectrometer (red).}
\label{fig:FRIBmass}
\end{figure*}

$r$-process masses will be measured using two complementary techniques. Time-of-flight mass measurements (see section \ref{sec:exptechniques:tof}) with the existing S800 spectrometer and, in the future, the new High Rigidity Spectrometer HRS currently under development, will cover broad regions of nuclides simultaneously with essentially no limit on short half-lives, and will therefore enable a complete mapping of the mass surface into the $r$-process relatively quickly but with accuracies that for some key nuclides will not be sufficient for $r$-process studies. Complementary to this technique will be high precision mass measurements using Penning-trap mass spectrometry (see section \ref{sec:exptechniques:trap}) using the LEBIT facility \citep{Ringle2006,Bollen2007}. Upgrades of LEBIT are underway \citep{Redshaw2013}, in particular a single ion capability (SIPT) that can peform mass measurements with production rates as low as one per day. This will enable to push high precision mass measurements to more exotic $r$-process isotopes, though it is limited to somewhat longer lived isotopes compared to the time-of-flight method. The Penning Trap measurements will provide high accuracy for the most important $r$-process nuclei within the technique's reach, and will provide essential calibration data for the time-of-flight approach, which will be able to push mass measurements to a broader range of more exotic nuclei. Figure \ref{fig:FRIBmass} shows the current and future reach of these mass measurement techniques at FRIB. As it can be seen, a large number of atomic masses will be within reach with these techniques. 

Indirect determinations of neutron capture rates will be performed at FRIB using transfer reactions and the $\beta$-Oslo technique \citep{Spyrou2014,Liddick2016} (see section \ref{sec:exptechniques:ncap}). FRIB's ReA3 reaccelereated beam capability will be critical to directly measure cross sections for important ($\alpha$,n) reactions in the weak $r$-process (see section \ref{sec:exptechniques:an})

\subsubsection{RIBF}
The Radioactive Isotope (RI) Beam Factory (RIBF) \cite{MotobayashiSakurai12,Okuno12} at RIKEN Nishina Center is regarded as the first among a new generation of RI-beam facilities and has been in operation since 2007. Heavy ion beams ranging from ${\rm ^{2}H}$ to ${\rm ^{238}U}$ are accelerated up to 345~MeV/u in the accelerator complex comprised of four ring cyclotrons, the SRC (Superconducting Ring Cyclotron) with K=2600~MeV, the IRC (Intermediate-stage Ring Cyclotron) with K=980~MeV, the fRC (fixed-frequency Ring Cyclotron) with L=570~MeV, and the RRC (RIKEN Ring Cyclotron) with L=540~MeV, together with three injectors. As of summer 2016, intensities of heavy-ion beams reach 49~pnA for ${\rm ^{238}U}$, 530~pnA for ${\rm ^{48}Ca}$, and 1000~pnA for light ions including ${\rm ^{18}O}$. The world-record intensity RI-beams are produced and separated with BigRIPS\cite{Kubo2012} through projectile fragmentation and in-flight fission reactions of the primary beams. $r$-process nuclei are produced mainly via the in-flight fission reaction of ${\rm ^{238}U}$.

There are on-going experimental activities related to $r$-process nucleosynthesis using RI-beams from BigRIPS: Half-life measurements with the EURICA setup \cite{nishimura11,xu14,lorusso15,wu17} have been most successful (see Section~\ref{sec:exptechnique:decay} for details). Decay activities have been extended to measurement of $\beta$-delayed neutron emitters with the BRIKEN setup in 2016 \cite{Tarifeno2017} (see Sec.~\ref{sec:exp:3He}). In the  future, masses in the same region of the nuclear chart will be measured with newly-constructed devices: the Rare RI Ring \cite{Ozawa12,Yamaguchi13} (see Section~\ref{sec:MR_TOF}) and the MRTOF \cite{schury13}. Both devices have capabilities to measure masses of nuclei whose half-lives are far shorter than 100~ms. 
     
  A new attempt to determine barrier heights and fragment distributions in fissions of heavy unstable nuclei is being initiated with the SAMURAI spectrometer. The experiment will provide novel and crucial information related to fission recycling (see also Section~\ref{section:FissionExperiment}).

\subsubsection{Chinese facilities}
HIRFL (Heavy Ion Research Facility in Lanzhou) is a major  user facility for nuclear physics research in China \cite{xia2002,zhou2016}. HIRFL consists of ECR (Electron Cyclotron Resonance) ion sources, a Sector Focus Cyclotron (SFC,$k$=69), a Separated Sector Cyclotron (SSC,$k$=450), and a Cooler Storage Ring (CSR). The CSR is a double cooler-storage-ring system including a main ring (CSRm) and an experimental ring (CSRe), coupled together by a recoil separator (RIBLL2) and a beam transport line. HIRFL can provide beams of all stable isotopes from hydrogen through uranium with a wide energy range. The maximum energies available at HIRFL for proton, carbon and uranium beams are 2.8 GeV, 1000 MeV/u and 100 MeV/u, respectively. Radioactive ion beams (RIBs) are produced  via fragmentation reactions or direct reactions using two recoil separators (RIBLL1 and RIBLL2), separately, for measurements of masses, decay properties, and critical nuclear reactions for astrophysics. 

BRIF (Beijing Radioactive Isotope Facility) is a newly constructed ISOL facility based on an existing 15 MV tandem accelerator at China Institute of Atomic Energy \cite{liu2011}. A proton beam with an energy of 100 MeV and an intensity up to 100 $\mu$A impinges an ISOL target such as UCx, and neutron-rich isotopes are produced by fission. The desired isotope with one unit of positive charge is selected by a separator with a mass resolution of 20,000. It is converted into a negative ion to be post-accelerated by the tandem accelerator. After the tandem, a super-conducting LINAC sector is being planned to further boost the beam energy by 2 MeV/q.  After the facility is commissioned, nuclear decay spectroscopy and nuclear reaction studies relevant to the $rp$-process or the $r$-process will be carried out. 

HIAF (High Intensity heavy ion Accelerator Facility)  is a new project recently approved by the Chinese central government. This new facility consists of a low energy LINAC, a synchrotron, a recoil separator, and an experimental ring, to be commissioned by 2023. For ${}^{238}$U$^{34+}$, the maximum available beam energy is 17 MeV/u out of the LINAC with an intensity of up to 30 p$\mu$A. The maximum beam energy out of the synchrotron is 0.8 GeV/u. This facility will focus on the production of super-heavy nuclei and neutron-rich heavy isotopes, using intense low energy heavy ion beams, for mass measurements, decay spectroscopy, and studies of nuclear reactions.  

Beijing ISOL (known as CARIF before 2012) is another project under review \cite{liu2011}. This project will accelerate the neutron-rich fragments produced by a reactor or an ISOL target up to a few hundred of MeV/u to produce even more neutron-rich nuclei via fragmentation reactions for $r$-process research and the synthesis of super-heavy isotopes. 

\subsubsection{FAIR}

At the International Facility for Antiproton and Ion Research (FAIR) at GSI Darmstadt/ Germany \cite{gutbrod06,kester16} radioactive ions  will be produced via fragmentation of ions as heavy as uranium. FAIR is currently under construction and the first stage is expected to be operational in 2021. FAIR is a major upgrade of the existing GSI facility. The philosophy from the very beginning was to enable fixed-target experiments as well as ring experiments with highest energies and intensities in a multi-user mode. The solution was an accelerator complex, which starts with different ion sources followed by a universal linear accelerator and a synchrotron. The heart of the GSI facility is the SIS18, a synchrotron with a magnetic rigidity of 18~Tm. The center of the first stages of the FAIR complex will be the synchrotron SIS100 (100~Tm), which will receive the beam from SIS18. Later stages of FAIR will host the synchrotron SIS300 (300~Tm). The intensity of the radioactive ion beam will be up to 10$^4$ times higher than currently achievable at GSI. The beams can be delivered to fixed-target experiments like CBM \cite{CBM11} or ring-based experiments like PANDA \cite{PANDA09}. The experiments relevant for nuclear astrophysics are either under the umbrella of NUSTAR \cite{muenzenberg10} or APPA \cite{stoehlker15} and can be fixed-target or ring-based \cite{litvinov13}.

FAIR will offer unique, unprecedented opportunities to investigate many of the astrophysically important reactions. The high yield of radioactive isotopes, even far away from the valley of stability, allows the investigation of isotopes involved in processes as exotic as the $r$ or $rp$ processes. 

The proposed R$^3$B setup \cite{R3B05}, a universal setup for kinematically complete measurements of
Reactions with Relativistic Radioactive Beams will cover experimental reaction studies with exotic nuclei far off stability. 
R$^3$B is a versatile reaction setup with high efficiency, acceptance, and resolution for reactions with high-energy radioactive beams. The setup will be located at the High Energy Cave, which follows the 
high-energy branch of the new fragment separator (Super-FRS). The experimental configuration is based on
a concept similar to the existing LAND setup at GSI introducing substantial improvements with
respect to resolution and an extended detection scheme. It comprises the additional detection efficiency of light
(target-like) recoil particles and a high-resolution fragment spectrometer. The setup is adapted to the
highest beam energies (corresponding to 20~Tm magnetic rigidity) provided by the Super-FRS
capitalizing on the highest possible transmission of secondary beams. The experimental setup is suitable
for a wide variety of scattering experiments, such as heavy-ion induced electromagnetic excitation,
knockout and breakup reactions, or light-ion (in)elastic and quasi-free scattering in inverse kinematics,
thus enabling a broad physics program with rare-isotope beams \cite{R3B05}.
Applying the Coulomb dissociation method \cite{Baur96,Bertulani10} R$^3$B contributes already now to almost every astrophysical scenario \cite{reifarth16}. In particular neutron-capture reactions \cite{reifarth14} can be investigated via the time-reversed $(\gamma,n)$ reactions contributing to the understanding of the $r$ process. With the expected increase in the production of radioactive species at FAIR, even more exotic reactions can be investigated. 

The suite of rings available already now and getting constructed over the course of the next 10 years will allow measurements of masses and different decay properties of exotic nuclei on the $r$-process path \cite{litvinov13,lestinsky16}. The high-energy storage ring (HESR) offers an interesting complement to the 1-pass R$^3$B setup mentioned above. With a magnetic rigidity of 50~Tm the HESR allows to store fully stripped uranium ions up to 5~AGeV, which covers the desired energy range for Coulomb excitation experiments. The HESR has a circumference of 574~m and features very long straight sections of about 100~m, which enable the detection of reaction products outside of the beam pipe even at very small angles (e.g. neutrons). The HESR is ideally suited for stacking over long periods, which is extremely useful when isotopes with very low production yield
are investigated. Depending on the desired reaction mechanism, the interaction zone would then be a gas jet target featuring hydrogen, helium or a high-Z gas for Coulomb breakup studies.

\subsection{ISOL facilities}
In the ISOL (Isotope Separator OnLine) technique thick targets of different materials (up to a few 100 g/cm$^{2}$) are bombarded with high-energetic proton beams of $\mu$A intensity (beam energies can be in the range of 100 - 2000 MeV/u). The proton-induced spallation of the target material produces exotic isotopes which are extracted via the ion source. The ISOL method is driven by target chemistry, so the most intense beams that can be extracted are those which have the lowest ionization potentials. These so-called surface-ionized species (e.g. alkali and earth-alkali metals) are a specialty of ISOL facilities. Other elements with high ionization potentials have quite low extraction yields. For cleaning of the beam and suppression of isobaric contaminants, additional purification steps are required. The various techniques range from the use of a neutron converter (to suppress neutron-deficient species via neutron-induced fission), element-selective laser ionization, and the use of quartz tubes or cold transfer lines etc. One limitation of the ISOL technique is the diffusion time of the radioactive species out of the target material. This restricts this method so far to isotopes with half-lives longer than $\approx$10~ms.

\subsubsection{CERN-ISOLDE}
The ISOLDE facility is located within the European Particle Physics Laboratory CERN in Geneva, Switzerland \cite{kugler00}. It was one of the earliest facilities in the world to apply the ISOL method for the production of rare isotope beams. A pulsed proton beam is nowadays delivered by CERN's Proton-Synchrotron Booster as a primary beam to two separate ISOLDE target stations. The characteristics of the currently used  proton beam are a kinetic energy of 1.4 GeV and an average intensity of up to 2~$\mu$A. More then 1200 radioactive nuclides are produced after the bombardment of various target materials. The demands of the specific experiments require different layouts to be used for target materials, geometries, and ion sources. For the specific cases concerning the $r$-process, the preferred target material is uranium-carbide (UC$_x$).
There are three different types of ion sources that are in use at ISOLDE: a surface, plasma, or laser ion source \cite{marsh13}. Whatever the ionization method, the target/ion-source unit is floated to a potential of up to 60 keV. The beams are then sent through primary mass separators, either the General Purpose Separator or the High Resolution Separator, with mass resolving powers of $m / \Delta m =$ 1000 and 5000 \cite{giles03}, respectively.

\subsubsection{TRIUMF-ISAC and ARIEL}
The Isotope Separator and Accelerator (ISAC) facility \cite{dilling14a,dilling14c} at TRIUMF, Canada’s particle accelerator centre in Vancouver, has been in operation since 1998. It provides a wide variety of intense beams of exotic nuclei by impinging a 480~MeV proton beam from the cyclotron with up to 100~$\mu$A on various target materials. After passing through the mass separator (M/$\Delta$M$\approx$1000) the radioactive beams are guided into the experimental halls.

ISAC provides the world-wide highest intensity of e.g. the 8.75-ms halo nucleus $^{11}$Li (up to 56,000 pps), $^{21}$Na (up to 1.1$\times$10$^{10}$ pps), or $^{23}$Mg (up to 2.6$\times$10$^{9}$ pps) (as measured at the low-energy ISAC yield station). The preferred target-ion source combination for the most neutron-rich $r$-process nuclei, e.g. Rb, Sr, Cd, In, Cs, Ba, is presently a uranium carbide (UC$_x$) target in combination with a surface-ion source. In order to enhance the signal-to-background ratio and suppress unwanted surface-ionized species, laser-ion sources like TRILIS or the new ion-guide laser-ion source (IG-LIS, \cite{raeder14}) are the perfect choice. The IG-LIS has proven to be a very powerful tool for low-intensity beams since it efficiently suppresses the background of surface-ionized species which can be orders of magnitude more intense. With this setup, spectroscopy of the 82~ms isotope $^{132}$Cd (N=84) could be performed with only 0.15~pps, the lowest yield ever measured at ISAC.

The ISAC-I facility uses beam energies up to 40~keV, and can re-accelerate beams for reaction studies in a first step up to 1.8~AMeV. Further acceleration using the ISAC-II super-conducting linac leads to beams of up to 16.5~AMeV for transfer reaction studies for light-mass nuclei and a few AMeV for nuclei above A=150.   

The neutron-rich astrophysical program at TRIUMF with focus on $r$-process isotopes consists presently of the TITAN Penning trap system \cite{kwiatkowski14} for mass measurements, the GRIFFIN $\gamma$-ray spectrometer and its ancillary detectors for spectroscopic studies \cite{svensson14}, and the new EMMA recoil mass spectrometer, which has been commissioned recently \cite{davids14}.

The new Advanced Rare IsotopE Laboratory (ARIEL) \cite{dilling14b} consists of an independently operating electron linac that accelerates electrons in the present design up to 35~MeV at a power of 100~kW for isotope production via photofission of uranium, as well as an additional proton beam line from the 520 MeV cyclotron. With this added driver beam capacity the amount of beamtime available at ISAC can almost be tripled within the next few years. One advantage in the use of photofission versus proton-induced spallation are the higher yields in the region around the two fission fragment peaks and the lower amounts of neutron-deficient contaminants.

ARIEL is presently under construction, and first beams from the existing ISAC production target will be sent by the end of 2019 through the new CANREB facilities, consisting of a high-resolution mass separator, a RFQ cooler and buncher, an electron beam ion source (EBIS) and a Nier spectrometer back, to be reaccelerated in the ISAC heavy-ion linac for experiments with clean beams. First beams from the photofission of uranium targets are expected by 2023.

\subsection{Gas-catcher based ISOL facilities}
Gas-catchers can be used instead of solid targets for ISOL beam production. They offer advantages in the efficient extraction of the radioactive species from the target, and a number of facilities exploit this technique. 

\subsubsection{IGISOL facility in Jyv\"askyl\"a}
The IGISOL (Ion Guide Isotope Separator On-Line) facility \cite{Moore2013} is located at the University of Jyv\"askyl\"a, Finland. IGISOL benefits from the universal and fast (sub-ms) ion-guide method to extract the reaction products \cite{Arje1985}. Thus, no separate ion source is needed and the production is not restricted to certain elements. The neutron-rich isotopes of interest are produced with proton-induced fission on $^{238}$U or $^{232}$Th employing either MCC30 or K130 cyclotrons of the JYFL Accelerator Laboratory. The reaction products are extracted out from the gas cell using a sextupole ion guide SPIG \cite{Karvonen2008}, accelerated to 30 keV, and mass-separated using a 55$^\circ$ dipole magnet ($M/\Delta M\approx 250$) before injecting them into a radio frequency cooler and buncher (RFQ) \cite{Nieminen2001}. The ion bunches from the RFQ enter the cylindrical double Penning trap mass spectrometer JYFLTRAP \cite{Eronen2012}, which consists of a purification trap employing mass-selective buffer gas cooling technique \cite{Savard1991} to select the ions of interest for high-precision mass measurements via time-of-flight ion cyclotron resonance technique \cite{Konig1995} in the second trap known as the precision trap. 

Overall, around 300 atomic masses have been measured at JYFLTRAP and most of those have been neutron-rich nuclei (see e.g. \cite{Kankainen2012, jyfltrapdatabase}). One of the advantages at JYFLTRAP is the possibility to purify the beam using a Ramsey dipolar excitations \cite{Eronen2008} by which even isomerically pure beams can be delivered for mass or post-trap spectrosocopy experiments (see e.g. \cite{Kankainen2013,Rodriguez2012}). To get an idea of the yields and possibilities for mass measurements at JYFLTRAP, a recent review on the mass measurements of fission fragments and related cross-section curves can be found from Ref.~\cite{Kankainen2012}. In addition, independent isotopic fission yields have been studied thoroughly at IGISOL \cite{Penttila2010,Penttila2014,Penttila2016}. To reach even more neutron-rich nuclei, neutron-induced fission is being explored and developed at IGISOL \cite{Gorelov2016}. 

\subsection{New production techniques}	\label{sec:newprodmech}
For a full understanding of the $r$-process it will be important to study nuclei that cannot be produced by fragmentation, fission, or spallation. Alternative production approaches will be required to produce these nuclei. 

The $r$-process that produces the heaviest elements found in nature, i.e. the third peak and the actinides, must proceed through very neutron-rich ($N>146$) nuclei that have more neutrons than $^{238}$U. This means these nuclei are out of reach for single step fragmentation or ISOL facilities.  Moreover, most theoretical studies estimate the $r$-process to cross a possible $N=184$ shell closure around $A\approx 280$ and end somewhere around $Z \approx 110$ and $A \approx 340$ \cite{Goriely2015}. Understanding the nuclear physics of the nuclei near the theoretical endpoint of the $r$-process is important for many reasons. One of them is the importance to understand $r$-process actinide production for using observed U and Th abundances as $r$-process chronometers. In this context it has been pointed out that the possible $N=184$ shell closure plays an important role \cite{schatz02}. Many $r$-process models, in particular in neutron star mergers, exhibit fission cycling, where nuclei at the endpoint of the $r$-process undergo fission, and the resulting fission fragments in the $A \approx 130-170$ mass region serve then as new seeds for a continued $r$-process. It is then critial to understand where exactly the $r$-process ends and what the resulting fission fragments are, not only to determine to which degree fission cycling occurs, but also to understand the produced abundances in the $A>130$ region, which would be affected by the relevant fission fragment distributions \citep{Beu08,Mendoza:2015}. Last but not least, the possible synthesis of long-lived superheavy elements in a hypothetical island of stability, has been a long standing question \citep{Goriely2015}.

The relevant nuclides with more neutrons than $^{238}$U, as well as neutron-rich superheavy elements, cannot be produced by fission, spallation, or fragmentation, the most commonly employed production mechanisms. Alternative approaches are therefore needed to address this particular problem. One approach are deep-inelastic collisions resulting in the exchange of multiple nucleons between two heavy, neutron-rich nuclei. Such multi-nucleon transfer (MNT) reactions \cite{Dasso94,hirayama2014,watanabe2015i} have recently attracted renewed attention as a possible path to the synthesis of new neutron-rich heavy nuclei. For example \cite{Vogt2015} recently revisited $^{136}$Xe$+^{238}$U induced reactions at INFN Legnaro using modern tools of high-resolution $\gamma$-spectroscopy in connection with the PRISMA spectrometer and showed that uranium isotopes out to $^{240}$U can be produced and studied. Recent theoretical predictions show that with $^{144}$Xe+$^{238}$U induced reactions new neutron-rich nuclei in the $Z=88-96$ region could be reached with reasonable production cross sections (assuming a sufficiently intense $^{144}$Xe beam could be produced by fission or fragmentation) \citep{Zhu2017}. 

Multi-nucleon transfer reactions are also of interest for enhancing the production of $N=126$ nuclei, which are critical for understanding the $r$-process. At RIKEN the KEK Isotope Separation System (KISS) facility is intended to take advantage of this production mechanism \cite{watanabe2013,hirayama2014}. KISS aims at extending $r$-process studies at RIKEN/RIBF to the $N=126$ nuclei relevant for the formation of the third $r$-process abundance peak at $A=195$. KISS is designed to provide radioactive nuclei produced via multi-nucleon transfer reactions \cite{Dasso94,watanabe2015i}. The reaction products are captured in a gas catcher, and reionized using a $Z$-selective laser-ion source. The system is being tested and will be in operation soon.

The “N=126 Factory” being developed at Argonne National Laboratory (ANL) is comparable to KISS. The N=126 Factory will similarly collect MNT products in a gas cell, but will differ significantly from KISS in its use of a large volume, helium-filled gas cell to stop and extract products in already-ionized form rather than laser-ionizing the neutralized products. The former scheme has already been used with great success at ANL’s CARIBU facility. Upon leaving the gas cell, the guided radioactive beam will be separated using a mass analyzing magnet before being cooled and bunched in a radio-frequency quadrupole ion trap. As beam purity is critical for high precision mass measurements, a multi-reflection time-of-flight mass spectrograph (MR-TOF) being developed at the University of Notre Dame will be used to separate isobars before their injection into the Canadian Penning Trap (CPT). The MR-TOF could also be used for mass measurements when either rates are too low or half-lives are too short for the Penning Trap approach \cite{schultz2016}. 

It remains to be seen if these techniques can be used to reach $r$-process nuclei beyond $N=126$ in the future. 





\section{Outlook}
\label{sec_summary}

This is an exciting time to study $r$-process nucleosynthesis.  The coming decade will experience an unprecedented confluence of major capability leaps in nuclear experimental, observational, theoretical, and computational science that will put a comprehensive theory for the origin of the elements, including the $r$-process, within reach. A new generation of rare isotope facilities such as RIKEN/RIBF, FRIB, FAIR, and others promise for the first time to produce and study a majority of the nuclei that are part of the $r$-process. Nuclear theory continues to make progress towards self-consistent microscopic predictions of $r$-process nuclear properties and the dense matter equation of state. Multi-messenger observations of the transient universe, including gravitational waves, neutrinos, and photons will provide direct information on explosive nucleosynthesis sites. Guided by today's large scale stellar surveys, a new generation of optical telescopes, including 4MOST and E-ELT will map the chemical enrichment history of our Galaxy. Computational capabilities have reached a point where multi-D simulations of supernovae, neutron star mergers, and other $r$-process sites have enough fidelity to model nucleosynthesis and confront the predictions with observational data, and enable chemical evolution models that put nucleosynthesis into a cosmic context. 

This era has already begun with the historic observation of the NS merger GW170817 that establishes mergers as an important $r$-process site, the RIKEN/RIBF rare isotope facility coming online, first $r$-process calculations based on DFT nuclear theory data, and the observation of $r$-process enriched metal-poor stars in dwarf galaxies. At this stage, progress is reflected in the large number of new open questions that have emerged: Is GW170817 a typical merger?  What is the rate of these mergers?  What elements do they create and how much do they contribute to the total inventory of $r$-process elements?  Can we identify particular $r$-process elements in future mergers?  Are there additional $r$-process sites and what are they? Why is the elemental abundance pattern of the heavy $r$-process so robust, and why does this robustness not extend to the actinides? How can one enrich a dwarf galaxy with $r$-process elements? 

Nuclear data will be essential for answering many of these questions, in particular to predict the specific elements that are created in an observed astrophysical environment, and to connect observed abundances and kilonova features back to astrophysical conditions and constraints on the nucleosynthesis site(s). 
With a new generation of rare isotope beam facilities many $r$-process nuclides will come into experimental reach for the first time. A broad range of beam capabilities -- fast, stopped, reaccelerated, and stored beams -- will be needed to address the broad range of nuclear physics that enters $r$-process models, including properties and reactions of extremely neutron-rich nuclei and the nuclear equation of state. Most of the techniques and instrumentation that will be needed have been developed over the last decade. Major experimental challenges remain to constrain neutron capture rates far from stability, and to study the production of neutron-rich actinides and superheavy elements in the $r$-process. 

Most of the structural input pertaining to nuclei that enter $r$-process element
abundance calculations come from theoretical models.  There has been exciting progress in global theoretical modeling throughout the nuclear chart. The grand perspective is to  use one consistent theoretical framework to compute all nuclear properties needed for $r$-process network calculations. A microscopic tool that is well suited to provide quantified microphysics is nuclear density functional theory employing a validated in-medium effective interaction (energy density functional), which can be used in calculations for both finite nuclei and nuclear matter (including the equation of state for neutron stars).  This approach is capable of predicting a variety of observables needed, and -- when aided by high-performance computing and statistical tools -- is able to assess the uncertainties on those observables. Such a capability is indispensable in the context of making reliable extrapolations in isospin and mass into the regions of $r$-process where experimental data are not available.

Multi-messenger observations will be key to confront nuclear-based astrophysical models of the $r$-process, to identify new $r$-process sites, and to determine their frequency and ejected material. When Advanced LIGO and Advanced VIRGO reach their designed sensitivity we expect to detect neutron star mergers at a significantly faster rate and should accumulate observations of several to several tens of events. The importance for $r$-process science of an effective network of ground- and space-based observatories to follow up any gravitational wave trigger from a neutron star merger was impressively demonstrated in the case of GW170817. Major progress is also on the horizon for observations of stellar spectra with major new capabilities under development, for example the E-ELT 39m telescope in Europe. Key elements for addressing $r$-process science are high-resolution spectrographs covering blue wavelengths, and large throughput using fiber-fed instruments that enable simultaneous observations of a large number of stars. The goal is a much enlarged non-biased stellar sample of neutron capture element abundances that can be confronted with advanced chemical evolution models. High resolution observations will also enable the extraction of isotopic abundances for some elements, such as Li, C, N, O, Mg, Ba, (Nd, Sm) and Eu for a large number of stars. This is currently only possible in small samples and will be one of the great advances in observations in the next decade that will help trace the $r$-process. Advances in atomic physics and stellar photosphere modeling will be needed to take full advantage of these new observations. 

Computational models of nucleosynthesis sites, for the $r$-process primarily core collapse supernovae and neutron star mergers, are essential for understanding the $r$-process. In all types of scenarios, realistic neutrino interactions and a broader range of dense matter equations of state are key for modeling the $r$-process.  For supernova models it will be important to reliably predict the electron fraction of the neutrino-driven wind. Neutron star merger simulations have advanced dramatically, driven by advanced computational capabilities that enable the necessary 3D simulations, and the GW170817 gravitational wave detection that provides new motivation for modeling these events.  However much remains to be done. Future high resolution magneto-hydrodynamical models with appropriate treatment of neutrino effects will 
provide tighter constraints on outflow properties and ejecta masses and their dependency on binary system parameters. The nucleosynthesis yields as well as light curve predictions will benefit from models that consistently connect the different phases of mass ejection. Such models are a prerequisite for the interpretation of observations and they may serve as input in galactic evolution models.

Realistic sensitivity studies that quantify the connection between nuclear physics and astrophysical observables are essential for guiding nuclear physics efforts, and to arrive at a full understanding of the mechanism of element formation in the cosmos. Such studies need to be expanded to a broader range of realistic models, and need to target a full range of possible observables. 

Last but not least, as this workshop summary demonstrated, addressing the $r$-process problem requires close links, collaboration, and rapid interaction between experimental and theoretical nuclear physics, astronomy, and computational modeling. Much has been achieved in this respect through coordinated efforts such as the Joint Institute for Nuclear Astrophysics, but more needs to be done to efficiently and rapidly synthesize progress in the various subfields expected in the coming decade into a comprehensive theory of element formation that is consistent with the full body of experimental and observational data.

\ack
This research was enabled by a workshop jointly organized by the Joint Institute for Nuclear Astrophysics Center for the Evolution of the Elements (JINA-CEE), supported by the National Science Foundation under grant no. PHY-1430152, and the International Collaborations in Nuclear Theory program.  

W.N. is supported by the U.S. Department of Energy under Award Numbers DOE-DE-NA0002847 (the Stewardship Science Academic Alliances program), DE-SC0018083 (NUCLEI SciDAC-4 collaboration), and 
DE-SC0013365 (Michigan State University);

C. H. is supported by the U.S. Department of Energy under Award Numbers DOE-DE-FG02-87ER40365 (Indiana University) and DE-SC0018083 (NUCLEI SciDAC-4 collaboration);

H.S. is supported by the US National Science Foundation under grant no. PHY-1102511;

A.A. was supported by the Helmholtz-University Young Investigator grant No. VH-NG-825, Deutsche Forschungsgemeinschaft through SFB 1245, and European Research Council through ERC Starting Grant No.677912 EUROPIUM

I.D. is supported by the Canadian NSERC Discovery Grants SAPIN-2014-00028 and RGPAS 462257-2014.

J.M-T. is supported by a Mexican grant under the project UNAM-DGAPA/PAPIIT IV100116.

R.R. is supported by the European Research Council under the European Unions's Seventh Framework Programme (FP/2007-2013) / ERC Grant Agreement n. 615126.

B.C. acknowledges support from the ERC Consolidator Grant (Hungary) funding scheme (project RADIOSTAR, G.A. n. 724560).

B.W.O. was supported by the National Aeronautics and Space Administration (NASA) through grant NNX15AP39G and Hubble Theory Grant HST-AR-13261.01-A, and by the NSF through grant AST-1514700.

Support for Francois Foucart was provided by NASA through Einstein Postdoctoral Fellowship grants numbered PF4-150122 awarded by the Chandra X-ray Center, which is operated by the Smithsonian Astrophysical Observatory for NASA under contract NAS8-03060

X.D.T. and M.W. are supported by the Chinese National Key Research and Development program (MOST 2016YFA0400501 and 2016YFA0400504). X.D.T. also acknowledges the supports from the National Natural Science Foundation of China under Grant No. 11475228 and 11490564 and 100 talents Program of the Chinese Academy of Sciences.

This material is based upon work supported by the Department of Energy/National Nuclear Security Administration under Award Number(s) DE-NA0003221 and DE-NA0003180.  The work of S. L. was supported by the National Science Foundation under Grants No. PHY 1102511 (NSCL)”

This work was supported by the US Department of Energy through the Los
Alamos National Laboratory and has been assigned report number LA-UR-
18-22069. Los Alamos National Laboratory is operated by Los Alamos
National Security, LLC, for the National Nuclear Security
Administration of US Department of Energy (Contract DEAC52-06NA25396).

A.A., M.B., and J.K. are supported by the U.S. National Science Foundation under grant number PHY-1713857.

O.J. acknowledges support by the European Research Council through grant ERC AdG 341157-COCO2CAS and by the Max-Planck–Princeton Center for Plasma Physics (MPPC).

J.A.C. is supported by the U.S. Department of Energy, Office of Nuclear Physics, under Contract No. DE-AC02-06CH11357.

A.K. acknowledges the support from the Academy of Finland (grants No. 275389, 284516 and 312544).
\clearpage
\bibliography{references}

\end{document}